\documentclass[useAMS,usenatbib,fleqn]{mnras}

\usepackage[utf8]{inputenc}
\usepackage{graphicx,color}
\usepackage{mathtools}	
\usepackage{amsmath}
\usepackage{amssymb}
\usepackage{bm}
\usepackage{fixltx2e} 
\usepackage{pstricks} 
\usepackage{arydshln} 
\usepackage[normalem]{ulem}

   \usepackage[british]{babel}             
   \usepackage{newtxtext}                  
   \usepackage[slantedGreek]{newtxmath}    
   \usepackage[T1]{fontenc}                

\def\apj{ApJ}

\def\mnras{MNRAS}
\def\aap{A\&A}
\def\apjl{ApJ}

\def\apjs{ApJS}
\def\pasa{PASA}

\def\aj{AJ}

\def\aapr{ARA\&A}

\newcommand{\bfv}{\bm{v}}

\newcommand{\bfx}{\bm{x}}

\newcommand{\1}{_\mathrm{1}}					   	
\newcommand{\core}{_\mathrm{1,c}}					   	
\newcommand{\env}{_\mathrm{e}}					   	
\newcommand{\2}{_\mathrm{2}}					   	

\newcommand{\gas}{_\mathrm{gas}}

\newcommand{\oneenv}{_\mathrm{1,e}}

\newcommand{\soft}{_\mathrm{soft}}
\newcommand{\softf}{_\mathrm{soft,0}}

\newcommand{\Gn}{\mathrm{G}}
\newcommand{\init}{_\mathrm{i}}
\newcommand{\final}{_\mathrm{f}}

\newcommand{\Rsol}{\,\mathrm{R_\odot}}
\newcommand{\Msol}{\,\mathrm{M_\odot}}
\newcommand{\Rsun}{\,\mathrm{R_\odot}}
\newcommand{\Msun}{\,\mathrm{M_\odot}}

\newcommand{\amb}{_\mathrm{amb}}

\newcommand{\rmD}{\mathrm{\Delta}}
\newcommand{\rmdelta}{\mathrm{\delta}}
\newcommand{\CE}{_\mathrm{CE}}

\newcommand{\refine}{_\mathrm{ref}}

\newcommand{\rmd}{\mathrm{d}}

\newcommand{\unb}{_\mathrm{unb}}

  \newcommand{\cm}{\,{\rm cm}}

  \newcommand{\erg}{\,{\rm erg}}
  
  \newcommand{\gcmcmcm}{\,{\rm g\,cm^{-3}}}

  \newcommand{\kms}{\,{\rm km\,s^{-1}}}

  \newcommand{\yr}{\,{\rm yr}}     
  \newcommand{\da}{\,{\rm d}}     
  \newcommand{\dynecmcm}{\,{\rm dyn\,cm^{-2}}}     
  \newcommand{\au}{\,{\rm au}}     
  
\title[Energy Budget in Common Envelope Evolution] 
{Energy Budget and Core-Envelope Motion in Common Envelope Evolution}
\author[L.~Chamandy et al.]{Luke Chamandy,$^{1}$\thanks{lchamandy@pas.rochester.edu}
Yisheng Tu,$^{1}$ \thanks{ytu7@u.rochester.edu}
Eric G.~Blackman, $^{1}$\thanks{blackman@pas.rochester.edu}
Jonathan Carroll-Nellenback,$^{1}$ \thanks{jcarrol5@ur.rochester.edu}
\newauthor 
Adam Frank,$^{1}$\thanks{afrank@pas.rochester.edu}
Baowei Liu$^{1}$ \thanks{baowei.liu@rochester.edu}
and Jason Nordhaus$^{2,3}$ \thanks{nordhaus@astro.rit.edu}
\\
$^{1}$Department of Physics and Astronomy, University of Rochester, Rochester NY 14627, USA\\
$^{2}$National Technical Institute for the Deaf, Rochester Institute of Technology, NY 14623, USA\\
$^{3}$Center for Computational Relativity and Gravitation, Rochester Institute of Technology, NY 14623, USA
}

\begin{document}


\maketitle

\begin{abstract}
We analyze a 3D hydrodynamic simulation of common envelope evolution 
to understand how energy is transferred between various forms and whether
theory and simulation are mutually consistent given the  setup.
Virtually all of the envelope unbinding in the simulation
occurs before the end of the rapid plunge-in phase, 
here defined to coincide with the first periastron passage.
In contrast, the total envelope energy is nearly constant during this time 
because positive energy transferred to the gas from the core particles is counterbalanced by 
the negative binding energy from the closer proximity of the inner layers to the plunged-in secondary.
During the subsequent slow spiral-in phase, 
energy continues to transfer to the envelope from the red giant core and secondary core particles. 
We also propose that relative motion between the centre of mass of the envelope 
and the centre of mass of the particles could account for the offsets of planetary nebula central stars
from the nebula's geometric centre.
\end{abstract}

\defcitealias{Chamandy+18}{Paper~I}

\section{Introduction}
\label{sec:intro}

In a binary stellar system, 
common envelope evolution (CEE) occurs  when the envelope of a primary star, usually a giant, engulfs a smaller companion.
Many astrophysical phenomena are believed to be preceded by one or more common envelope (CE) phases.
Examples include asymmetric and bipolar planetary nebulae (PNe) and pre-PNe (PPNe), 
black hole (BH)-BH and neutron star (NS)-NS mergers, 
high- and low-mass X-ray binaries, and likely type Ia supernovae (SNe) \citep[see][for a recent review]{Ivanova+13a}.
Many observed binary systems have such small binary separations that they must be post-CE systems.

The so-called ``energy formalism'' (EF) was developed to predict the fate of a given binary system undergoing CEE 
and is useful for population synthesis studies
\citep{Vandenheuvel76,Tutukov+Yungelson79,Livio+Soker88,Dekool90,Dewi+Tauris00}.
In this prescription, the two possible fates of CEE are merger or envelope ejection, 
depending on the value of an efficiency parameter $\alpha\CE$, which is poorly constrained
and cannot be reliably estimated from simulations if the envelope is not completely unbound (i.e. ejected).  
Thus far, 3D hydrodynamical simulations have yet to result in an ejected envelope 
unless an additional energy source (recombination energy) is introduced.
However, the role of recombination is not yet universally agreed upon, 
in part because the released energy may be radiated away 
before it can be absorbed to contribute much to envelope ejection.
In general, the absence of envelope ejection in simulations 
may also involve some combination of limitations of the theory (unjustified approximations, missing physics)
or limitations of the simulations (unrealistic initial conditions, small duration, limited resolution, missing physics).

Due to the  complexity and 3D morphology of CEE, global 3D models are useful.
Early 3D hydrodynamical simulations of CEE were performed by 
\citet{Livio+Soker88}, followed by \citet{Rasio+Livio96}, 
who used smoothed particle hydrodynamics (SPH) 
and by \citet{Sandquist+98} and \citet{Sandquist+00},
who used a finite difference code with nested grids.
These papers analyzed  the global energy budget
and measured the amount of bound mass versus time.

More recent papers exploring the energy budget and mass unbinding include \citet{Passy+12b}, 
using a SPH code,  \citet{Ricker+Taam12}, using adaptive mesh refinement (AMR), 
and \citet{Ohlmann+16a}, using a moving mesh code.
Our initial conditions herein closely match those of the latter to facilitate comparisons.

\citet{Iaconi+17} reviewed  all previous simulations and compared SPH and AMR results, 
including a compilation of  unbound mass for each simulation.
Both \citet{Iaconi+17} and \citet{Iaconi+18} found that the unbound mass 
can increase as the resolution is enhanced in both AMR and SPH simulations.
\citet{Iaconi+18} also  found that the final fraction of unbound mass
is generally larger for less massive envelopes or  more massive secondaries.

The main goal of this work is to analyze 
the various energy terms in our simulation as accurately as possible as an example
to assess whether simulation and theory are mutually consistent given the choices of the setup.
In doing so, we also shed light on the envelope ejection process.
Specifically we address: how does the energy transition from one form to another with time?
What are the expectations for envelope removal and energy transfer from analytic theory based on the EF?
Do these expectations agree with simulation results?
What strategies should be prioritized to achieve envelope ejection in future simulations?

In Sec.~\ref{sec:simulation_overview} we describe the simulation methods and setup.
We analyze the global energy budget  in Sec.~\ref{sec:energy_budget}.
In Sec.~\ref{sec:unbinding}, we explore how and to what extent the envelope becomes unbound.
Sec.~\ref{sec:particle_CM_motion} focusses on the relative motion between the gas and particles,
its effect on envelope unbinding, and implications  for explaining  observed offsets
of some PN central stars from the geometric centres of their  nebulae.
In Sec.~\ref{sec:energy_formalism}, we apply the EF
to interpret our simulation results. 
We conclude in Sec.~\ref{sec:conclusions}.

\section{Simulation overview}
\label{sec:simulation_overview}
The simulation that we analyze here is Model~A of \citet{Chamandy+18} (hereafter \citetalias{Chamandy+18}), 
which involves the interaction of a $2.0\Msol$ red giant (RG) primary with a $0.4\Msol$ point particle core
and a $1.0\Msol$ point particle representing a white dwarf (WD) or main sequence (MS) secondary.
Unlike Model~B of that paper, Model~A did not have a subgrid model for accretion onto the secondary
and Model~A is simpler in that respect.
See \citetalias{Chamandy+18} for a  detailed description of the simulation setup which we summarize here.

The 3-D hydrodynamic simulation utilized the 3D AMR multi-physics code AstroBEAR \citep{Cunningham+09,Carroll-Nellenback+13},
and accounts for all gravitational interactions (particle--particle, particle--gas, and gas--gas).
The RG density and pressure profiles were set up by first running a 1D simulation 
with the code Modules for Experiments in Stellar Astrophysics (MESA) \citep{Paxton+11,Paxton+13,Paxton+15},
selecting a snapshot corresponding to the RGB phase, 
and then adapting it to the resolution of our 3D simulation (\citetalias{Chamandy+18}, and also see \citealt{Ohlmann+17}).
A comparison of the radial profiles of the RGB star in our initial condition with those of the MESA model used
is presented in Appendix~\ref{sec:profile_comparison}.

The stars are initialized in a circular orbit at $t=0$ with orbital separation $a|_{t=0}=49.0\Rsol$, 
slightly larger than the RG radius of $R_1=48.1\Rsol$. 
To focus our computational resources on the post-plunge evolution, 
we chose not to start the simulation at an earlier phase of evolution, e.g. Roche lobe overflow.
The simulation  was terminated at $t=40\da$.
The mesh was refined at the highest level with voxel dimension $\delta=0.14\Rsol$ before $t=16.7\da$ and $\delta=0.07\Rsol$ thereafter 
\textit{everywhere} inside a large spherical region centered on the point particles.
The initial radius of this maximally resolved region was $r\refine= 72\Rsol$ and at all times $r\refine > 2.5a$.
The spline softening radius for both particles was set to $r\soft\approx17\delta$ for the entire simulation.
The base resolution used was $2.25\Rsun$, 
and a buffer zone of $16$ cells allowed the resolution to transition gradually between base and highest resolution regions.
To assess the effects of changing $\delta$, $r\refine$ or $r\soft$, we performed some additional runs.
The results of these convergence tests are presented in Appendix~\ref{sec:convergence}.

The box dimension is $1150\Rsun$ and no envelope material reaches the boundary by the end of the simulation.
We use extrapolating hydrodynamic boundary conditions  and there is a small inflow during the simulation
that is fully accounted for in the analysis.
The ambient pressure is $1.0\times10^{5}\dynecmcm$, 
so chosen to truncate the pressure profile of the primary and resolve the small pressure scale height at the surface.
To prevent a large ambient sound speed and 
an unreasonably small time step, the ambient density is set to $6.7\times10^{-9}\gcmcmcm$. This 
equals the density at the primary surface.
Effects of the ambient gas are discussed throughout the text (see also \citetalias{Chamandy+18})
and do not affect the main conclusions of this work.

\section{Energy budget}
\label{sec:energy_budget}
\begin{figure*}
  \includegraphics[width=\textwidth,clip=true,trim= 0 0 0 0]{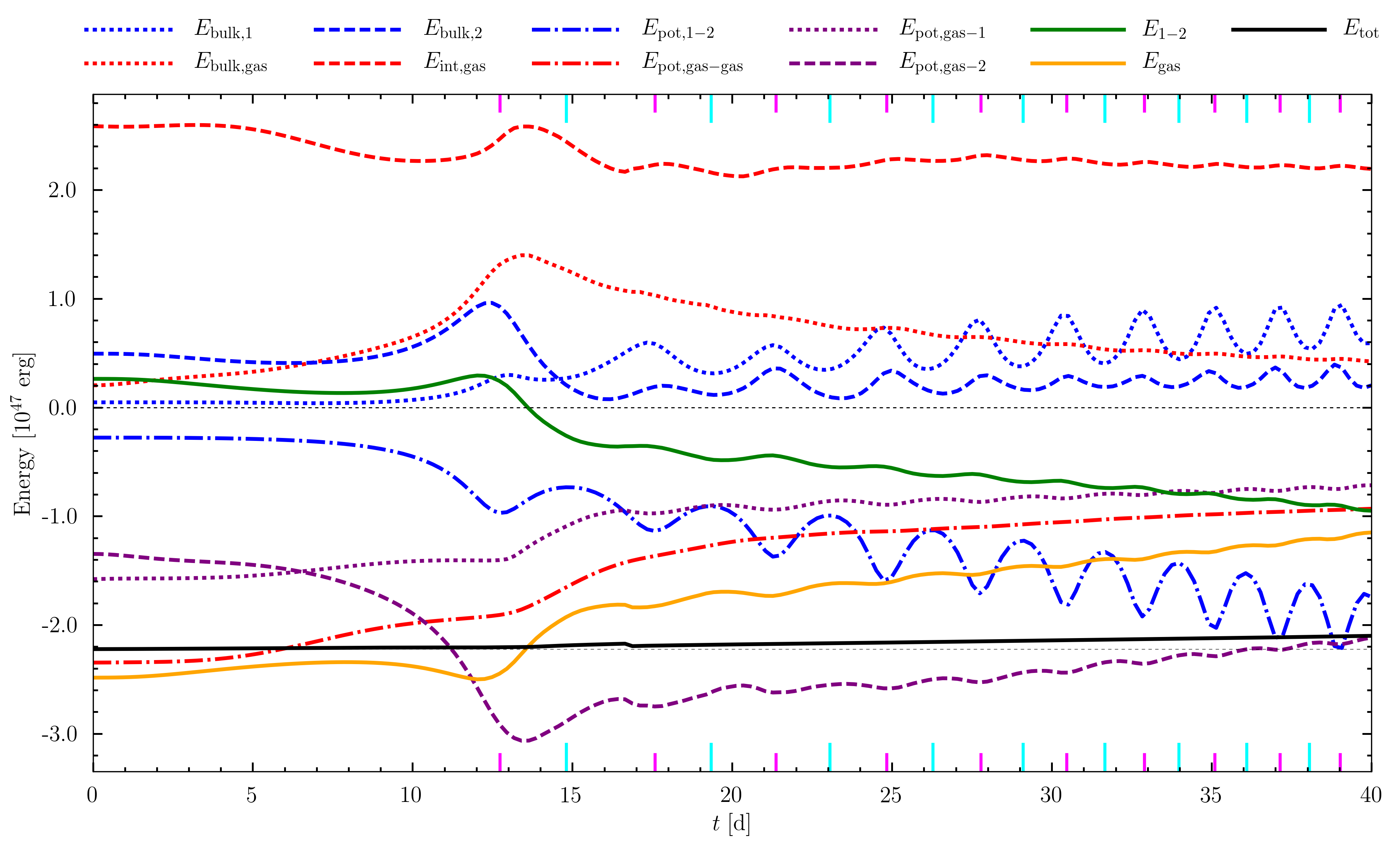}\\
  \includegraphics[width=\textwidth,clip=true,trim=   0 0 0 0]{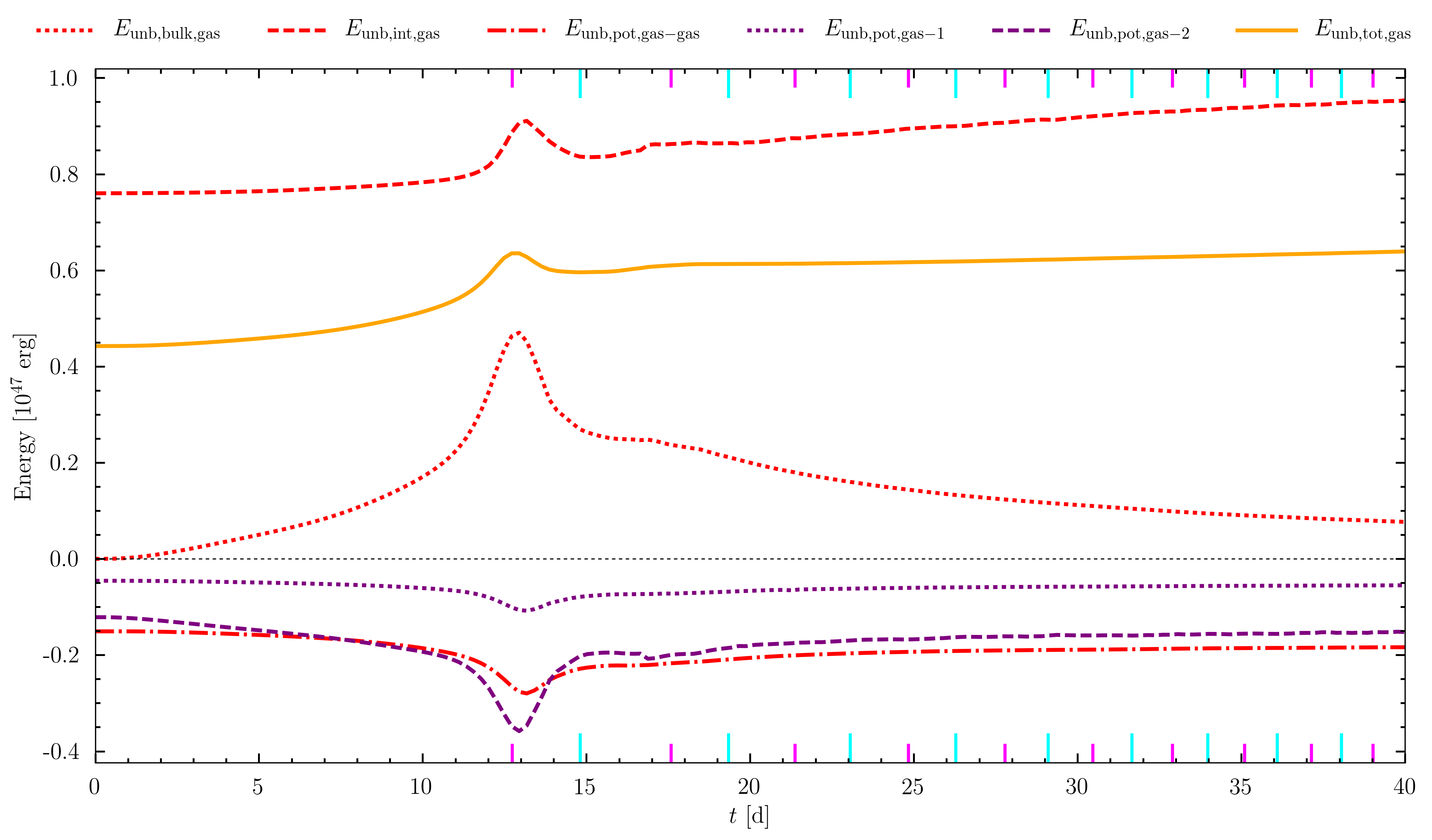}
  \caption{Top: Evolution of the various energy components integrated over the simulation domain
           (see Tab.~\ref{tab:energy_terms} for symbol definitions).
           Those terms involving only particles are plotted in blue, 
           only gas in red, and those involving gas and particles are plotted in mauve.
           Total energy is shown by a black line, with its initial value plotted as a grey dashed line for reference. 
           Green and orange solid lines show the total particle and gas energies, respectively, 
           with terms involving both particles and gas counting toward the total gas energy.
           A discontinuity at $t=16.7\da$ is caused by the change in the spline softening length of both particles from $2.4\Rsol$ to $1.2\Rsol$.
           The sampling rate of the data plotted is about one frame every $0.23\da$.
           Times of apastron and periastron passage are shown as long cyan and short magenta tick marks, respectively.
           \textit{Bottom:} As in the top panel but now showing the energy of the unbound gas only,
           where `unbound' is defined as $\mathcal{E}\gas\ge 0$.
           Note the difference in vertical axis range compared to the top panel.
           \label{fig:energy_time}
          }            
\end{figure*}

\begin{table*}
  \begin{center}
  \caption{Terms in the energy budget, in units of $10^{47}\erg$, integrated over the simulation domain. 
           Values are shown for the start of the simulation at $t=0$, end of plunge-in at $t=13\da$, 
           and end of the simulation at $t=40\da$.
           Also shown are the changes in the values between these times.
           For ease of presentation, expressions involving the gas--particle interactions neglect the modification 
           of the particle potential at distances from the particle less than the spline softening length,
           but the values quoted were obtained using the full expressions for the potential.
           \label{tab:energy_terms}
          }
  \begin{tabular}{lllrrrrrr}
    \hline
Energy component		&Symbol			&Expression		&$t=0$	&$t=13\da$&$t=40\da$&$\rmD E_{0-13\da}$&$\rmD E_{13-40\da}$&$\rmD E_{0-40\da}$\\
\hline                  	 
Particle~1 kinetic		&$E_\mathrm{bulk,1}$	&$\tfrac{1}{2}M\core v\core^2$	&$0.05$	&$0.30$	&$0.59$	&$0.25$		&$0.29$		&$0.54$		\\
Particle~2 kinetic		&$E_\mathrm{bulk,2}$	&$\tfrac{1}{2}M\2v\2^2$	&$0.49$	&$0.86$	&$0.20$	&$0.37$		&$-0.66$	&$-0.29$		\\
Particle-particle potential	&$E_\mathrm{pot,1-2}$	&$-\Gn M\core M\2/a$	&$-0.28$&$-0.96$&$-1.74$&$-0.69$	&$-0.78$	&$-1.46$		\\
\hline
Gas bulk kinetic		&$E_\mathrm{bulk,gas}$	&$\tfrac{1}{2}\int\rho(\bfx)v\gas^2(\bfx)\rmd V$	&$0.20$	&$1.35$	&$0.42$	&$1.15$	&$-0.93$&$0.22$	\\
Gas internal			&$E_\mathrm{int,gas}$	&$\tfrac{3}{2}\int P(\bfx)\rmd V$			&$2.59$	&$2.53$	&$2.19$	&$-0.06$&$-0.33$&$-0.39$\\
Gas-gas potential		&$E_\mathrm{pot,gas-gas}$&$\tfrac{1}{2}\int\Phi\gas(\bfx)\rho(\bfx)\rmd V$	&$-2.35$&$-1.90$&$-0.93$&$0.45$	&$0.97$	&$1.41$	\\
Gas-particle~1 potential	&$E_\mathrm{pot,gas-1}$	&$-\Gn M\core\int(\rho(\bfx)/|\bfx-\bfx\core|)\rmd V$	&$-1.58$&$-1.40$&$-0.72$&$0.18$	&$0.68$	&$0.86$	\\
Gas-particle~2 potential	&$E_\mathrm{pot,gas-2}$	&$-\Gn M\2\int(\rho(\bfx)/|\bfx-\bfx\2|)\rmd V$	&$-1.35$&$-2.99$&$-2.12$&$-1.64$&$0.87$	&$-0.77$\\
\hline
Particle total			&$E_\mathrm{1-2}$	&$E_\mathrm{bulk,1}+E_\mathrm{bulk,2}+E_\mathrm{pot,1-2}$	&$0.26$&$0.20$&$-0.95$&$-0.06$	&$-1.15$&$-1.21$\\
Gas total			&$E\gas$	&$E_\mathrm{bulk,gas}+E_\mathrm{int,gas}+\sum_j E_\mathrm{gas-j}$&$-2.49$&$-2.40$&$-1.15$&$0.08$&$1.25$	&$1.33$	\\
\hline
Total				&$E_\mathrm{tot}$	&$E_\mathrm{1-2}+E\gas$					&$-2.22$&$-2.20$&$-2.10$&$0.02$	&$0.10$	&$0.12$	\\
\hline
  \end{tabular}
  \end{center}
\end{table*}

We write the total energy $E_\mathrm{tot}=E_{1-2}+E_\mathrm{gas}$.
Here $E_{1-2}$ is particle orbital energy and is equal to  
the sum of particle kinetic energy $E_\mathrm{bulk,1}+E_\mathrm{bulk,2}$ 
and mutual potential energy of particles $E_\mathrm{pot,1-2}$. 
Complementally, the \textit{binding} energy $E\gas$ is defined as the sum of the gas-particle potential energy  $E_\mathrm{pot,gas-1}+E_\mathrm{pot,gas-2}$
plus the gas-only contribution $E_\mathrm{bulk,gas}+E_\mathrm{int,gas}+E_\mathrm{pot,gas-gas}$.
Since the binding energy is negative,
transferring energy from $E_{1-2}$ to $E\gas$ would make $E\gas$ less negative, and the gas less bound.
In what follows, `increase' of energy means toward more positive (or less negative)  values,
while `decrease' means toward less positive (or more negative) values.  

To alleviate  confusion from previous literature, 
we  include gas-particle potential energy and gas bulk kinetic energy in $E_\mathrm{gas}$, 
\textit{not} in $E_{1-2}$.
By doing so, we can characterize the problem in terms of transfer between ``particle energy'' and ``gas energy''.
The theory  is  addressed further in Sec.~\ref{sec:energy_formalism} but here we focus on simulation results.

In the top panel of Fig.~\ref{fig:energy_time} we show the time-evolution of each energy component integrated over the simulation domain.
Apastron and periastron passages are labeled on the time axis  by long cyan or short magenta  tick marks respectively.
Expressions for the various contributions and their values at $t=0$, $t=13\da$ and $t=40\da$, 
are given in Tab.~\ref{tab:energy_terms}.
Time $t=13\da$ is  approximately that  of first periastron passage and
conveniently  delineates the transition  between the end of the plunge-in phase and the beginning of the slow in-spiral.%
\footnote{These phases are \textit{loosely} equivalent to the dynamical plunge-in 
and self-regulating spiral-in phases discussed in \citet{Ivanova+13a}.}
The inter-particle separation evolves from $a=49.0\Rsun$ at $t=0$ 
to $a=14.1\Rsun$ at $t=13\da$ and $a=7.8\Rsun$ at $t=40\da$ 
\citepalias{Chamandy+18}.

A key result from Fig.~\ref{fig:energy_time} (top) and Tab.~\ref{tab:energy_terms}
is that the potential energy term $E_\mathrm{pot,gas-2}$
is important even when the secondary is situated outside the RG surface at $t=0$
and by the end of plunge-in at $t=13\da$ comprises almost half of 
the gas potential energy.
We note that for a more realistic initial condition 
for which the initial separation is equal to the Roche limit separation of $\sim109\Rsun$, 
this term would be less important initially, 
and the initial energy of the particles and envelope would be less negative by about $0.4\times10^{47}\erg$
(this is discussed in detail in Sec.~\ref{sec:energy_formalism}).

The net energy transferred to $E\gas$ from $t=0$ until $t=13\da$ is negligible
even though almost all of the envelope unbinding 
occurs during this time (Sec.~\ref{sec:unbinding}).
The reason is that although the plunge-in of the secondary violently disrupts and energizes the outer layers of the envelope, 
it moves the secondary deeper in the envelope and more tightly binds it.
The gain in gas kinetic energy is offset by the potential energy becoming more negative, 
and therefore negligible net exchange between particle energy 
and gas energy from the start of the simulation up to the end of plunge-in.
 
Complementally, owing to the  continuous and highly variable gravitational force 
exerted on the particles by the gas, 
the total particle kinetic energy increases by almost as much 
as their mutual potential energy decreases between $t=0\da$ and $t=13\da$, 
resulting in almost zero net change in particle orbital energy.

Subsequently, after $t\approx15\da$, energy is transferred from  particle energy 
to gas energy at a roughly constant rate as the particles spiral in closer together.
We estimate that up to $\sim0.3\times10^{47}\erg$ of the intial 
energy in the particles and envelope is transferred to the ambient gas during the simulation.
In the subsections below we expand on these points and discuss each of the curves in the top panel of Fig.~\ref{fig:energy_time} in detail.

\subsection{Total energy}
\label{sec:total_energy}
The total energy is plotted in solid black in Fig.~\ref{fig:energy_time} 
and changes by $5\%$ between $t=0$ and $t=40\da$
(a dotted horizontal grey line shows the initial value for reference).
The total energy rises gradually, 
except for a dip after $t=16.7\da$ when the softening radius around both particles
and the smallest resolution element $\delta$, were halved.
This discontinuity is expected because reducing the spline softening radius 
from $r\soft=r\softf$ to $r\soft=r\softf/2$ 
immediately strengthens the gravitational force for $r<r\softf$.
About $16\%$ of the net increase in energy during the simulation 
is caused by inflow of the ambient medium from the domain boundaries.
The remaining error in energy conservation might be caused numerically by the finite time step,
which leads to particle orbits that are not completely smooth,
or by  errors introduced by the multipole Poisson solver.
This small variation in the total energy does not affect the conclusions of the present study.
We note that our initial RGB star is more bound than the 1D MESA model 
by about $50\%$ (App.~\ref{sec:profile_comparison}), and this, 
along with other details of the setup, 
imply that our numerical results do \textit{not} represent
detailed observational predictions.

\subsection{Particle and gas contributions}
\label{sec:net_energy}
The solid green and solid orange lines in the top panel of Fig.~\ref{fig:energy_time} 
show the particle energy $E_\mathrm{1-2}$ and gas energy $E\gas$, respectively.
The quantity $E_\mathrm{1-2}$ is the sum of the quantities shown by the blue curves, 
namely the kinetic energies of both particles and their mutual potential energy.
$E_\mathrm{1-2}>0$ at $t=0$ even though the binary system is bound
because $E_\mathrm{1-2}$ does not include the contribution from $E_\mathrm{pot,gas-2}$.
$E\gas$ is equal to the sum of the quantities shown by the red and mauve curves pertaining 
to gas--only and gas--particle energy terms, respectively. 
The sum $E_\mathrm{1-2}+E\gas=E_\mathrm{tot}$ is shown in solid black.
The green and orange curves show how the orbital energy of the particles 
is gradually transferred to the gas once the plunge-in phase ends. 

To explain the energy evolution in greater detail, 
we now discuss the relationships between individual energy terms.

\subsubsection{Particles}
\label{sec:particles_energy}
Energy terms pertaining to the particles only (RG core primary, 
labeled with subscript `1' and hereafter referred to as `particle~1,' 
and secondary, labeled with subscript `2' and hereafter referred to as `particle~2') are shown in blue.
The kinetic energy of particle~1, $E_\mathrm{bulk,1}$ (dotted blue), first remains steady 
and then gradually rises as the inter-particle separation reduces.
It oscillates approximately synchronously  with the orbit, 
with maxima in kinetic energy coinciding with periastron passages.
The kinetic energy of particle~2, $E_\mathrm{bulk,2}$ (dashed blue), 
first increases during the plunge-in phase from $t=8$ and $t=13\da$.
This $E_\mathrm{bulk,2}$ then decreases as the secondary migrates from having orbited 
the larger RG (core+envelope) mass to orbiting primarily only the smaller mass of particle~1. 
Following this decrease, $E_\mathrm{bulk,2}$ then rises less rapidly than $E_\mathrm{bulk,1}$, 
as there is continued competition between reduced particle separation 
and reduced gas mass interior to the orbit.
Naturally, $E_\mathrm{bulk,2}$  oscillates in phase with  $E_\mathrm{bulk,1}$.

The potential energy of particles $E_\mathrm{pot,1-2}$ 
(which excludes that from gas--particle gravitational forces) is shown in dash-dotted blue,
and its mean value over an orbit period reduces by $1.7\times10^{47}\erg$ between $t=0$ and $t=40\da$.
$E_\mathrm{pot,1-2}$ steadily decreases  at $t=40\da$, 
even as the rate of change of the mean inter-particle separation $\dot{\overline{a}}(<0)$ 
(where bar denotes mean) reduces in magnitude \citepalias{Chamandy+18} so that $\ddot{\overline{a}}>0$.
This behaviour is expected from the $1/a$ Newtonian potential; 
for a circular orbit this gives $\dot{E}_\mathrm{pot,1-2}\propto \dot{a}/a^2$ 
so the decrease in $|\dot{a}|$ competes with the reduction in $a$. 
Whether $\ddot{E}_\mathrm{pot,1-2}$ is positive or negative 
depends on the details of  orbital evolution.

\subsubsection{Gas}
\label{sec:gas_energy}
Energy terms pertaining  to gas only are shown in red in Fig.~\ref{fig:energy_time}.
The total bulk kinetic energy of gas $E_\mathrm{bulk,gas}$ (dotted red)
rises during plunge-in as envelope material is propelled outward 
and then gradually reduces as gravity and shocks decelerate the envelope.  
The dashed red curve shows the internal energy of the gas $E_\mathrm{int,gas}$, 
of which about $0.8\times10^{47}\erg$ comes from the ambient medium. 
The latter has a pressure of $1\times10^5\dynecmcm$ and fills the simulation domain.  
The ambient medium  hardly contributes to $E_\mathrm{bulk,gas}$ however.
$E_\mathrm{int,gas}$ is initially fairly steady, 
but then incurs modest variations from gas expansion and compression.
Both $E_\mathrm{int,gas}$ and $E_\mathrm{bulk,gas}$ show small-amplitude oscillations 
with maxima approximately coinciding with periastron passages.

Each close encounter of the particles dredges up  material in dual spiral wakes.
During plunge-in, the gas acquires mostly bulk kinetic energy, 
but also  significant  internal energy, as  expected from the observed spiral shocks.
The subsequent slow decrease of $E_\mathrm{bulk,gas}$ is accompanied by an increase 
in  potential energy from gas self-gravity $E_\mathrm{pot,gas-gas}$ (dash-dotted red).
Of the total $E_\mathrm{pot,gas-gas}$, 
the gravitational interaction between the ambient medium and envelope 
and  that of the ambient medium with itself respectively 
contribute the relatively small amounts of $0.2\times10^{47}\erg$ and $0.1\times10^{47}\erg$.
Between $t=0$ and $t=40\da$, substantial work ($1.4\times10^{47}\erg$)
is done expanding the envelope against its own gravity.
In principle, unbinding the gas from the particles 
does not  require the gas to become unbound from \textit{itself}, 
but much of the kinetic energy acquired by the envelope in the simulations 
is drained into expansion  against its own self-gravity.

\subsubsection{Gas-particles interaction}
\label{sec:gas_particles_energy}
The mauve curves in Fig.~\ref{fig:energy_time} show the potential energy terms 
accounting for the gas--particle~1 gravitational force $E_\mathrm{pot,gas-1}$ (dotted mauve)
and gas--particle~2 gravitational force $E_\mathrm{pot,gas-2}$ (dashed mauve).
These terms must be included in assessing the extent to which the envelope is bound. 
The ambient medium contributes negligibly to them ($<0.2\times10^{47}\erg$ total).

Initially the inner layers of the RG are hardly affected by interaction with the secondary, 
and so $E_\mathrm{pot,gas-1}$ varies slowly
until the plunge-in, when the inner  envelope is strongly disrupted.
After the end of the plunge-in at $t=13\da$, 
$E_\mathrm{pot,gas-1}$  increases with time as 
work is done to expand the envelope against the gravitational force from particle~1, 
situated roughly at its centre.

The gravitational interaction between gas and particle~2 (dashed mauve) is a bit more subtle  
and not fully accounted for in previous analyses using the CE energy formalism. 
Even at $t=0$, $E_\mathrm{pot,gas-2}$ is important, being almost equal to $E_\mathrm{pot,gas-1}$
because particle~1 is closer to the bulk of the gas even though particle~2 is more massive.
However, as particle~2 plunges toward the envelope centre, 
$E_\mathrm{pot,gas-2}$ increases in magnitude by $1.6\times10^{47}\erg$
and by $t=13\da$ becomes the most important contribution to the gas potential energy.
From $t=0$ until the end of plunge-in,  $E\gas$ gains only $0.1\times10^{47}\erg$, or about $3\%$. 
This highlights that the liberation of orbital energy as particle~2 plunges in does not come ``for free'' 
because when  $M\2$ arrives  close to the envelope centre, 
the envelope is bound inside a much deeper potential well.
During plunge-in, the energy liberated by $E_\mathrm{pot,gas-2}$ becoming more negative
is transferred to the bulk kinetic energy of gas.
Part of this kinetic energy does work to expand the envelope against gravity, 
as evidenced by increases in the gas-particle~1 and gas-gas potential energies.

From $t=13\da$ to $t=40\da$, 
the envelope then expands to become less bound at the expense of the gas kinetic energy (dotted red)
and particle--particle potential energy (dash-dotted blue) 
and significant work is expended on moving gas outward against self-gravity 
and the gravitational forces of particle~2 and particle~1.

\section{Partial envelope unbinding}
\label{sec:unbinding}

\begin{figure}
  \includegraphics[width=\columnwidth,clip=true,trim= 0 0 0 0]{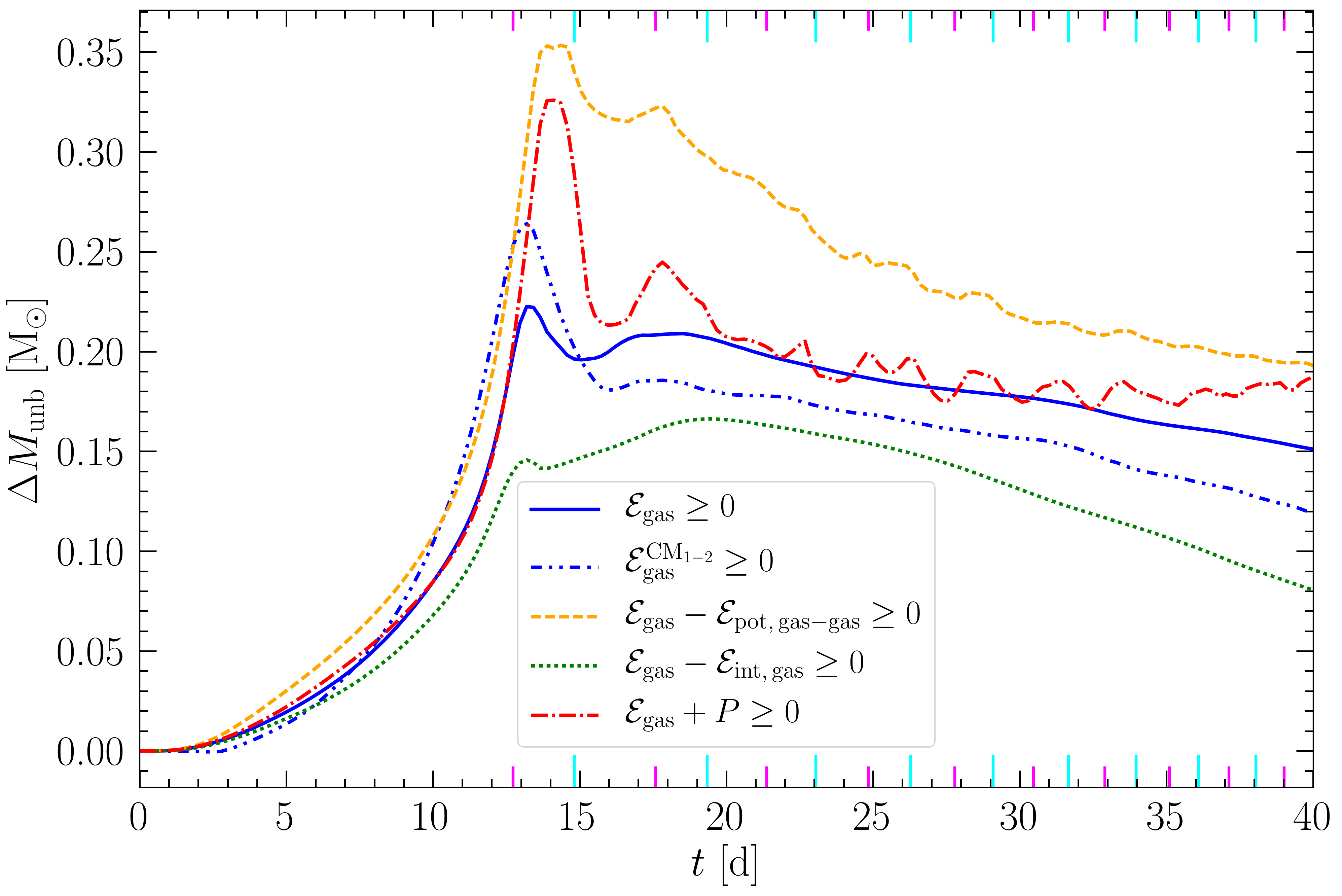}\\
  \caption{Change in unbound mass of the envelope with time according to the fiducial definition of `unbound,' $\mathcal{E}\ge0$ (solid blue),
           as well as various alternative definitions labeled in the legend (see Sec.~\ref{sec:unbinding}).
           \label{fig:unbound_mass}
          }            
\end{figure}

We delineate gas as `unbound' if its total energy density,
$\mathcal{E}\gas=\rmdelta E\gas/\rmdelta V\ge 0$,
where $\mathcal{E}\gas$ equals the sum of bulk kinetic, 
internal and potential (due to self-gravity and interactions with both particles) energy densities.
    
As we explain below,
virtually all of the unbinding of gas occurs 
between the start of the simulation and end of the plunge-in phase.
As will become apparent in Sec.~\ref{sec:spatial}, 
this happens in spite of the negligible total energy transfer between particle orbital energy
and gas binding energy (Sec.~\ref{sec:energy_budget}), 
but can be explained by recognizing
that the nature and relative importance of energy exchange
between  various forms depends strongly on position.
    
\subsection{Unbound mass}
\label{sec:unbound}
In Fig.~\ref{fig:unbound_mass} we plot the change in the unbound mass (defined by $\mathcal{E}\gas\ge0$)
as a function of time (solid blue), 
not including the unbound ambient gas that inflows through the domain boundaries.
Note that most of the ambient medium is already unbound at $t=0$ (about $1.7\Msun$).
Only a small amount of ambient mass is bound at $t=0$ ($<0.04\Msun$),
and while some of this bound mass may become unbound during the course of the simulation, 
its contribution to Fig.~\ref{fig:unbound_mass} would be negligible.
In principle there could also be a small negative contribution from unbound ambient mass becoming bound,
but data from 2D slices suggests that unbound ambient material, 
located at a distance $\gtrsim200\Rsun$ from particle~1, generally remains unbound.
Therefore, the change in unbound mass $\rmD M\unb$ plotted in Fig.~\ref{fig:unbound_mass}
corresponds quite closely to the unbinding of  bound envelope material.

The unbound mass increases from the start of the simulation until the end of plunge-in, 
then reduces slightly, recovers, 
levels off at $t\approx17\da$, and then decreases steadily after $t\approx19\da$.
The steady decrease is caused by energy transferred from the envelope to the ambient medium.
We estimate the total energy transfer in Sec.~\ref{sec:efficiency}.
As the ambient material has fairly large density and pressure in our simulation, 
this decrease is not seen in most other simulations.
In nature, a circumbinary torus formed during the Roche-lobe overflow (RLOF) phase
would be present and likely produce a similar effect.
A peak in the unbound mass near the first periastron and a subsequent levelling off is also seen
in the \textsc{phantom} simulation of \citet{Iaconi+17} 
(see their Fig.~9, top panel for results from the run with the most comparable setup to ours; see also \citealt{Iaconi+18}),
and other CEE simulations \citep{Sandquist+00,Passy+12b,Nandez+14}.
However, in  \citet{Ricker+Taam12}, unbinding peaks near the first periastron
but  continues for several orbits until the end of the simulation.
In  \citet{Sandquist+98}, unbinding occurs around the first periastron, stops, 
and then restarts much later in the evolution (visible as a relative drop in the bound mass as compared 
to the mass that leaves the grid, and most significant for their simulations 4 and 5).

By $t=13\da$ the total change in unbound mass 
is given by $\rmD M\unb\approx 0.22\Msun$ or about $14\%$ of the envelope mass,
and this reduces to $0.21\Msun$ or $13\%$ of the envelope mass
between $t\approx17\da$ and $t\approx19\da$.
This is comparable to the fraction of $13\%$ obtained by \citet{Iaconi+17}.
Using the moving mesh code \textsc{arepo} 
with initial conditions very close to ours,  \citet{Ohlmann+16a}
obtained a value of $8\%$ by the end of their simulation,
and found that most of this was ejected during the first $40\da$. 
The difference between their value and ours is likely 
caused by the slight differences in initial conditions 
(see App.~\ref{sec:ohlmann_comparison} for a detailed comparison,
and a  discussion of differences in the initial conditions).
We note that our initial RGB primary is significantly more strongly bound
than the MESA star from which it is modeled (see App.~\ref{sec:profile_comparison}),
and this alone would be expected to lead to an underestimate of the amount of unbinding.
However, as with all CEE simulations to date, 
other physics, such as convection and radiation transport, are also absent,
and initial conditions are not fully realistic.
We reiterate that the main goal, as stated in the introduction,
is not to predict observations precisely,
but rather to understand and account for energy transfer and envelope unbinding in the simulation,
given the choices of the setup.

Although the gas energy $E\gas$ 
hardly changes between $t=0$ and the end of plunge-in at $t\approx13\da$ 
(and likewise for $E_{1-2}$ since the two are complementary; see Sec.~\ref{sec:energy_budget}),
this is  when most of the envelope unbinding occurs.
Some of the gas gains energy to become less bound, while the remainder loses energy 
to become more bound and the net change is nearly zero.
This happens  as the secondary plunges toward the bulk of material at the centre,
strengthening  its overall pull on the envelope, but also
imparting an impulse to the gas it encounters locally. 
To understand this in more detail, we discuss the spatial variation of the energy density in Sec.~\ref{sec:spatial}.

The other lines in Fig.~\ref{fig:unbound_mass} 
represent changes in unbound mass using alternative definitions of `unbound'.
More liberal definitions plotted are $\mathcal{E}\gas-\mathcal{E}_\mathrm{pot,gas-gas}\ge 0$ 
(exclusion of self-gravity; orange dashed),
$\mathcal{E}\gas+P\ge0$ (replacement of internal energy density with enthalpy density; red dashed-dotted),
and $\mathcal{E}\gas^\mathrm{CM_{1-2}}\ge 0$, 
where the left-hand-side is the gas energy density in the frame of the particles' centre of mass
(blue dashed-double-dotted; we motivate this choice in Sec.~\ref{sec:particle_CM_motion}).
We also plot $\mathcal{E}-\mathcal{E}_\mathrm{int,gas}\ge 0$ 
(exclusion of the internal energy density; green dotted) for comparison. 
For each curve, the unbound mass at $t=0$, located in the ambient medium, 
is subtracted from the total unbound mass, as well as any unbound mass that has entered through the domain boundaries.

\cite{Ivanova+Nandez16} previously argued that energy deposition 
occurs only outside the orbit of the particles.  
Although the gas  between the particles is not undisturbed
as seen in 2D slices of density or energy (Sec.~\ref{sec:spatial} and \citetalias{Chamandy+18}), 
the effect on the gas  outside the orbit is likely stronger than inside it.
In calculating the unbound mass fraction it is then an interesting alternative to 
exclude the mass of gas within a sphere of radius $a(t)$ centred on particle~1.
About $23\%$ of the envelope mass resides within a distance $a$ from particle~1 at $t=13\da$, 
dropping to  $3\%$ at $t=40\da$.
At both times, most all of this interior mass is bound.
Therefore, the change in the unbound mass for the `exterior' envelope 
is basically unchanged from the values we calculated for the whole envelope (Fig.~\ref{fig:unbound_mass}).
The total envelope mass is $1.6\Msun$ and at the end of plunge-in at $t=13\da$, 
$18\%$ of the envelope mass exterior to the orbit is unbound,
compared with $14\%$ of the whole envelope.

\subsection{Spatial analysis}
\label{sec:spatial}

\begin{figure*}
  \begin{tabbing}
    \=\includegraphics[height=47mm,clip=true,trim=  50 40 200 160]{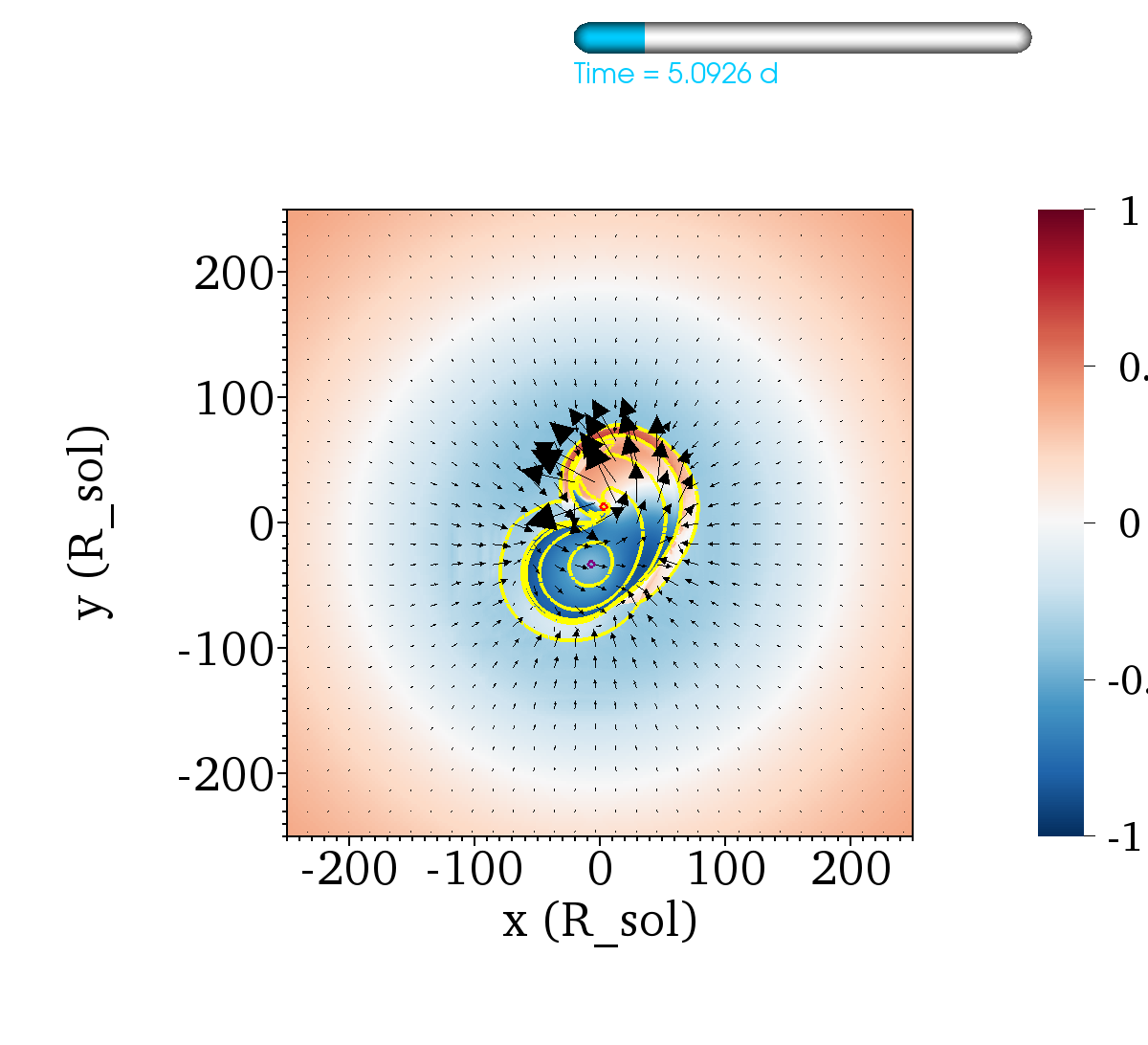}
    \includegraphics[height=47mm,clip=true,trim= 290 40 200 160]{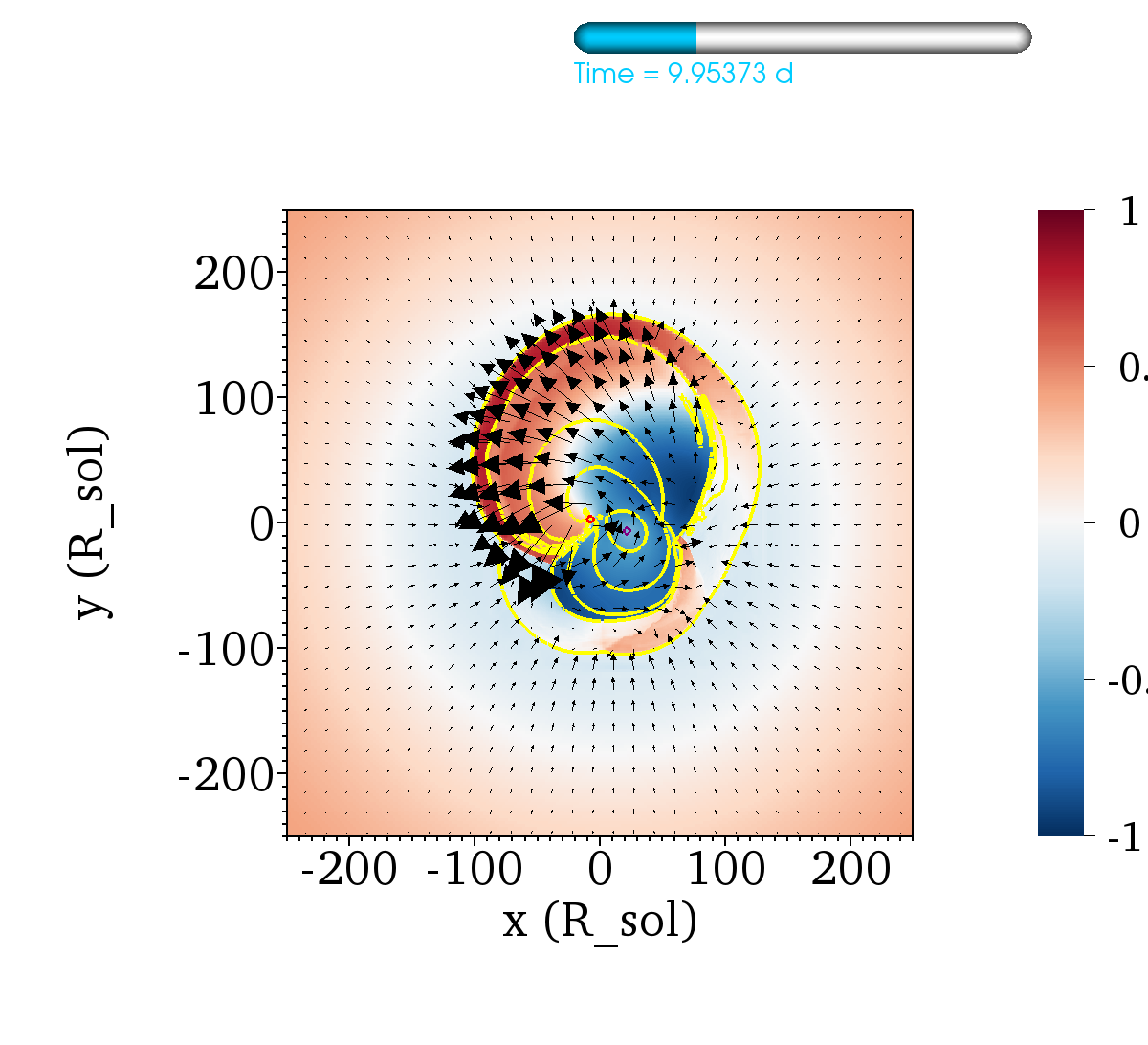}
    \includegraphics[height=47mm,clip=true,trim= 290 40 200 160]{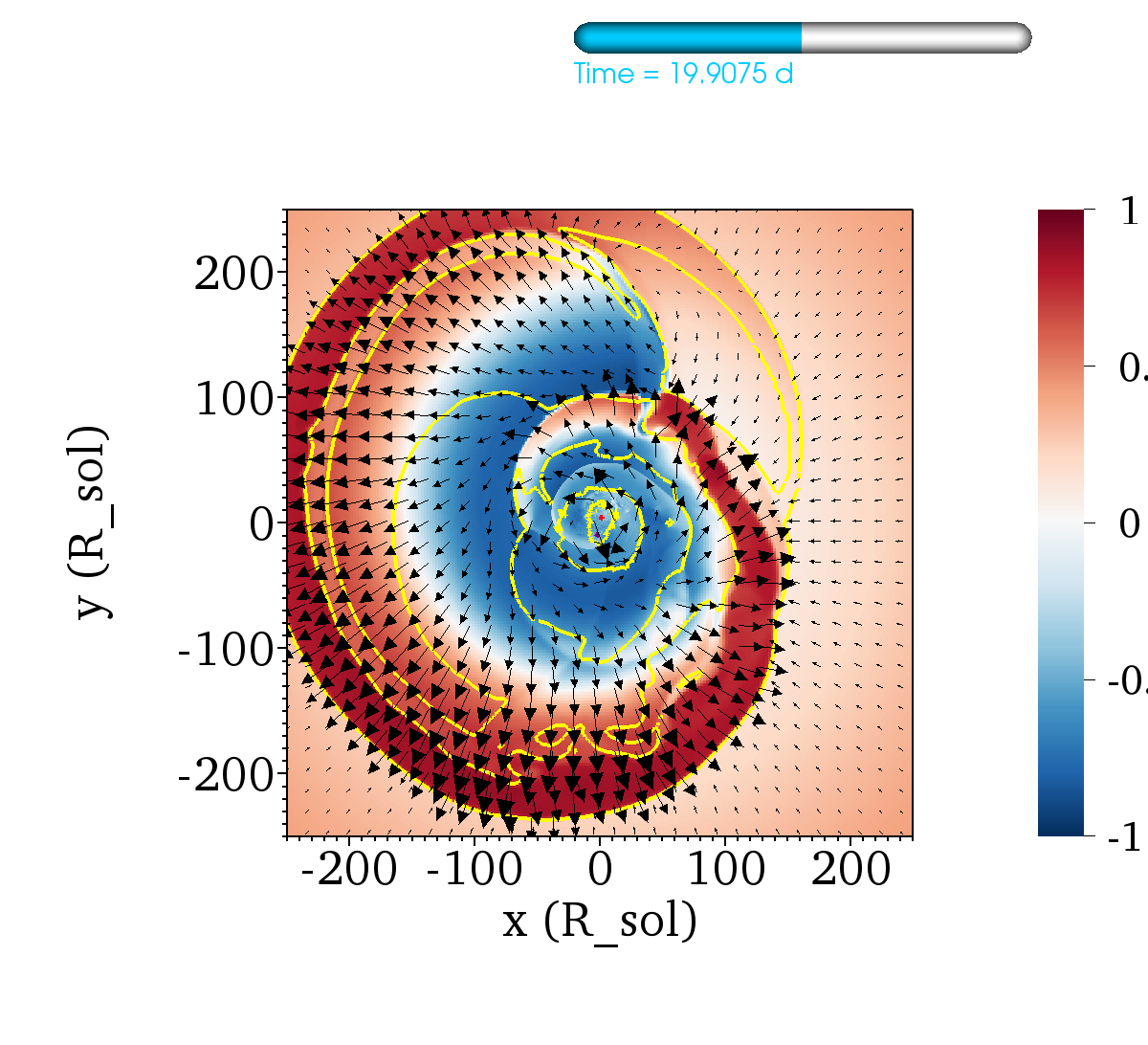}
    \includegraphics[height=47mm,clip=true,trim= 290 40  40 160]{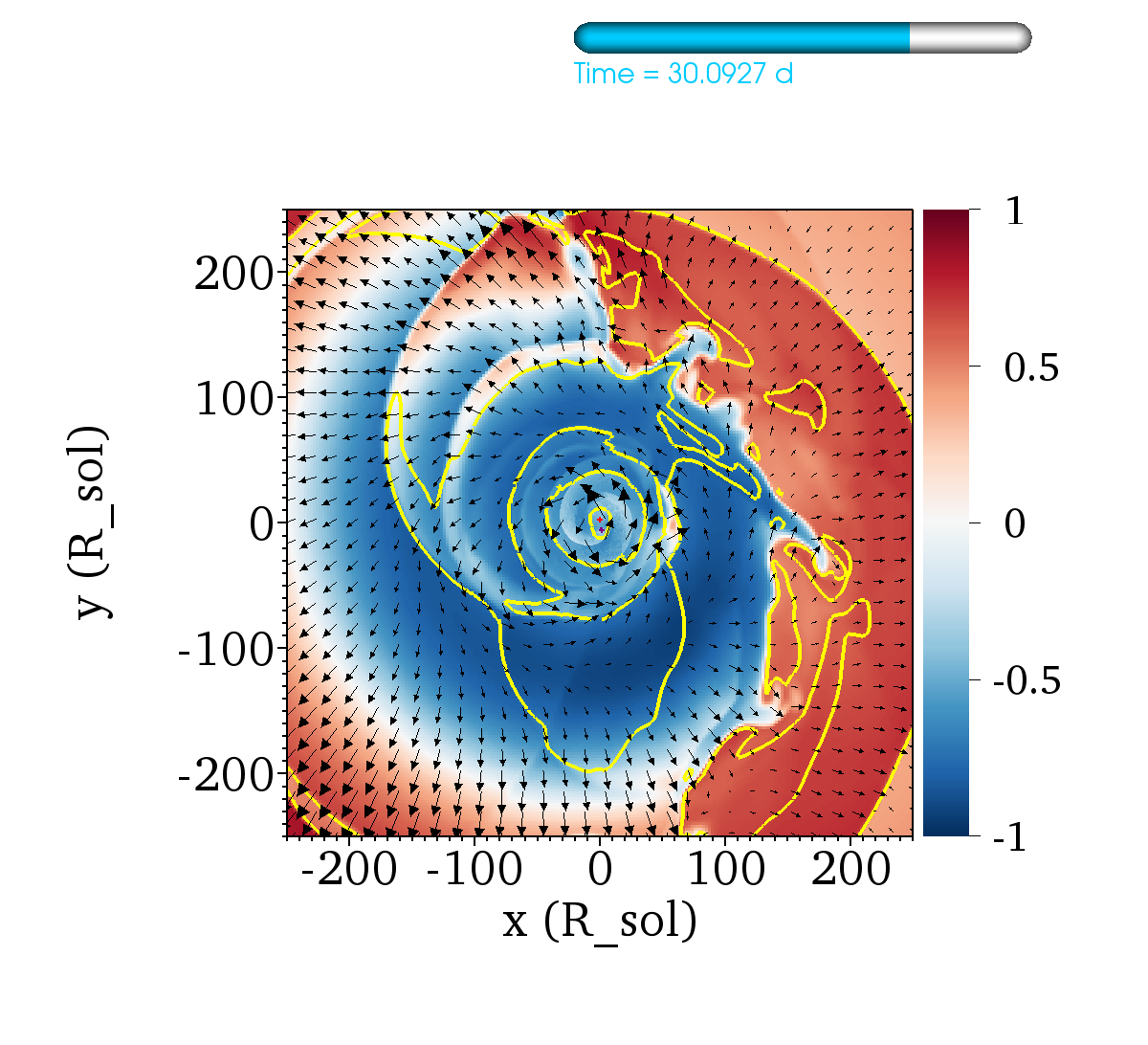}\\
    \>\includegraphics[height=47mm,clip=true,trim=  50 40 200 160]{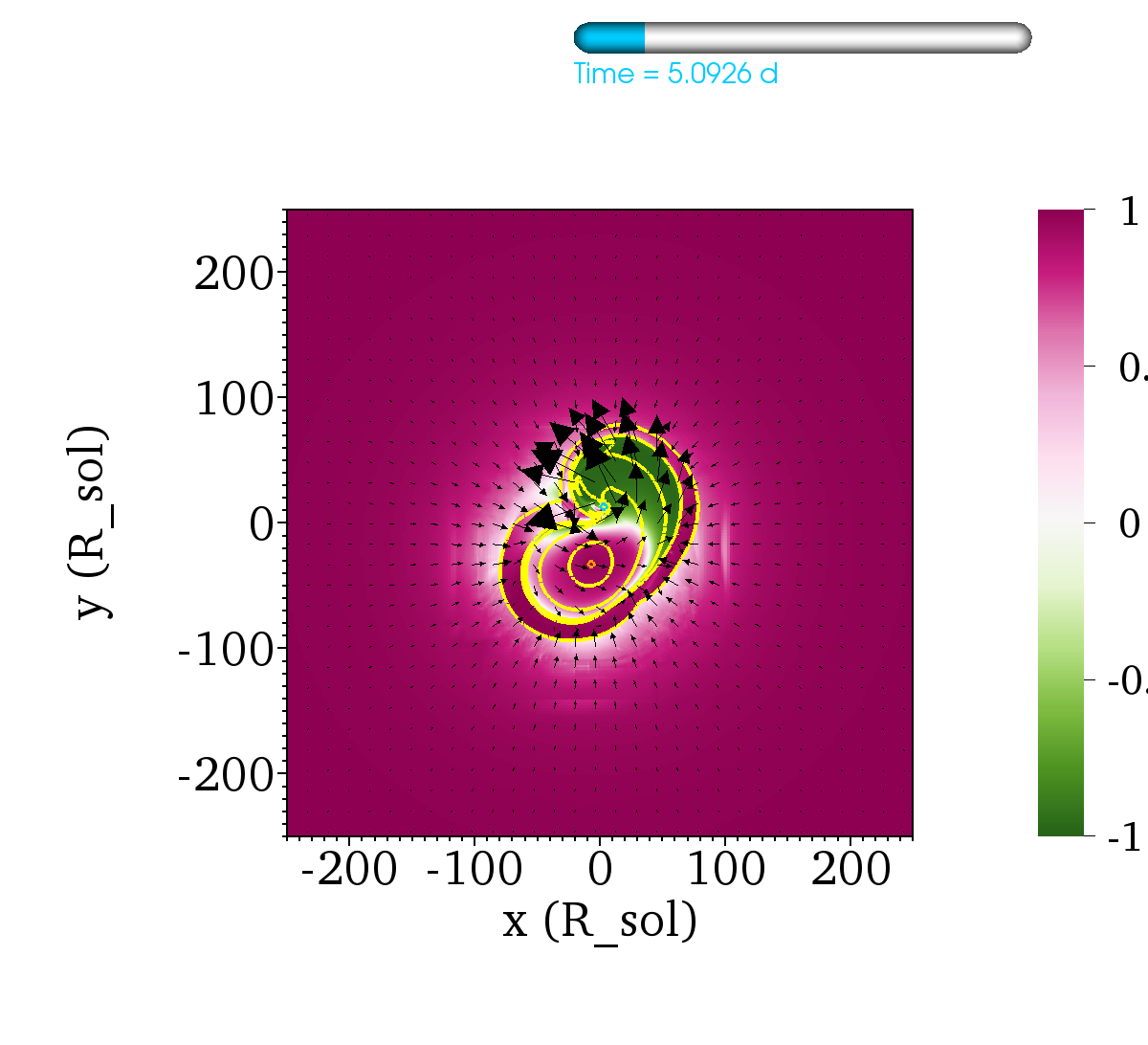}
    \includegraphics[height=47mm,clip=true,trim= 290 40 200 160]{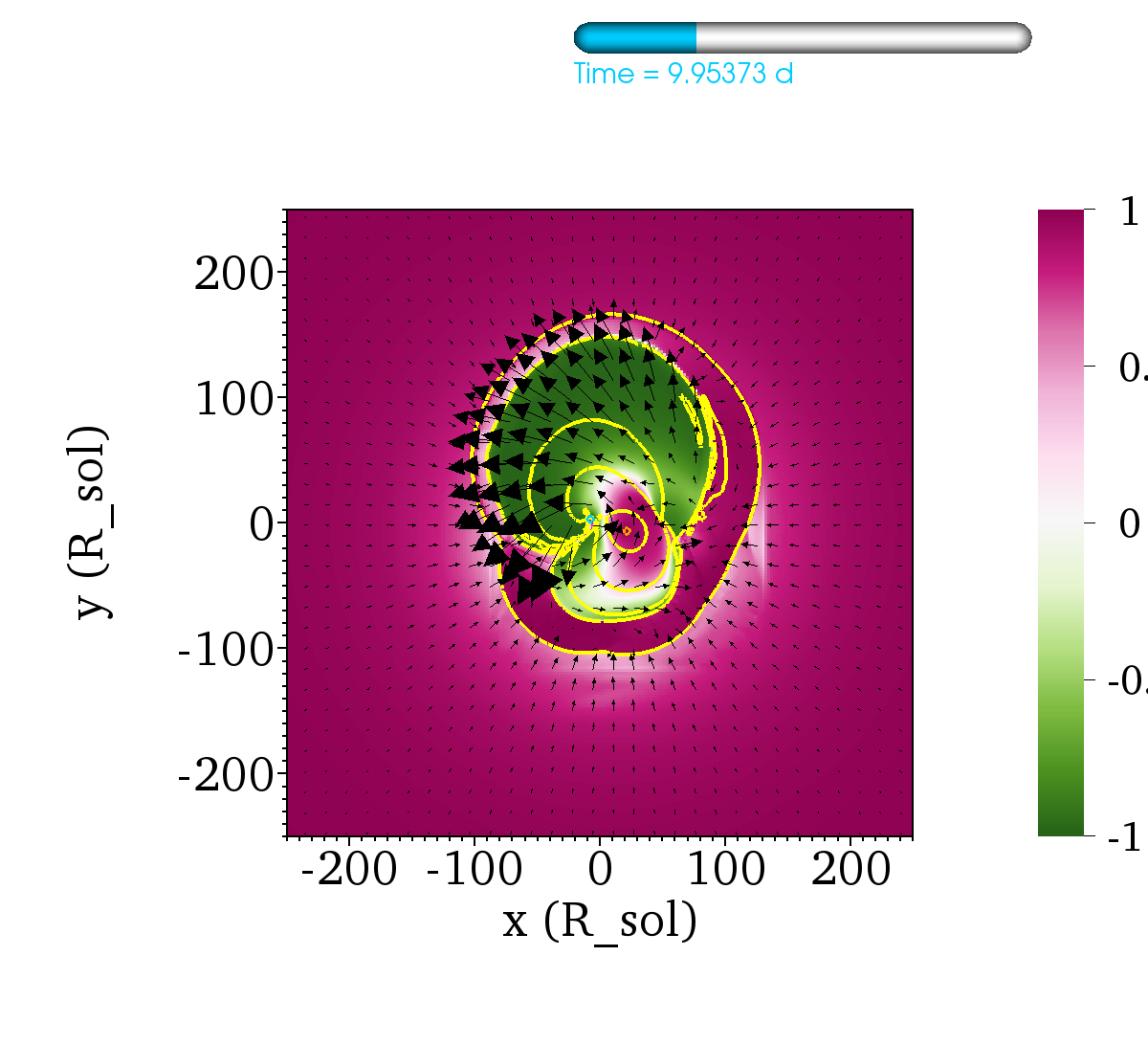}
    \includegraphics[height=47mm,clip=true,trim= 290 40 200 160]{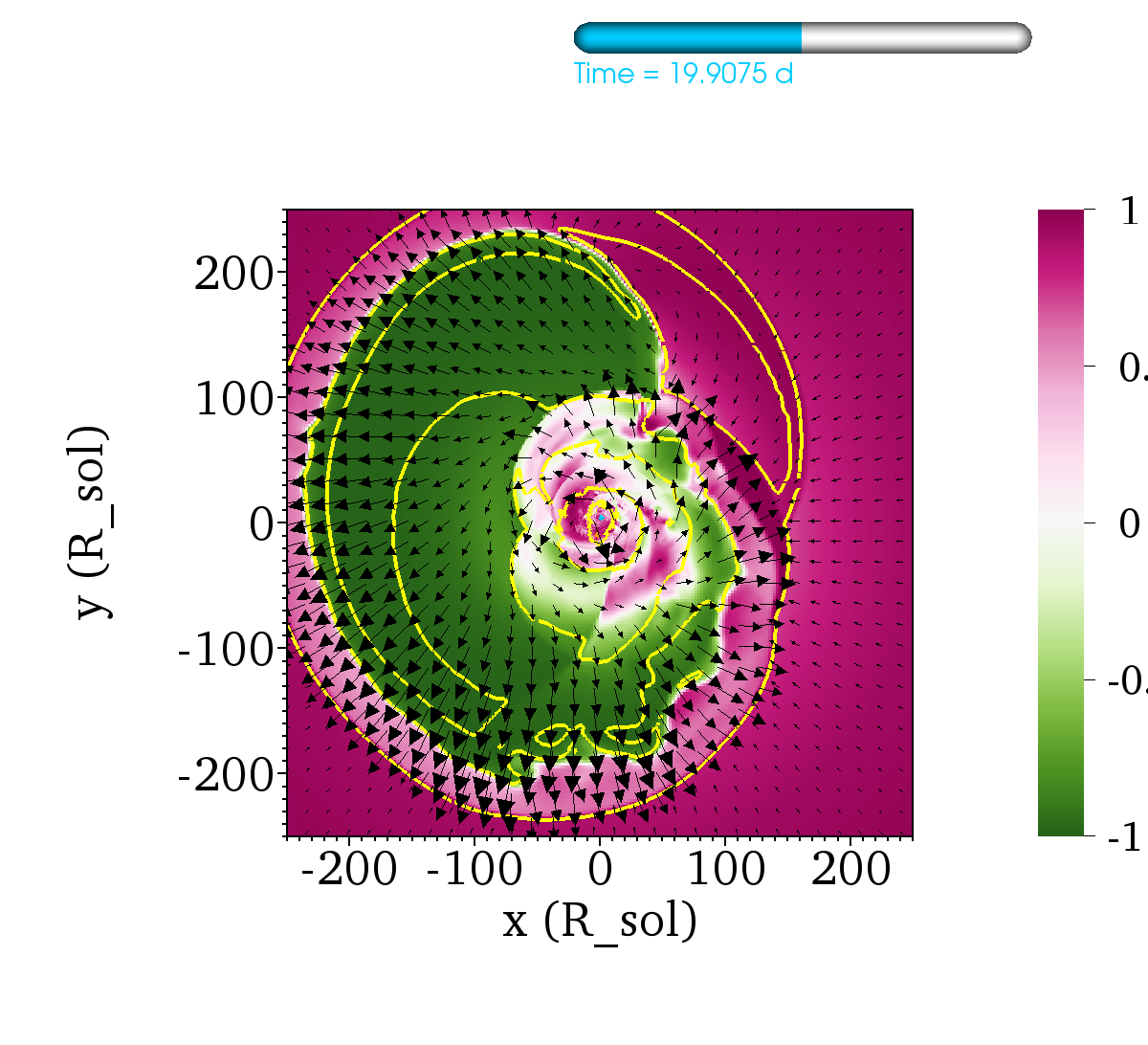}
    \includegraphics[height=47mm,clip=true,trim= 290 40  40 160]{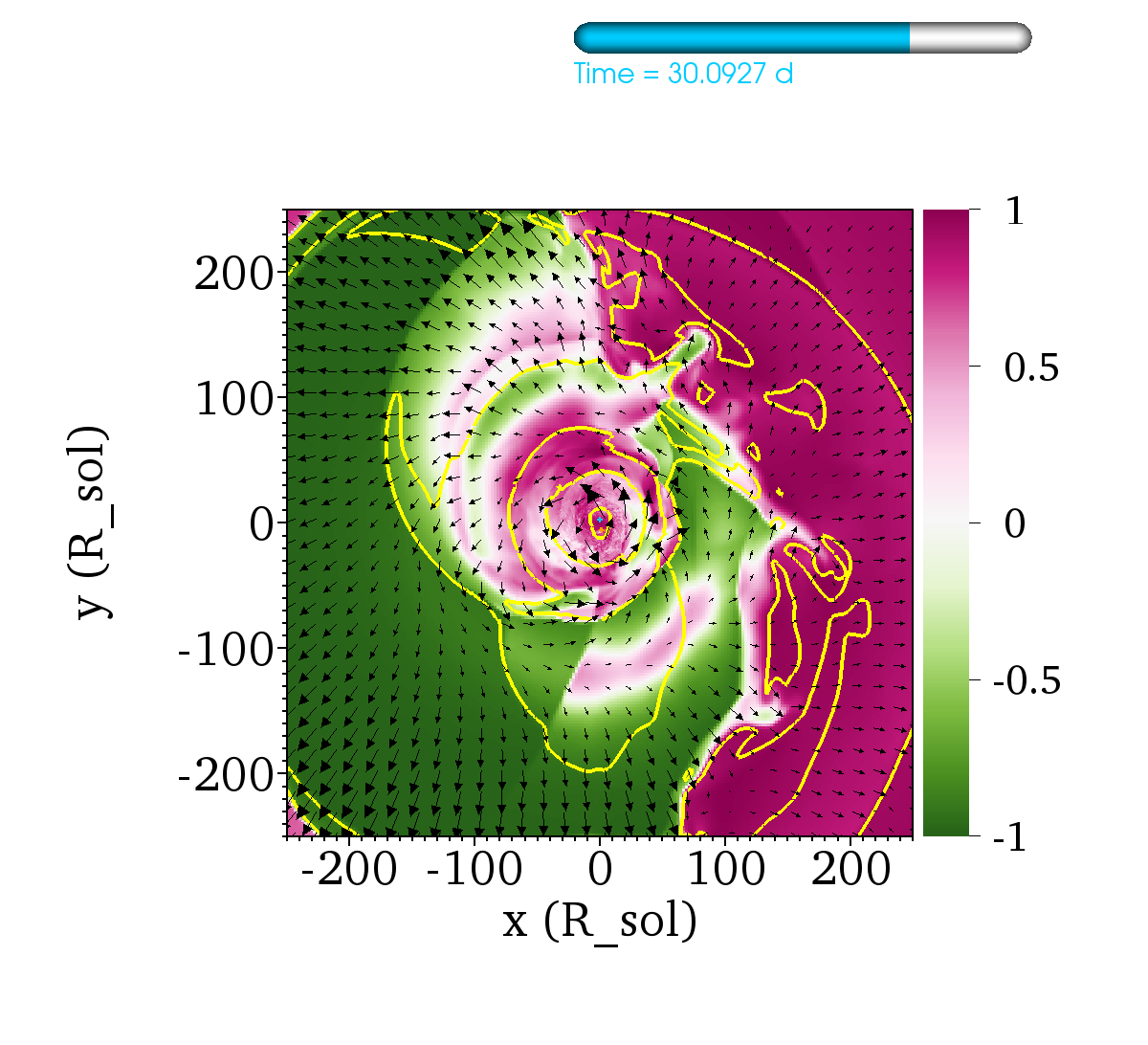}\\
    \>\includegraphics[height=47mm,clip=true,trim=  50 40 200 160]{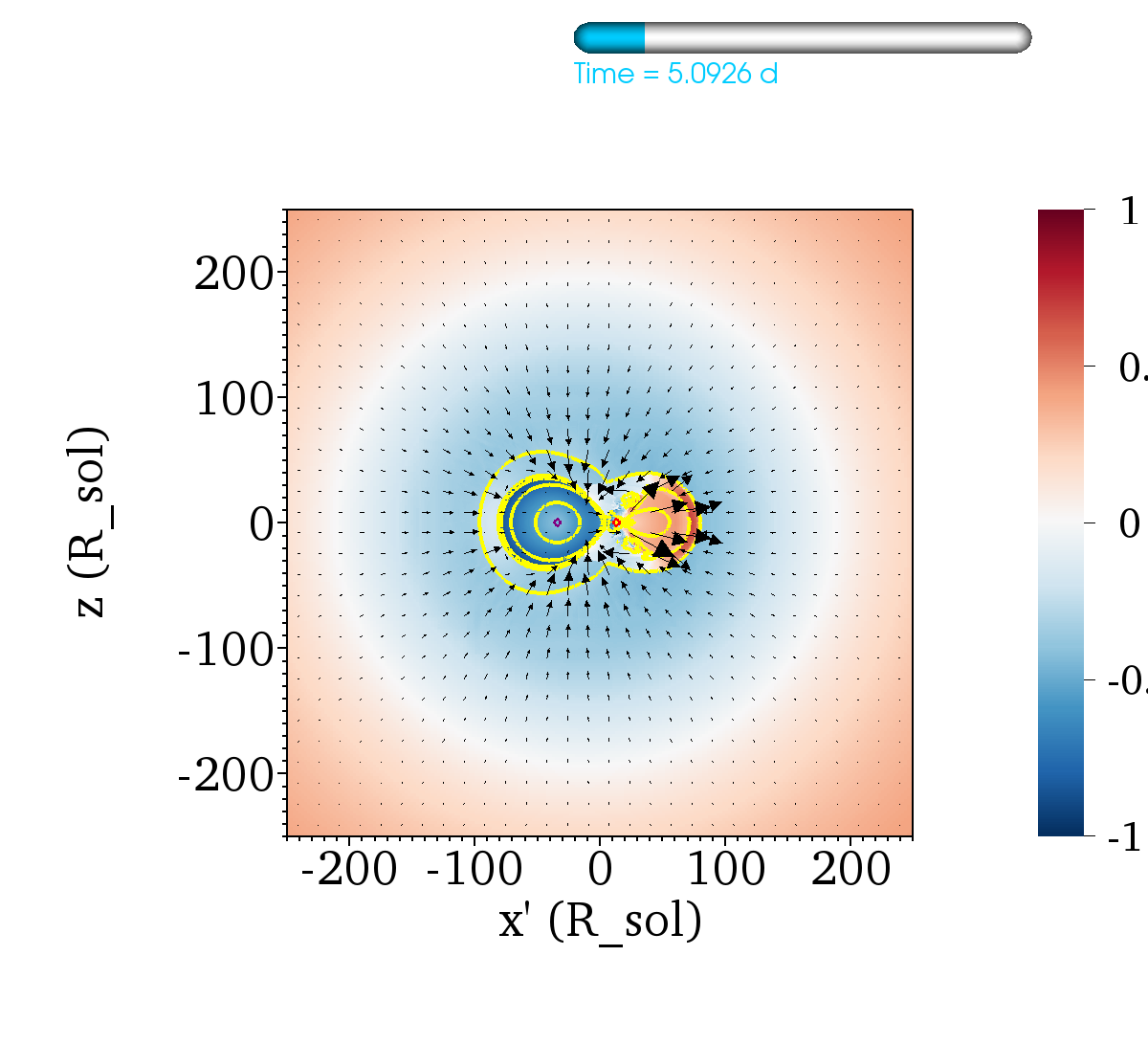}
    \includegraphics[height=47mm,clip=true,trim= 290 40 200 160]{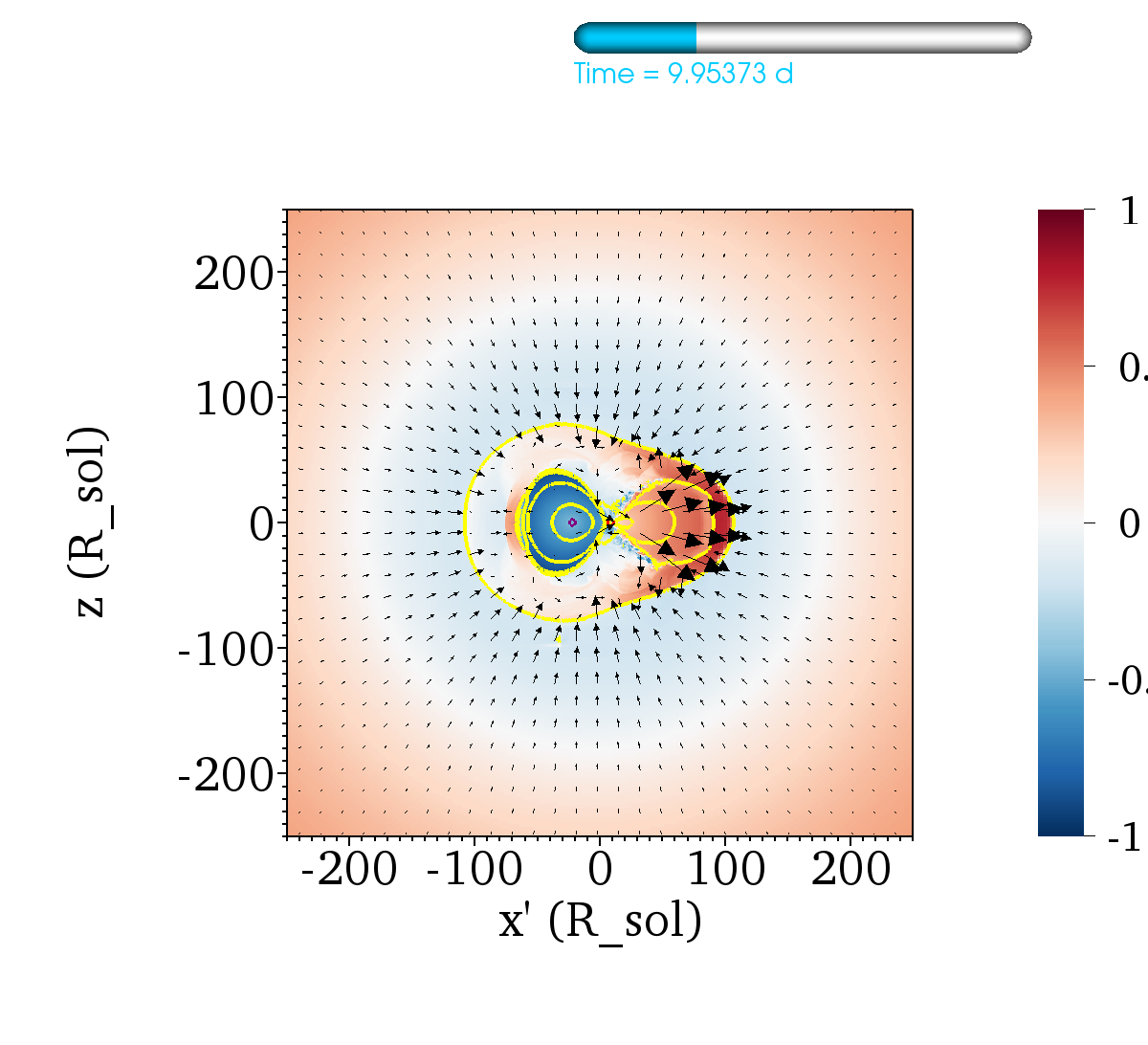}
    \includegraphics[height=47mm,clip=true,trim= 290 40 200 160]{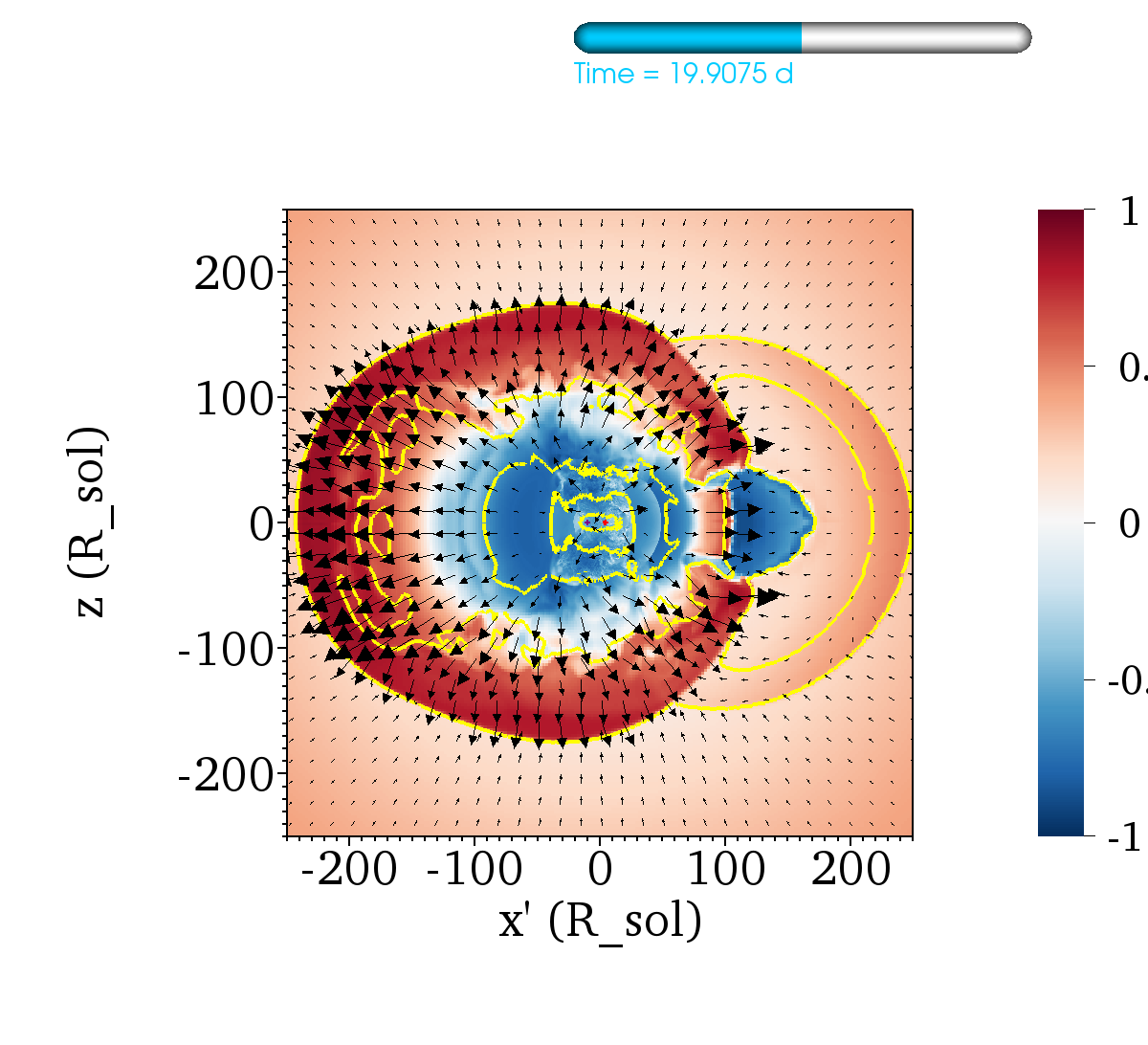}
    \includegraphics[height=47mm,clip=true,trim= 290 40 200 160]{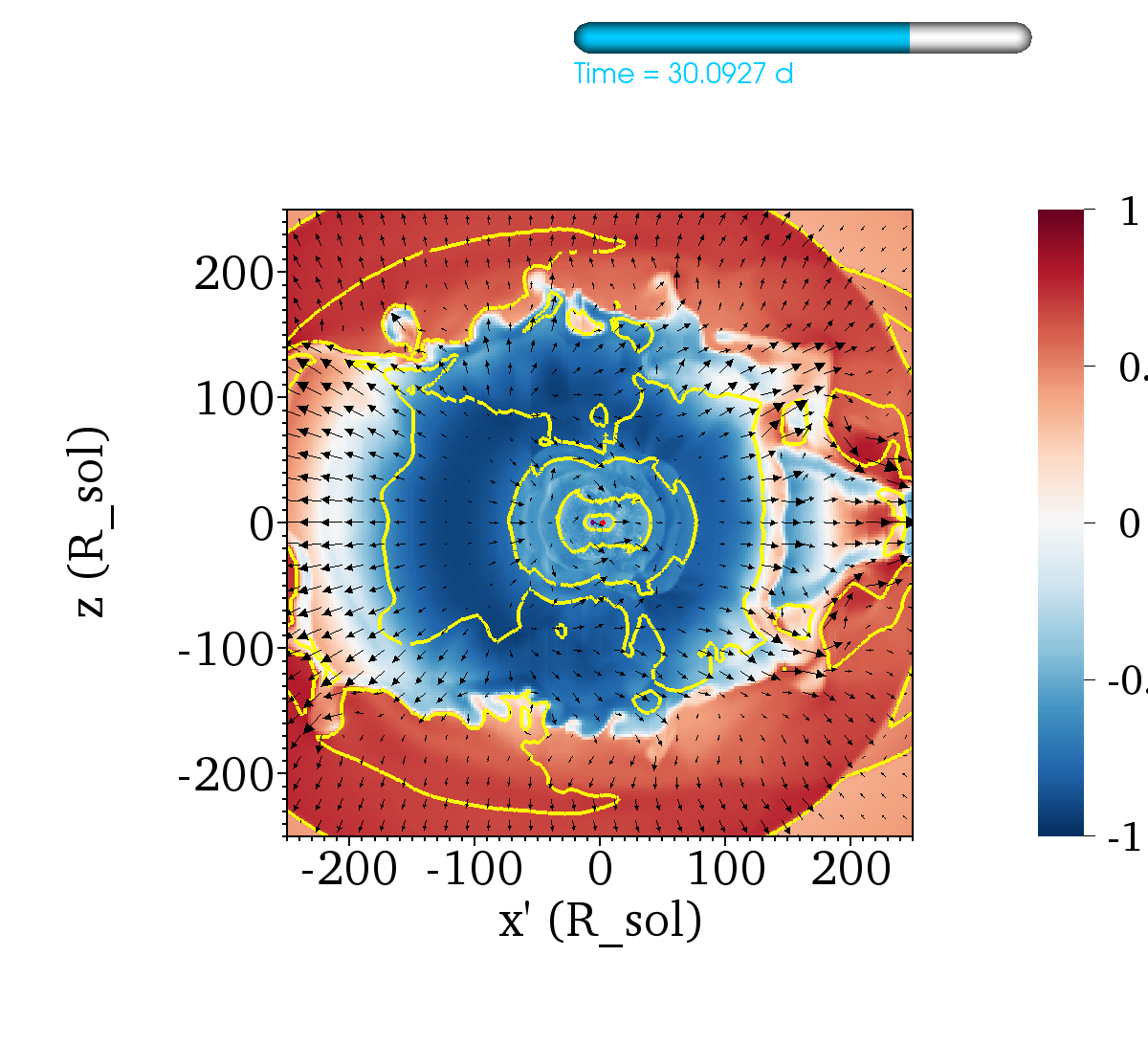}\\
    \>\includegraphics[height=47mm,clip=true,trim=  50 40 200 160]{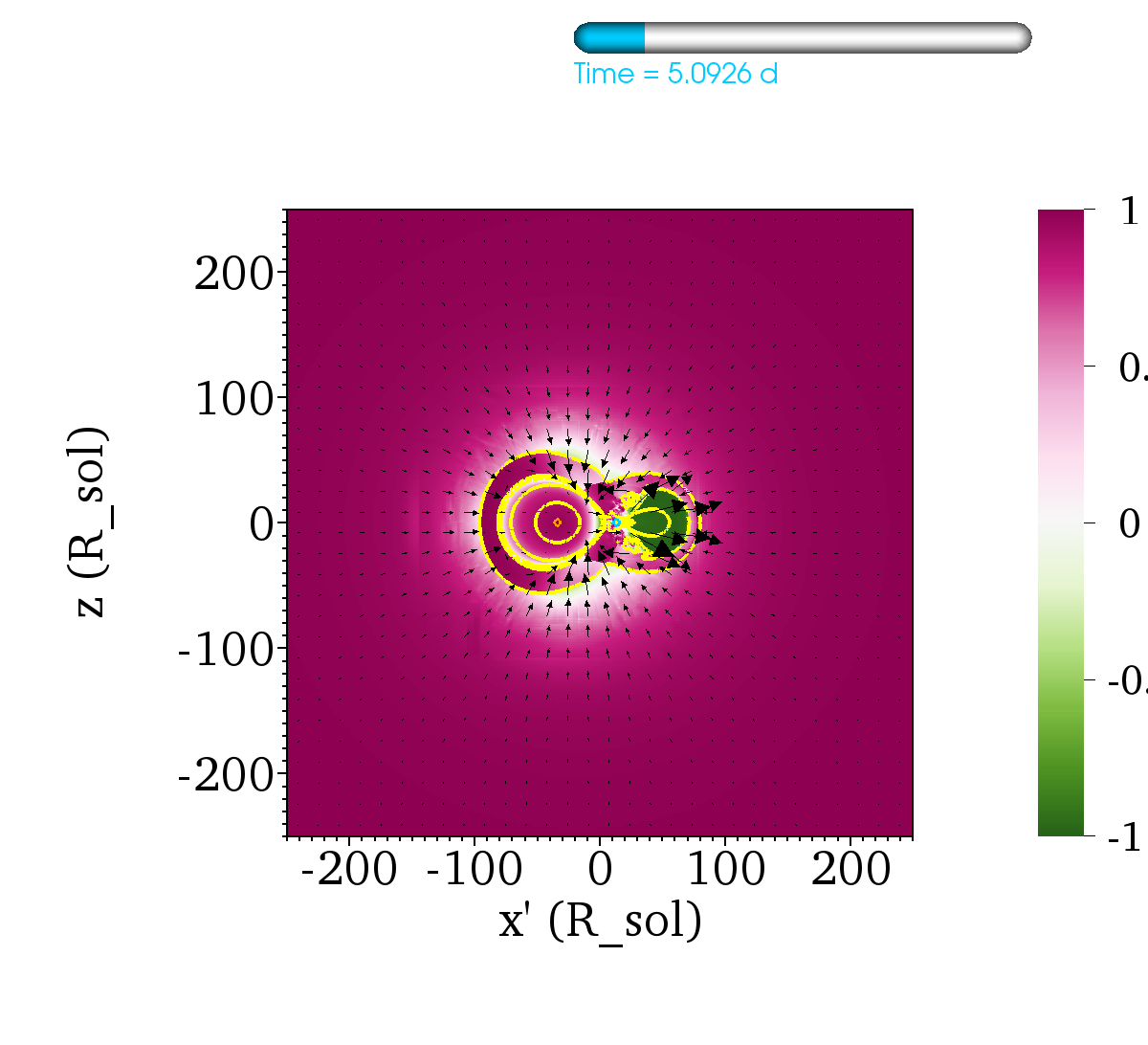}
    \includegraphics[height=47mm,clip=true,trim= 290 40 200 160]{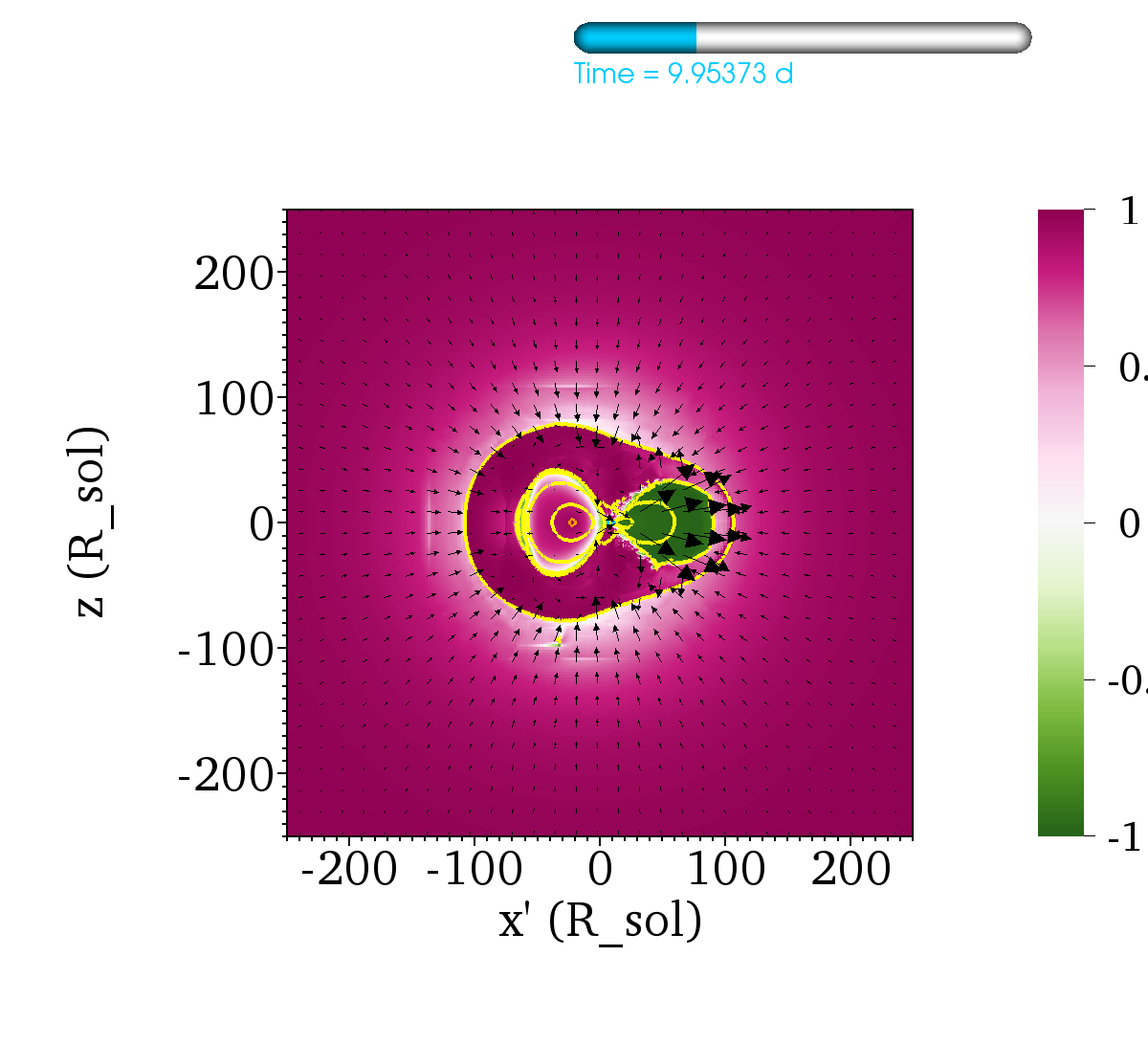}
    \includegraphics[height=47mm,clip=true,trim= 290 40 200 160]{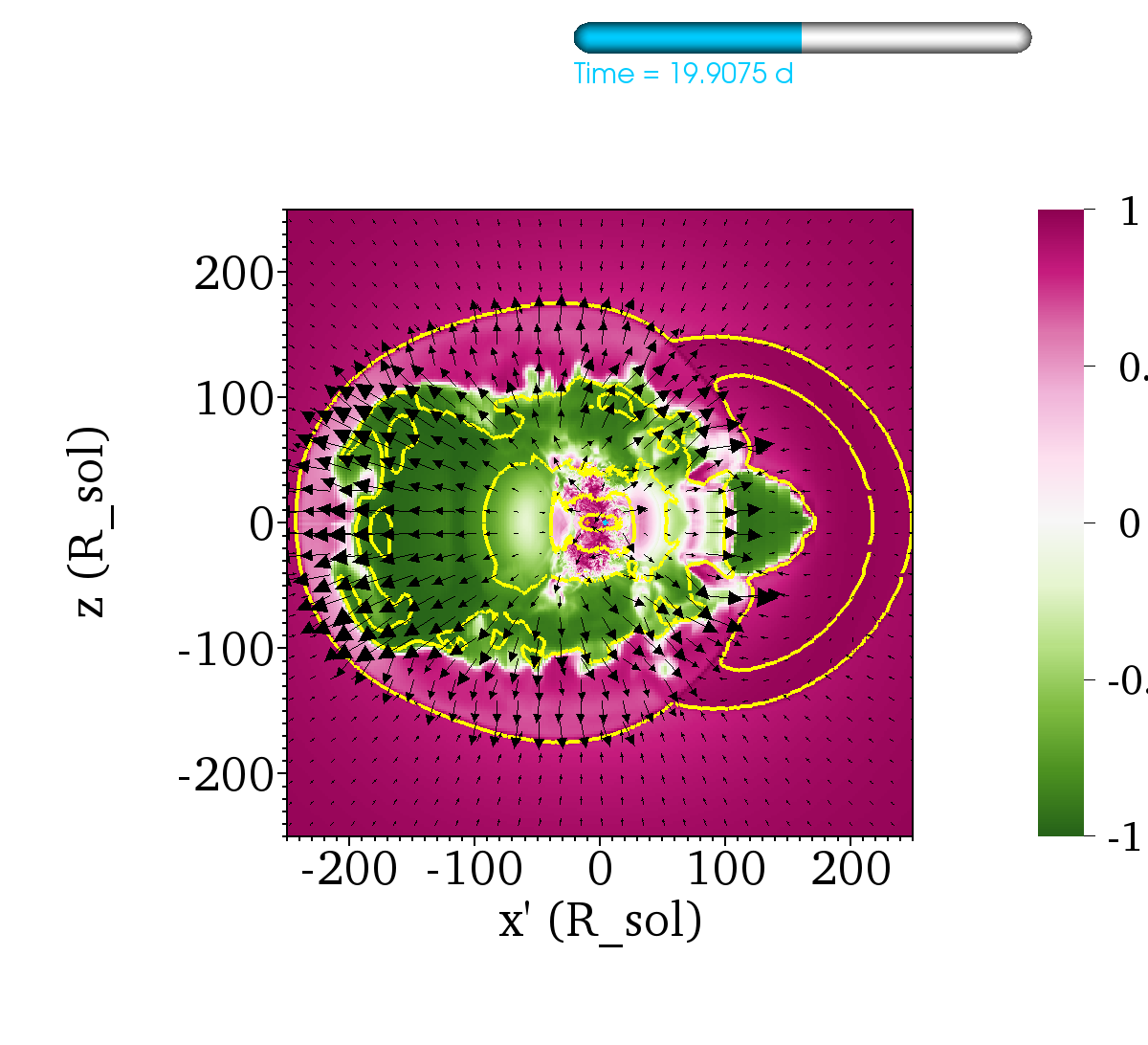}
    \includegraphics[height=47mm,clip=true,trim= 290 40 200 160]{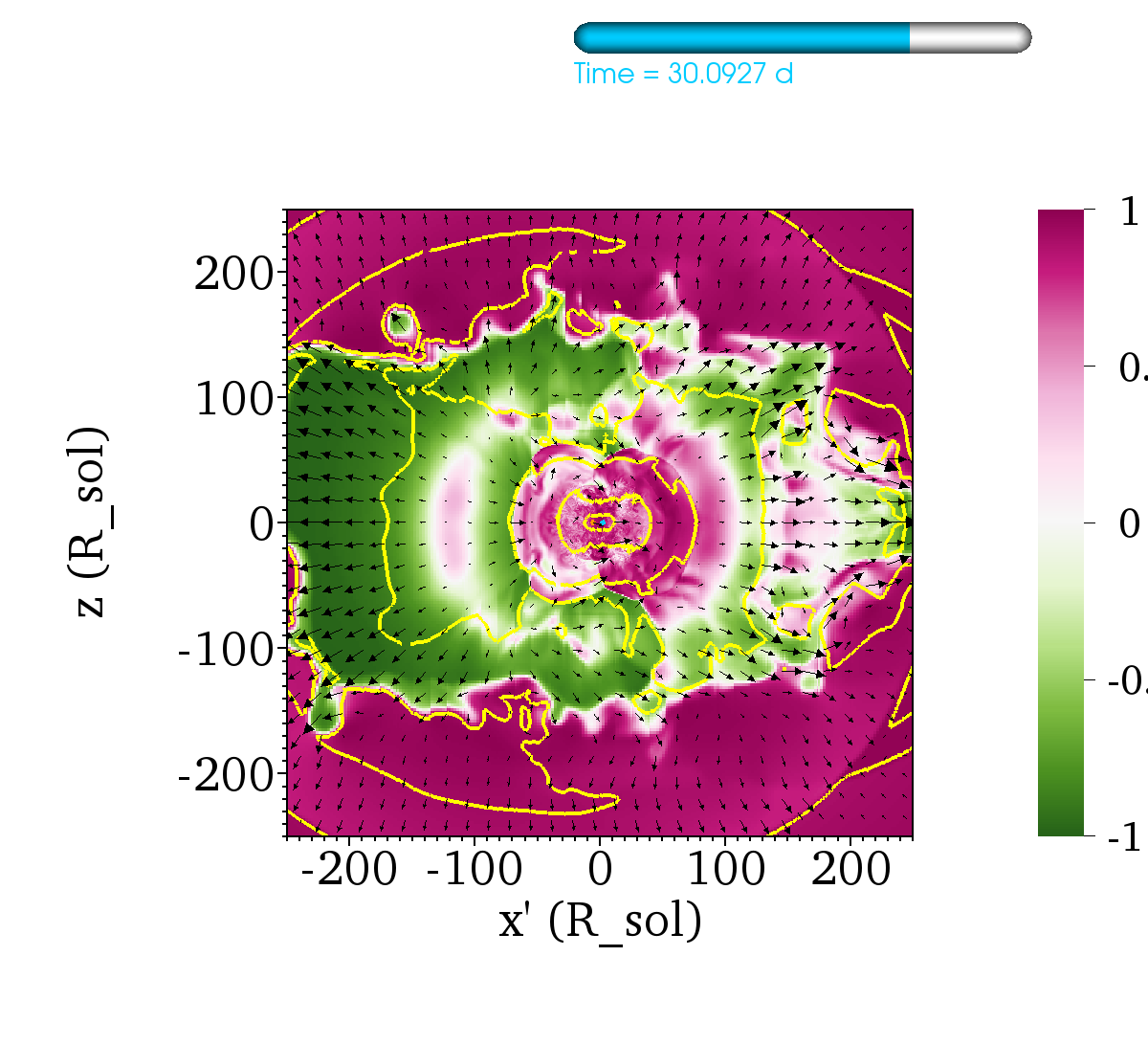}\\
  \end{tabbing}
  \caption{\textit{Top row:}
           Snapshots of the quantity 
           $\mathcal{E}\gas/\max(\mathcal{E}_\mathrm{bulk,gas}+\mathcal{E}_\mathrm{int,gas},-\mathcal{E}_\mathrm{pot,gas})$,
           where $\mathcal{E}_\mathrm{pot,gas}= \mathcal{E}_\mathrm{pot,gas-1} 
           +\mathcal{E}_\mathrm{pot,gas-2} +\mathcal{E}_\mathrm{pot,gas-gas}$,
           in the orbital plane at $t=5$, $10$, $20$ and $30\da$.
           This is the gas energy density, normalized to  the magnitude of the gas kinetic energy density
           (internal + bulk) or the total gas potential energy density, whichever is greater.
           A value of $1$ corresponds to a maximally unbound system, 
           while a value of $-1$ corresponds to a maximally bound system,
           for our standard definition of `unbound': $\mathcal{E}\gas\ge0$.
           At $t=0$, both particles are situated on the $x$-axis with particle~2 (circular red contour near centre) 
           located to the right of particle~1 (circular mauve contour near centre),
           and the orbit is anti-clockwise.
           Yellow contours show the density from $\rho=10^{-4}\gcmcmcm$ downward in logarithmic intervals of 1 dex.
           Vectors show the component of the velocity parallel to the orbital plane
           (representing the points located at the ends of the arrow tails).
           The frame of reference is that of the simulation, 
           and each plot is centred on the particle centre of mass. 
           \textit{Second row from top:}
           As for the top row but now showing the quantity 
           $(\mathcal{E}_\mathrm{int,gas}-\mathcal{E}_\mathrm{bulk,gas})/\max(\mathcal{E}_\mathrm{int,gas},\mathcal{E}_\mathrm{bulk,gas})$.
           This compares the internal and bulk kinetic energy densities. 
           Magenta means internal energy dominates (between the two forms), 
           green means bulk kinetic energy dominates, and white means they are approximately equal.
           Softening spheres are now represented by orange and cyan contours for particles~1 and 2, respectively.
           \textit{Second row from bottom:}
           As top row but now for the $x'$-$z$ plane, 
           which is orthogonal to the orbital plane and intersects both particles.
           \textit{Bottom row:}
           As second row from top but now for the $x'$-$z$ plane.
           \label{fig:spatial}
          }            
\end{figure*}

\begin{figure*}
  \begin{tabbing}
    \=\includegraphics[height=47mm,clip=true,trim=  50 40 200 160]{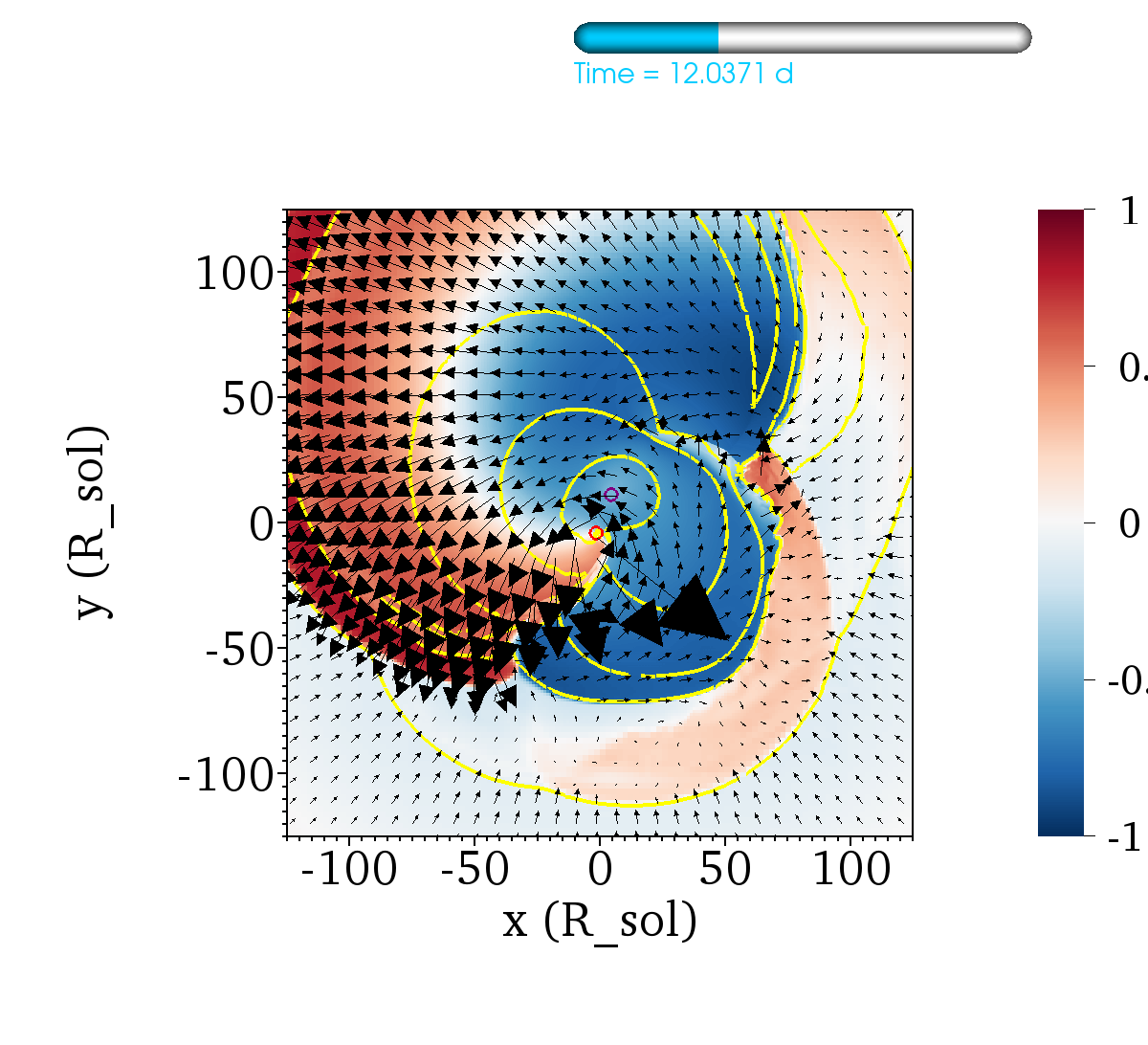}
      \includegraphics[height=47mm,clip=true,trim= 290 40 200 160]{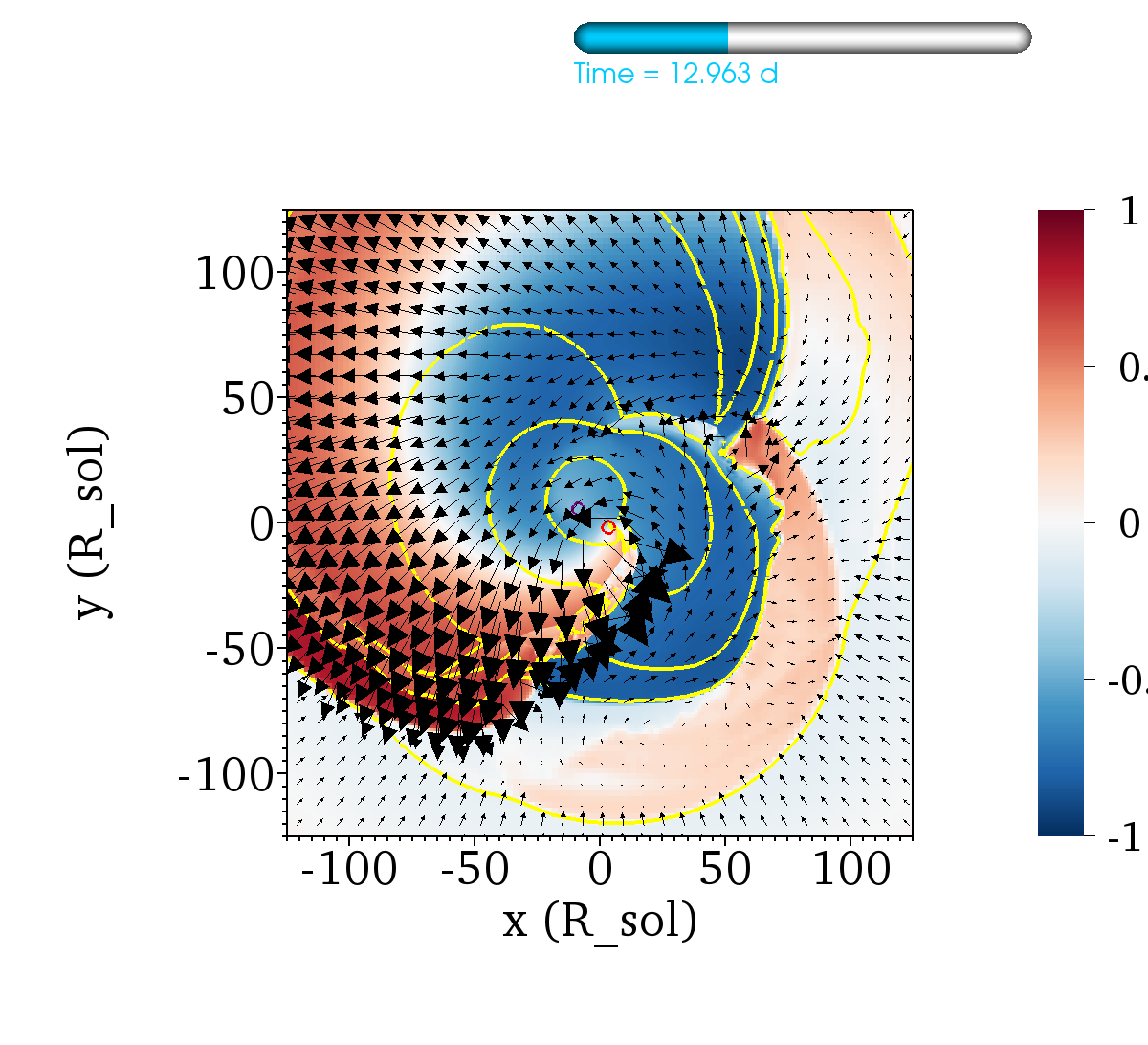}
      \includegraphics[height=47mm,clip=true,trim= 290 40 200 160]{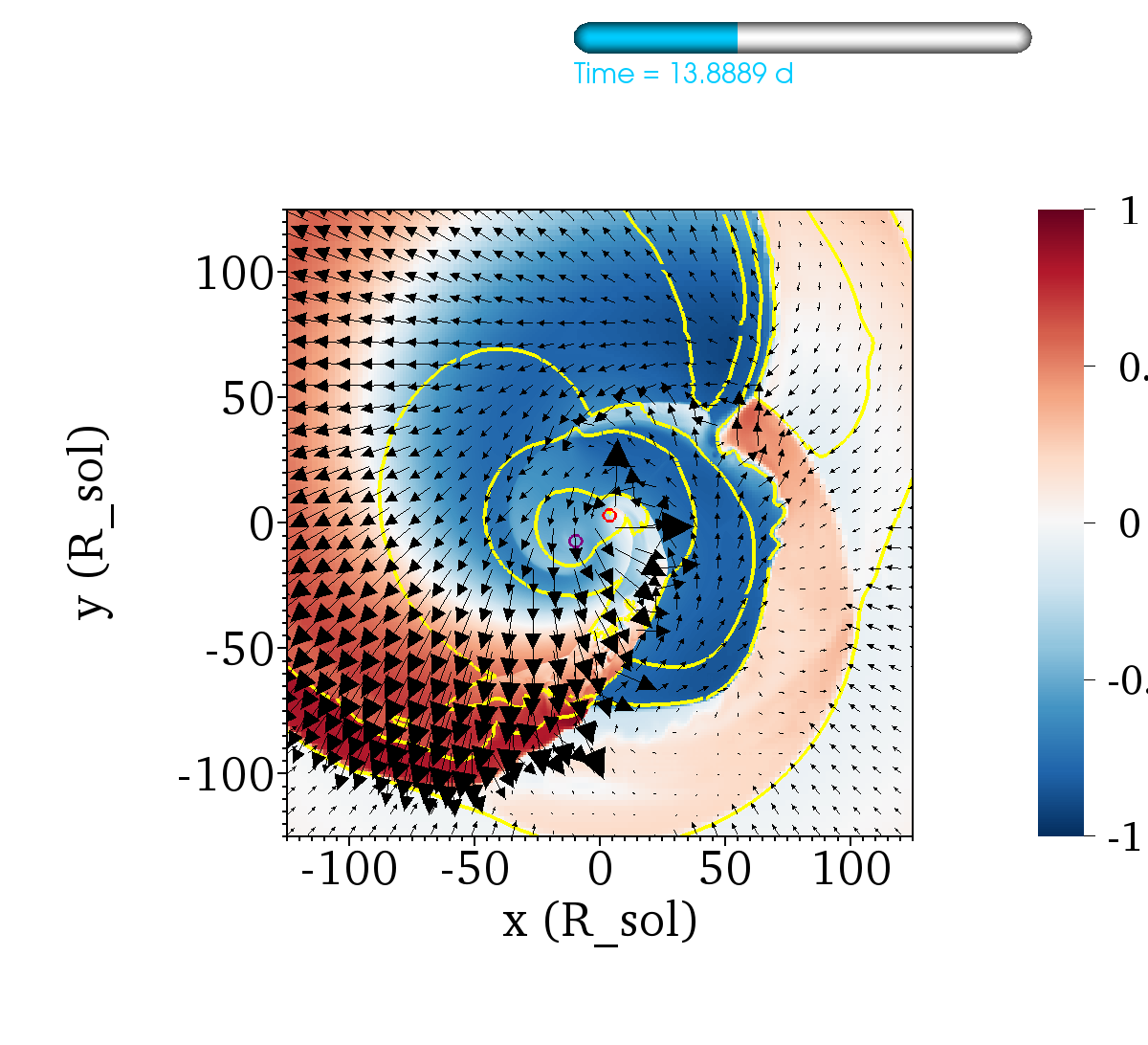}
      \includegraphics[height=47mm,clip=true,trim= 290 40  40 160]{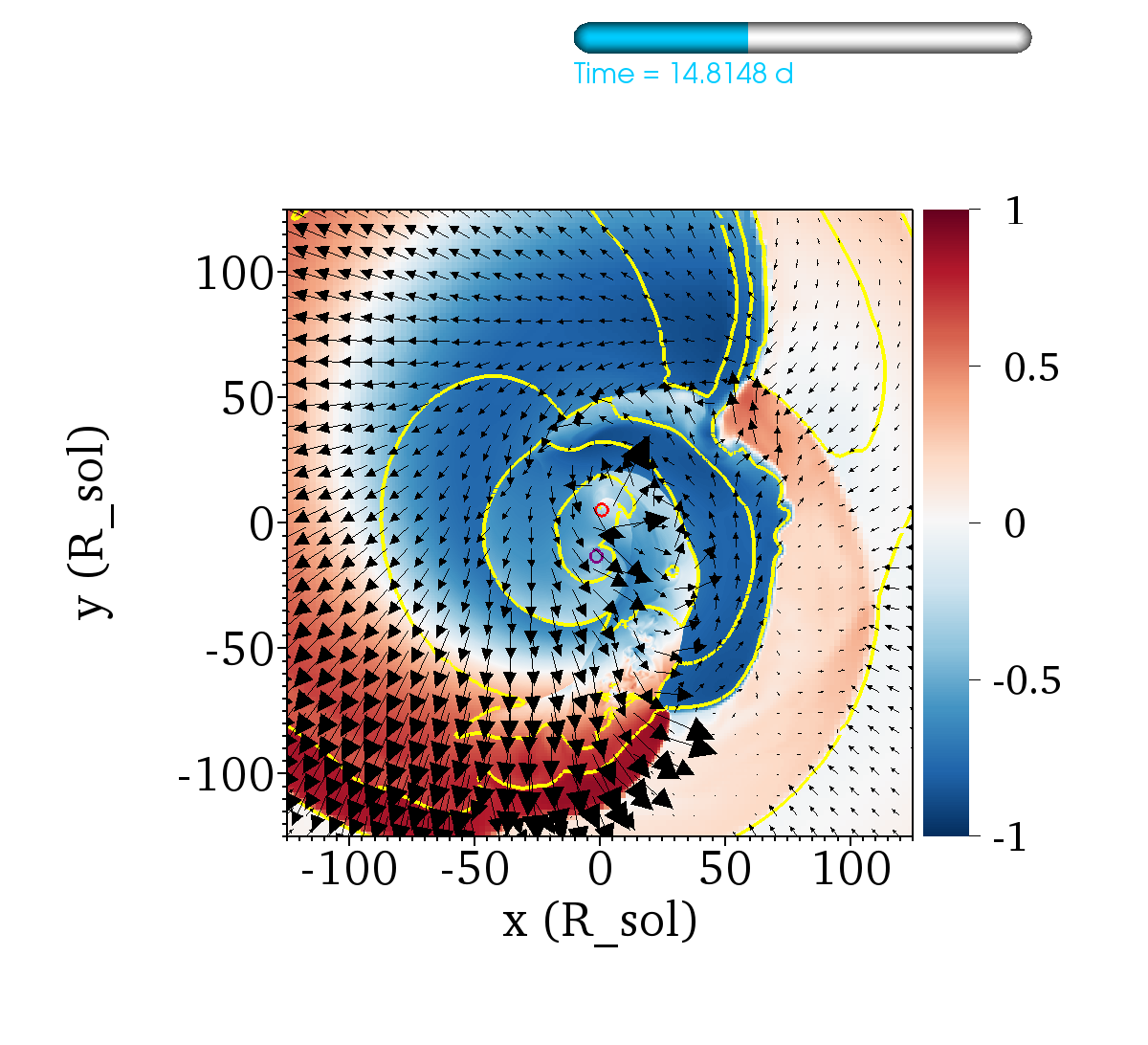}\\
    \>\includegraphics[height=47mm,clip=true,trim=  50 40 200 160]{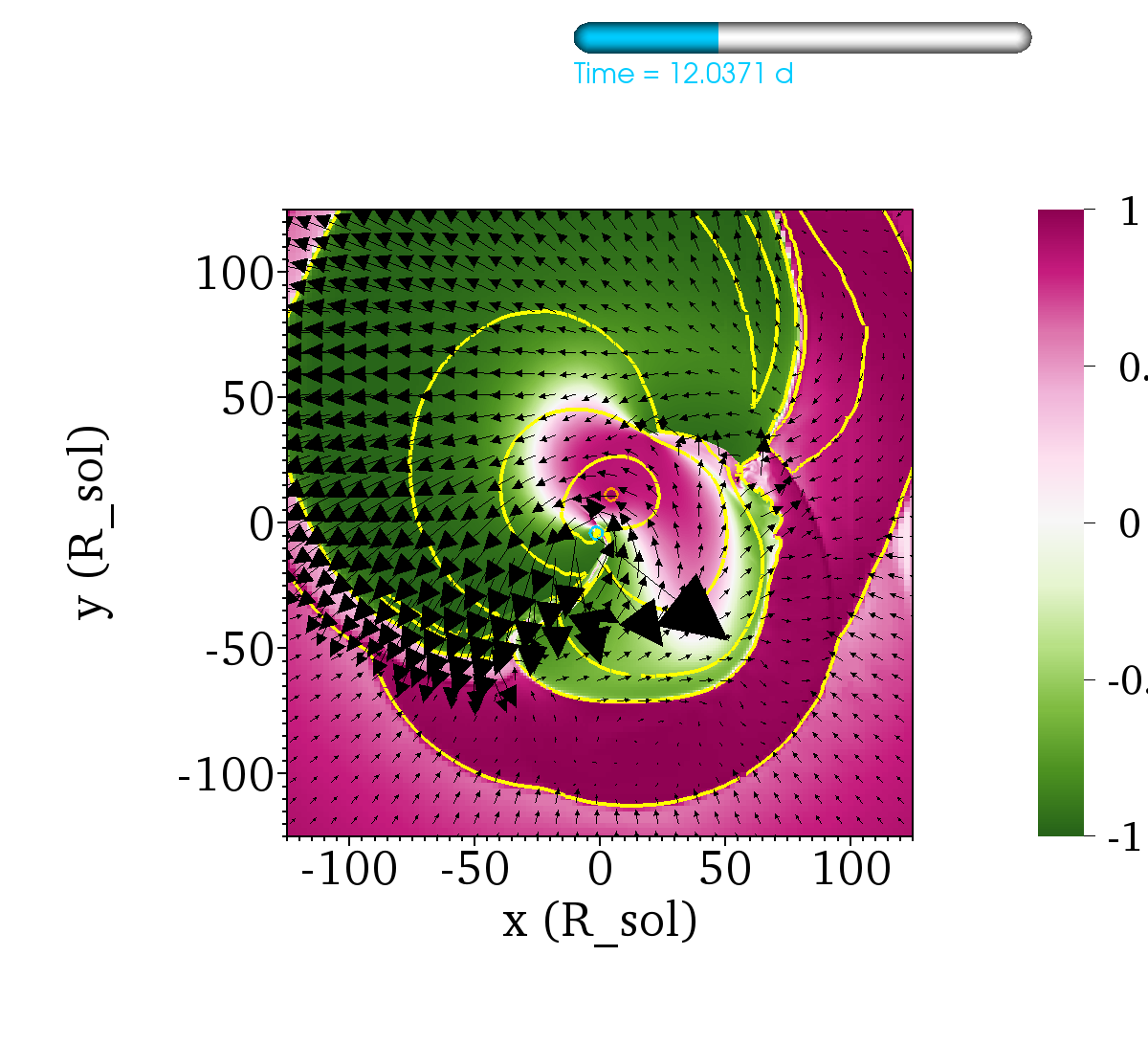}
      \includegraphics[height=47mm,clip=true,trim= 290 40 200 160]{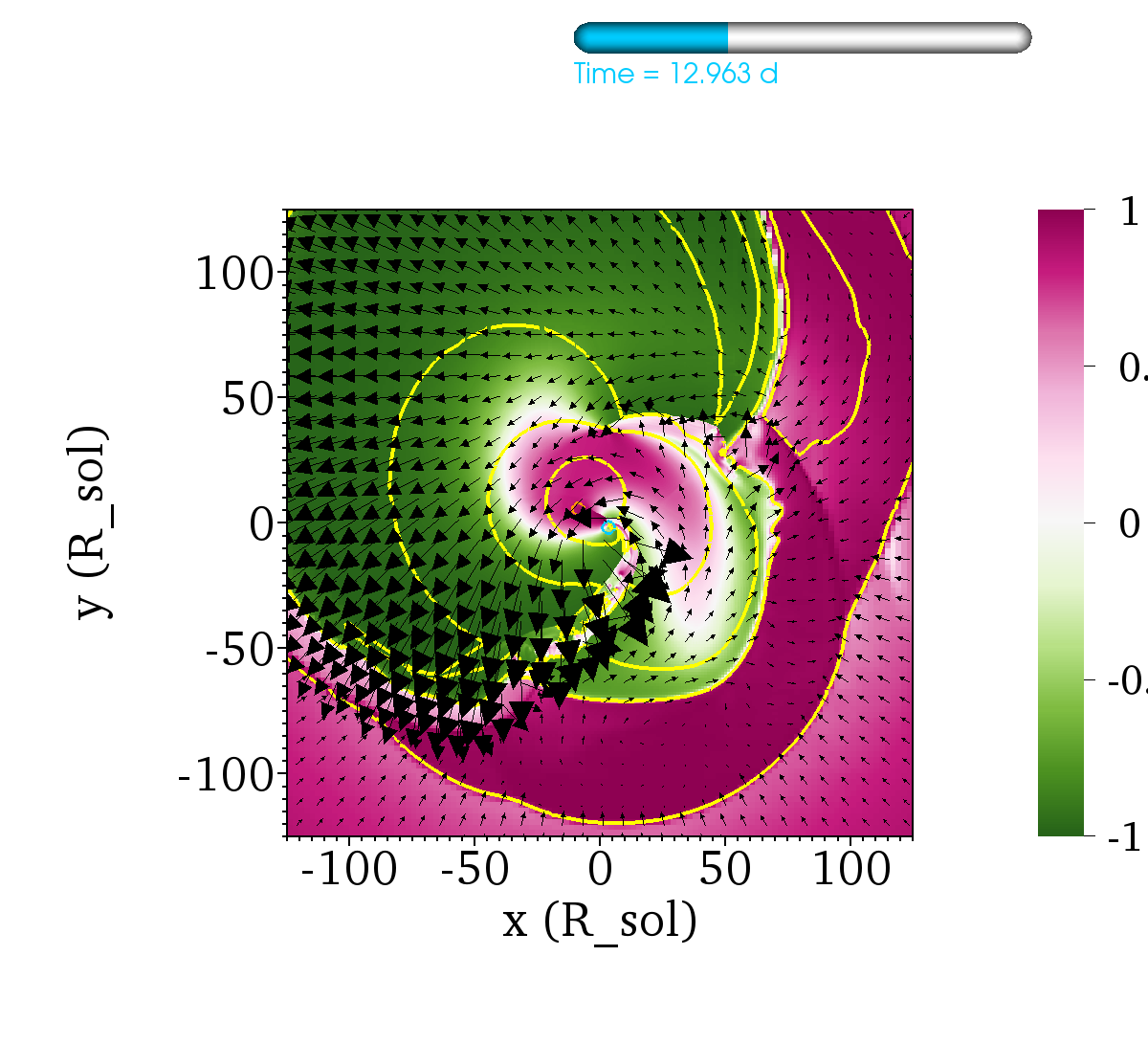}
      \includegraphics[height=47mm,clip=true,trim= 290 40 200 160]{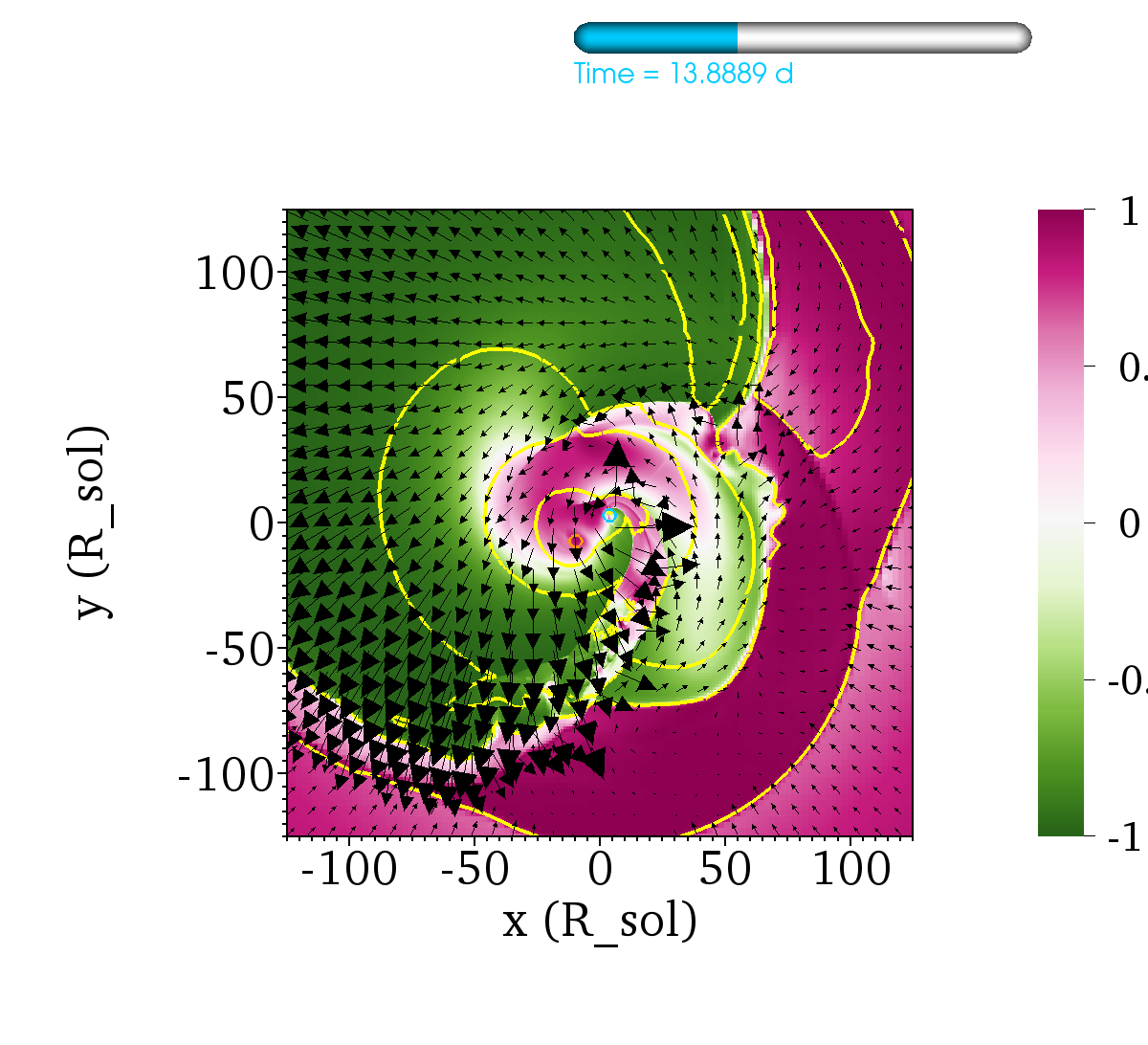}
      \includegraphics[height=47mm,clip=true,trim= 290 40  40 160]{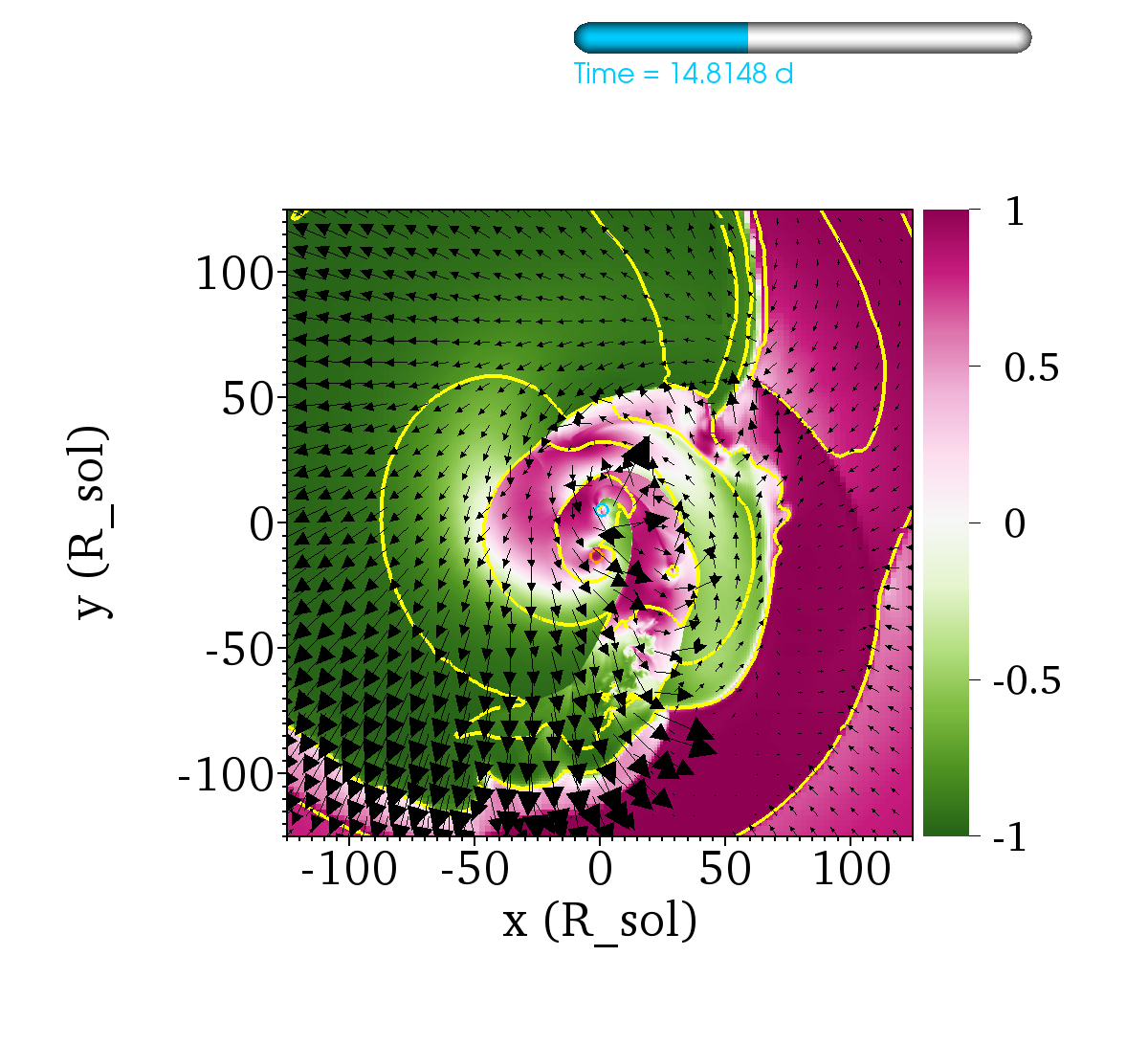}\\
    \>\includegraphics[height=47mm,clip=true,trim=  50 40 200 160]{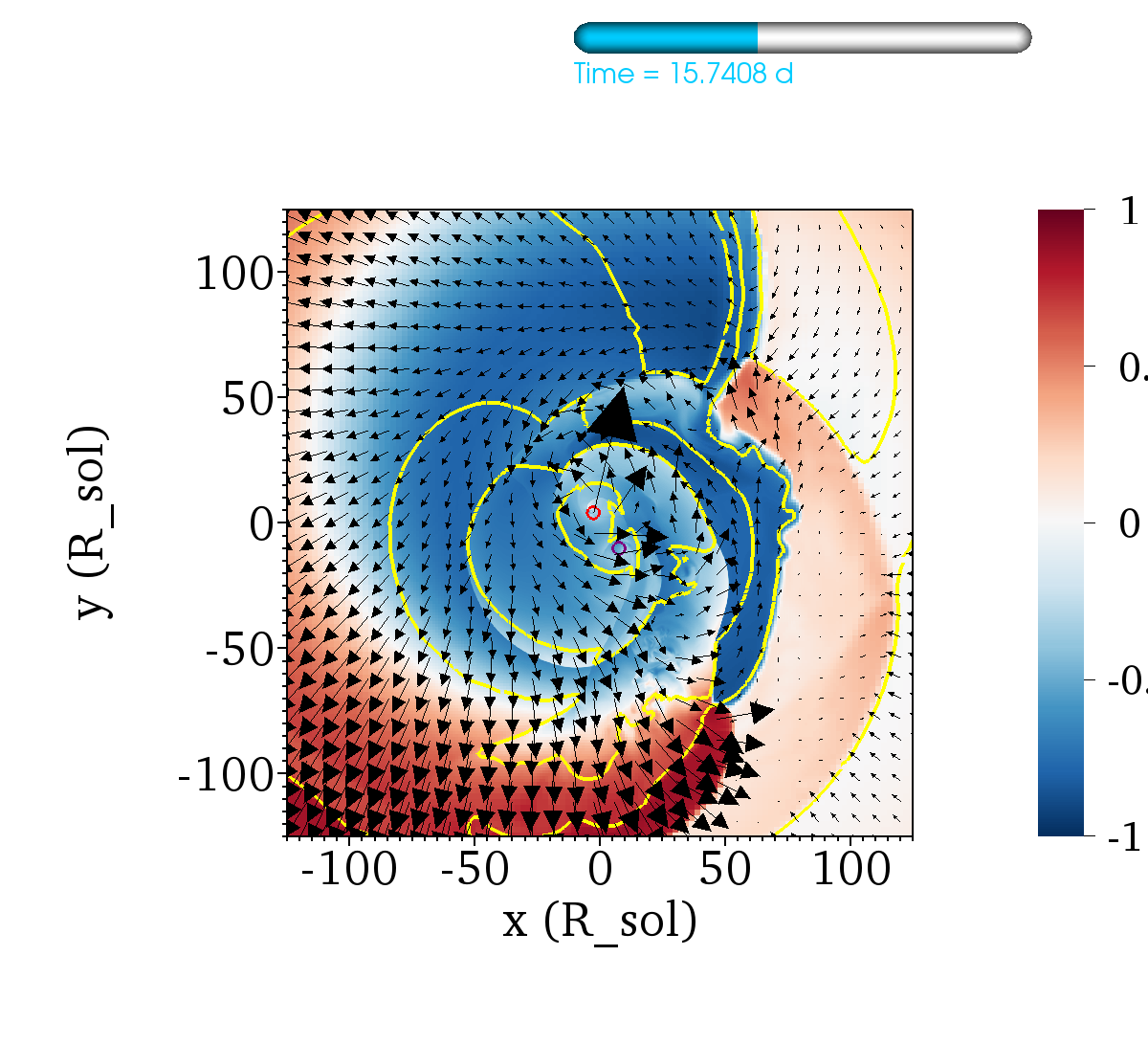}
      \includegraphics[height=47mm,clip=true,trim= 290 40 200 160]{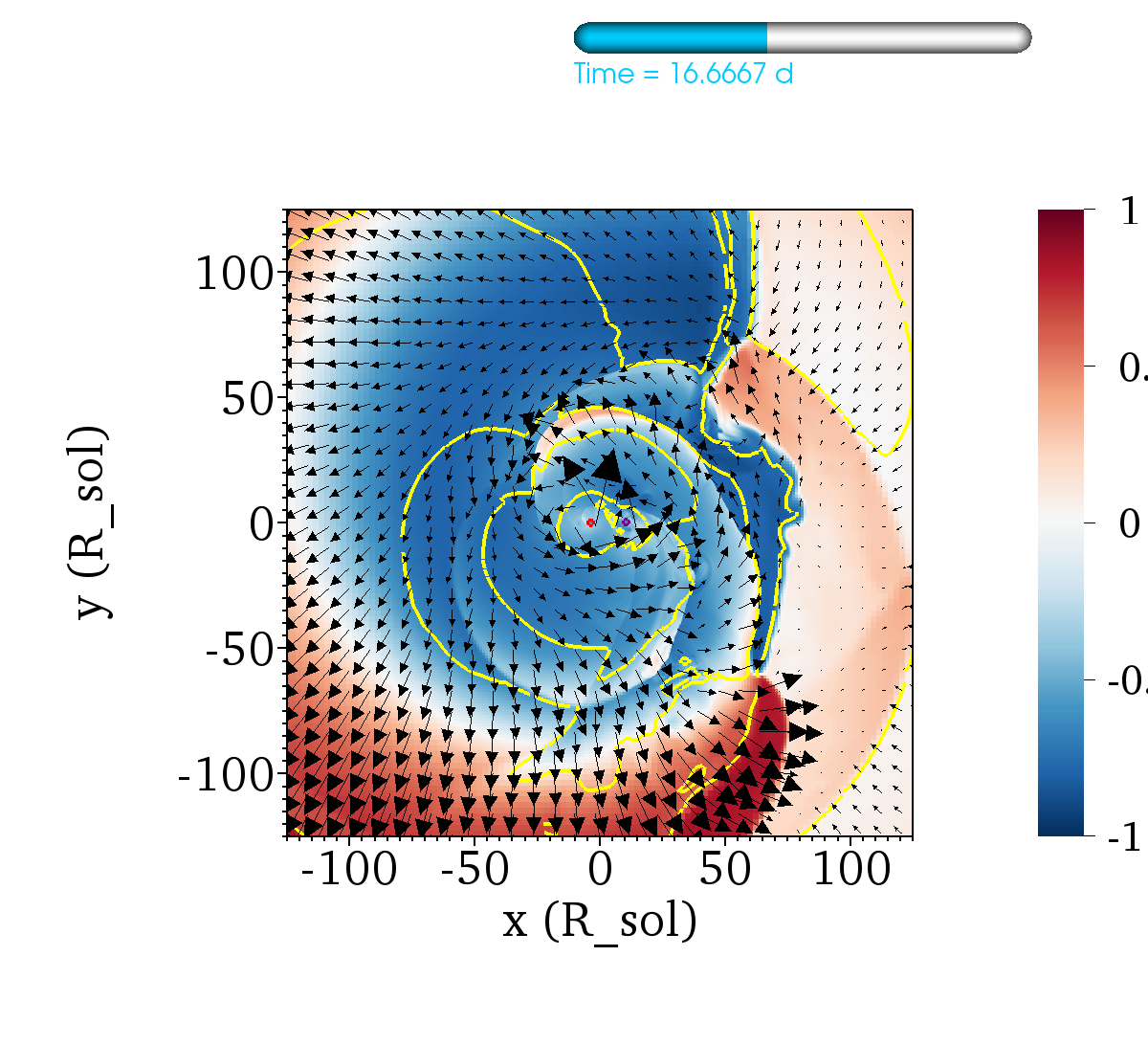}
      \includegraphics[height=47mm,clip=true,trim= 290 40 200 160]{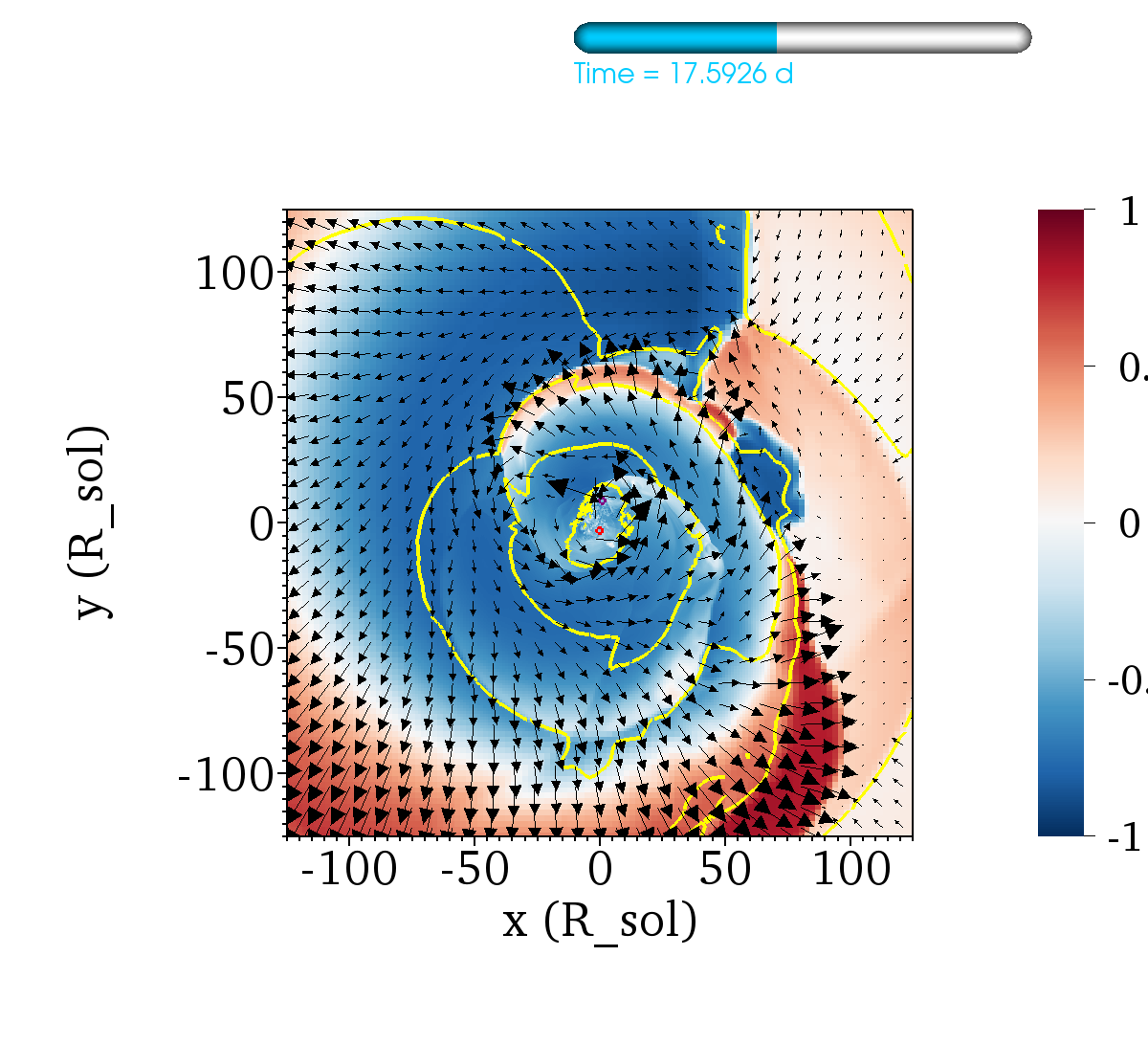}
      \includegraphics[height=47mm,clip=true,trim= 290 40 200 160]{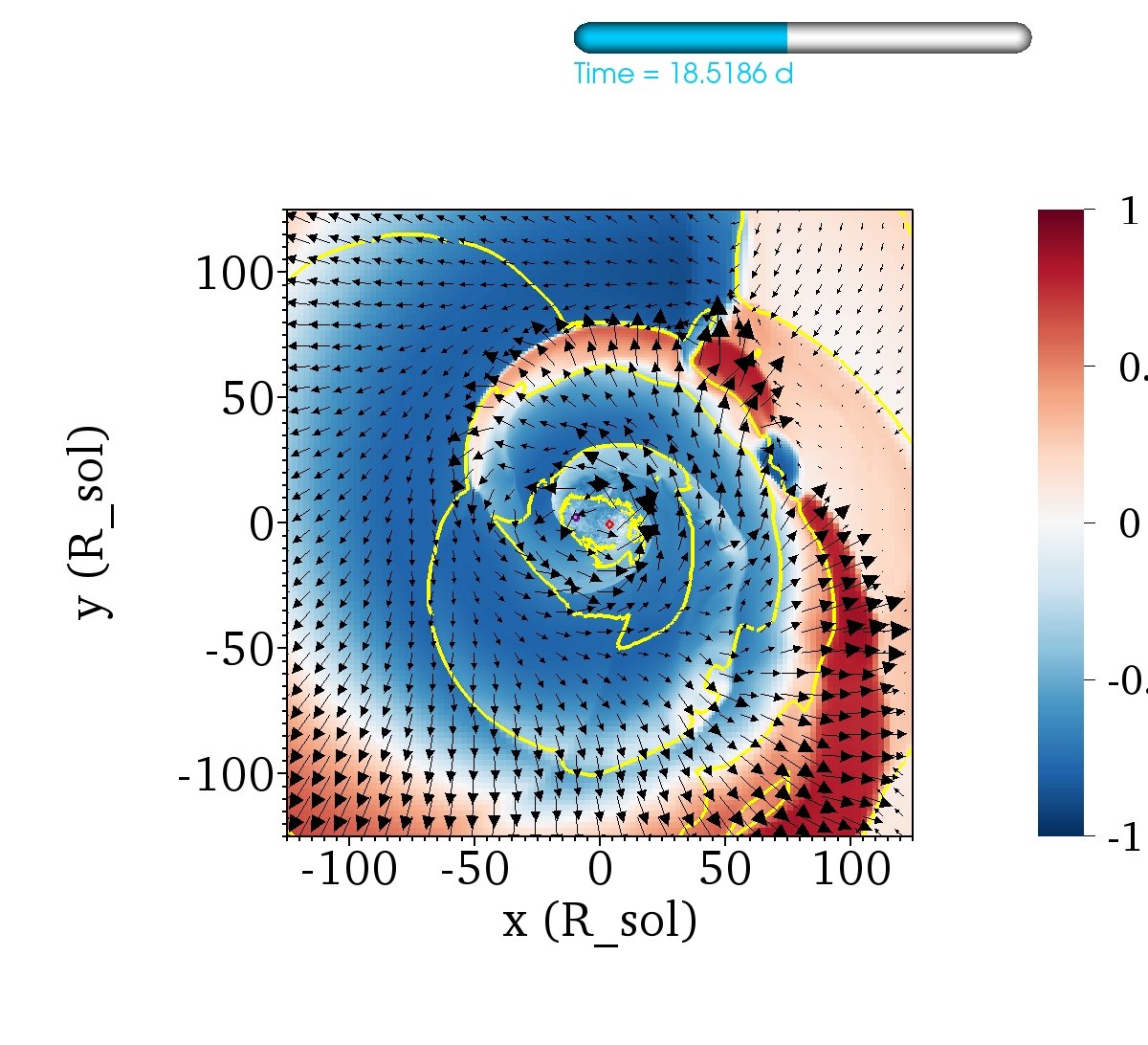}\\
    \>\includegraphics[height=47mm,clip=true,trim=  50 40 200 160]{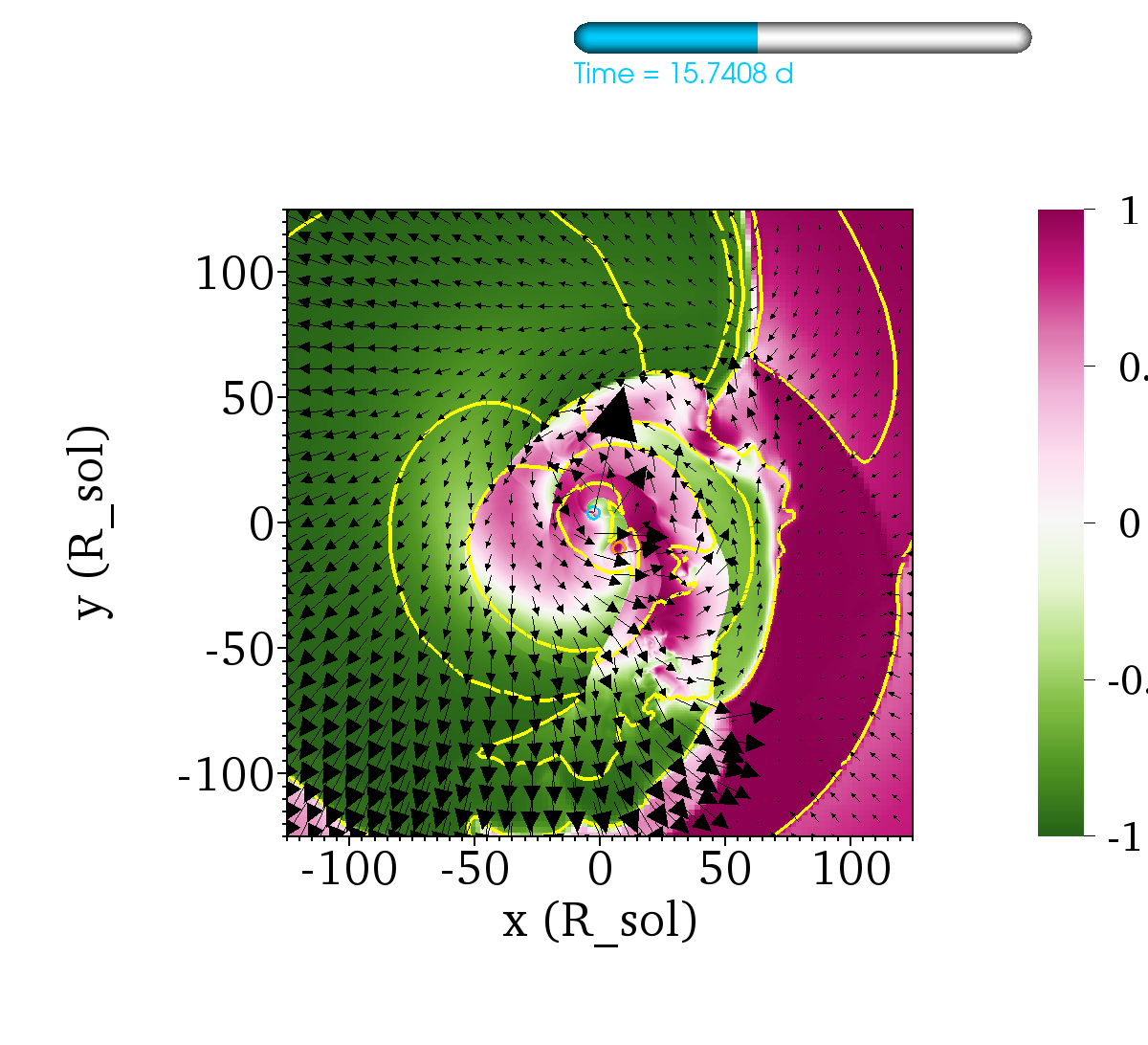}
      \includegraphics[height=47mm,clip=true,trim= 290 40 200 160]{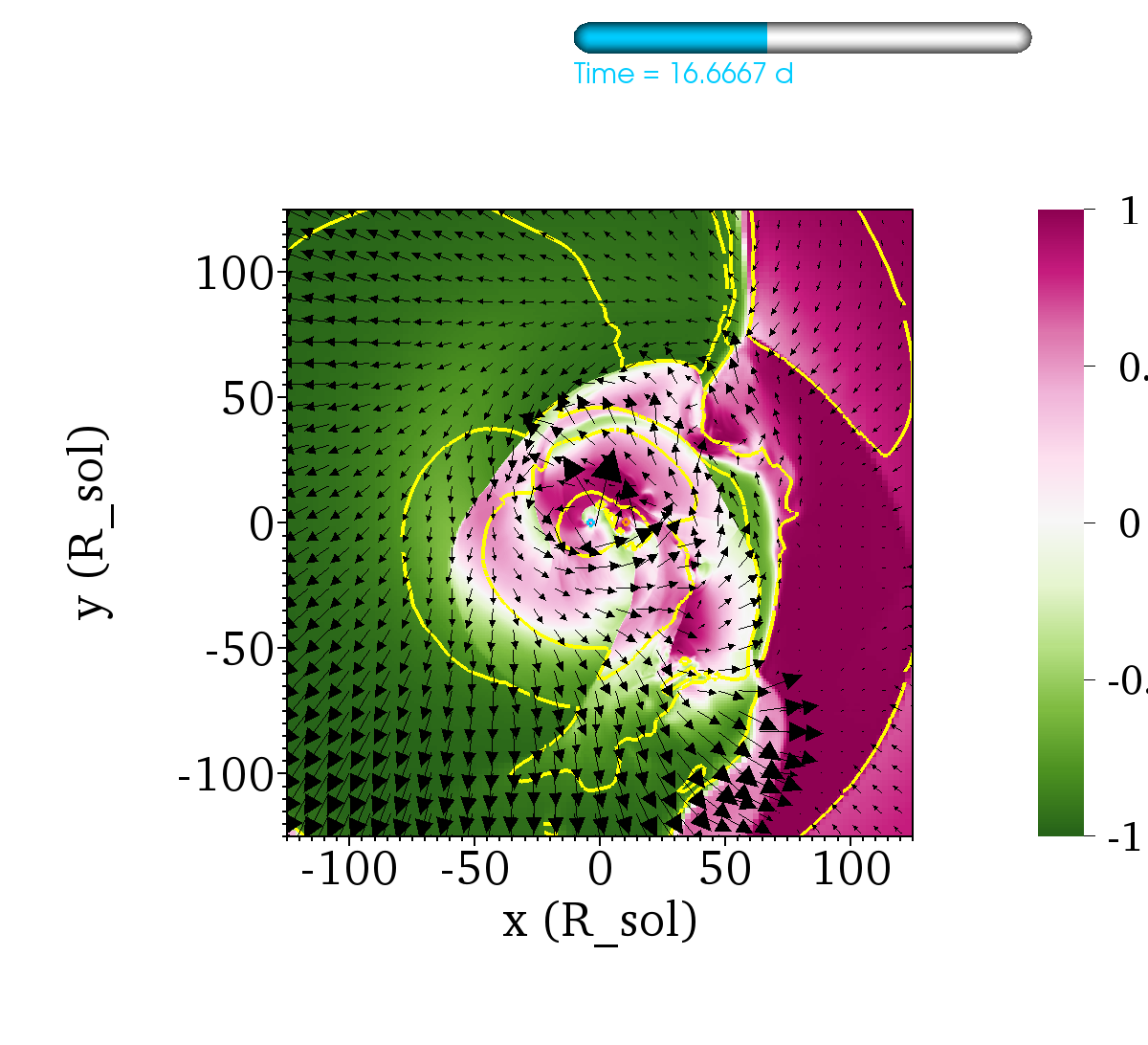}
      \includegraphics[height=47mm,clip=true,trim= 290 40 200 160]{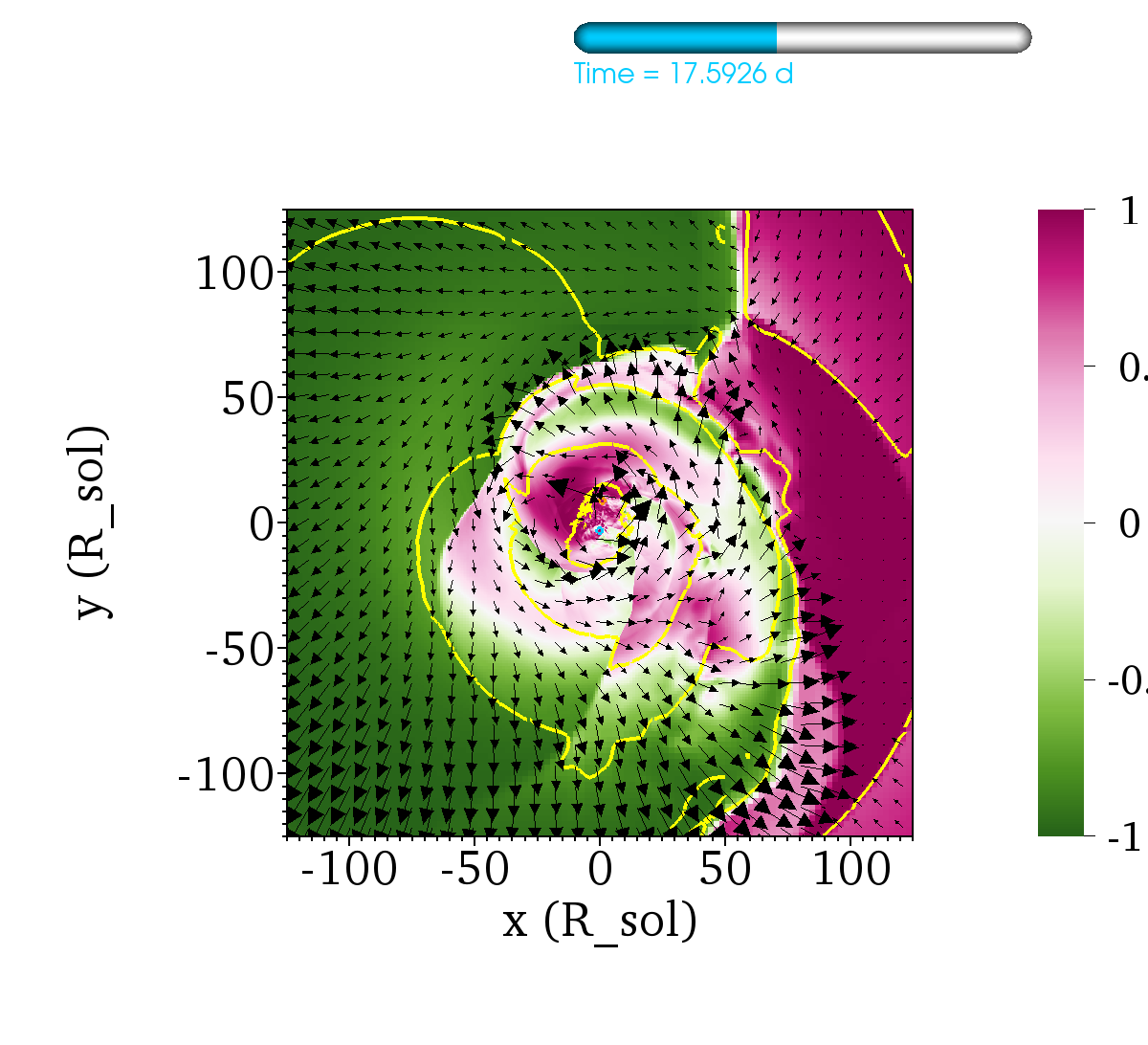}
      \includegraphics[height=47mm,clip=true,trim= 290 40 200 160]{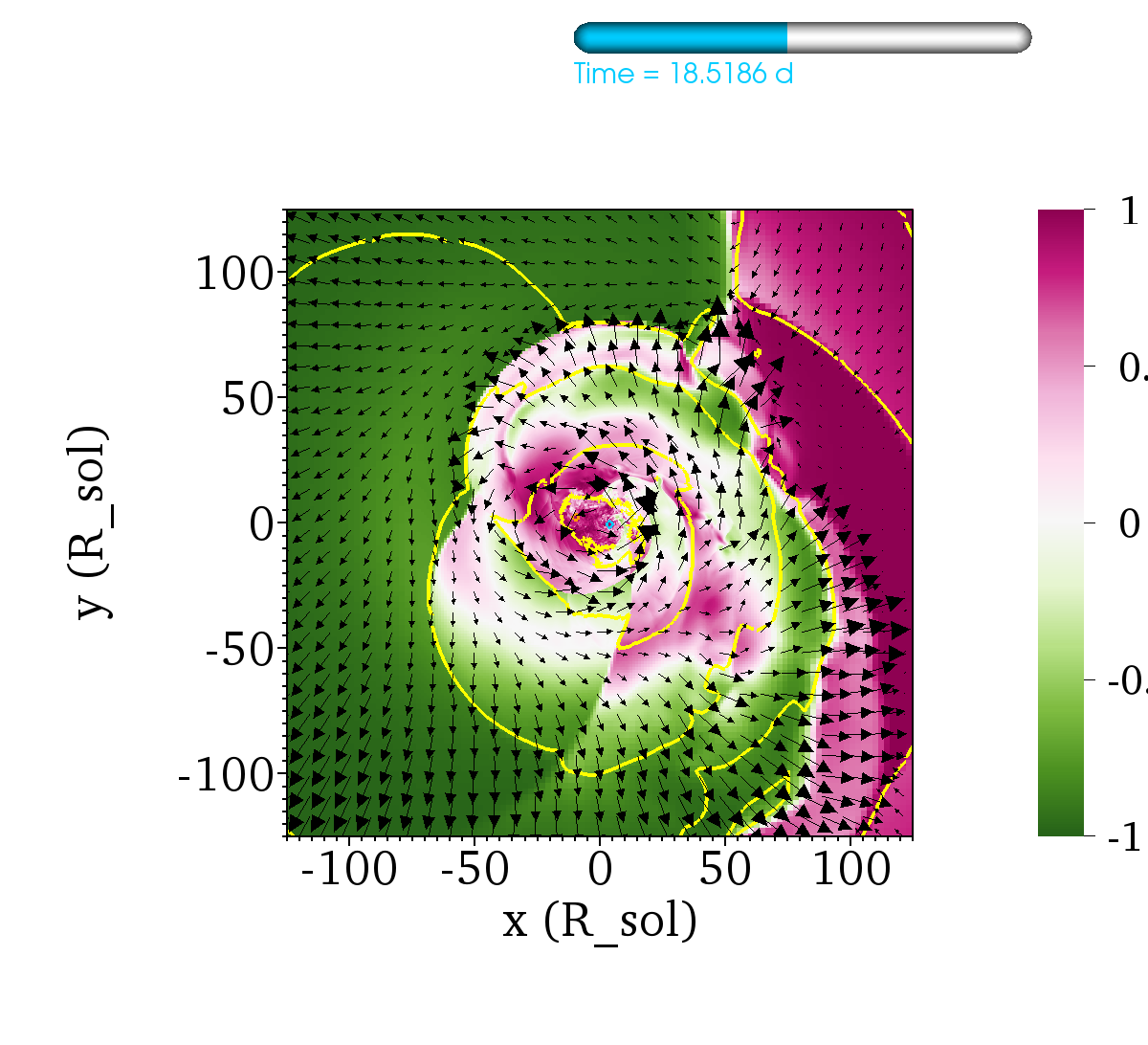}\\
  \end{tabbing}
  \caption{Similar to the top two rows of Fig.~\ref{fig:spatial} but zoomed in 
           and showing eight snapshots equally spaced in time (by about $0.9\da$) 
           between $t=12.0\da$ and $t=18.5\da$.
           Particles complete only a fraction of an orbital revolution between successive snapshots.
           In the top row, showing $t=12.0\da$ to $t=14.8\da$, 
           the most recently energized unbound material (red, near particle~2) becomes bound (blue) 
           when it loses kinetic energy as it collides with the bulk of the envelope material.
           In the second-from-bottom row, showing $t=15.7\da$ to $t=18.5\da$, 
           some of the bound material in the spiral wake of particle~2 becomes unbound 
           (above centre) as it moves into a lower density region.
           \label{fig:spatial_zoom}
         }            
\end{figure*}

\begin{figure}
    \includegraphics[width=\columnwidth,clip=true,trim= 50 40 40 160]{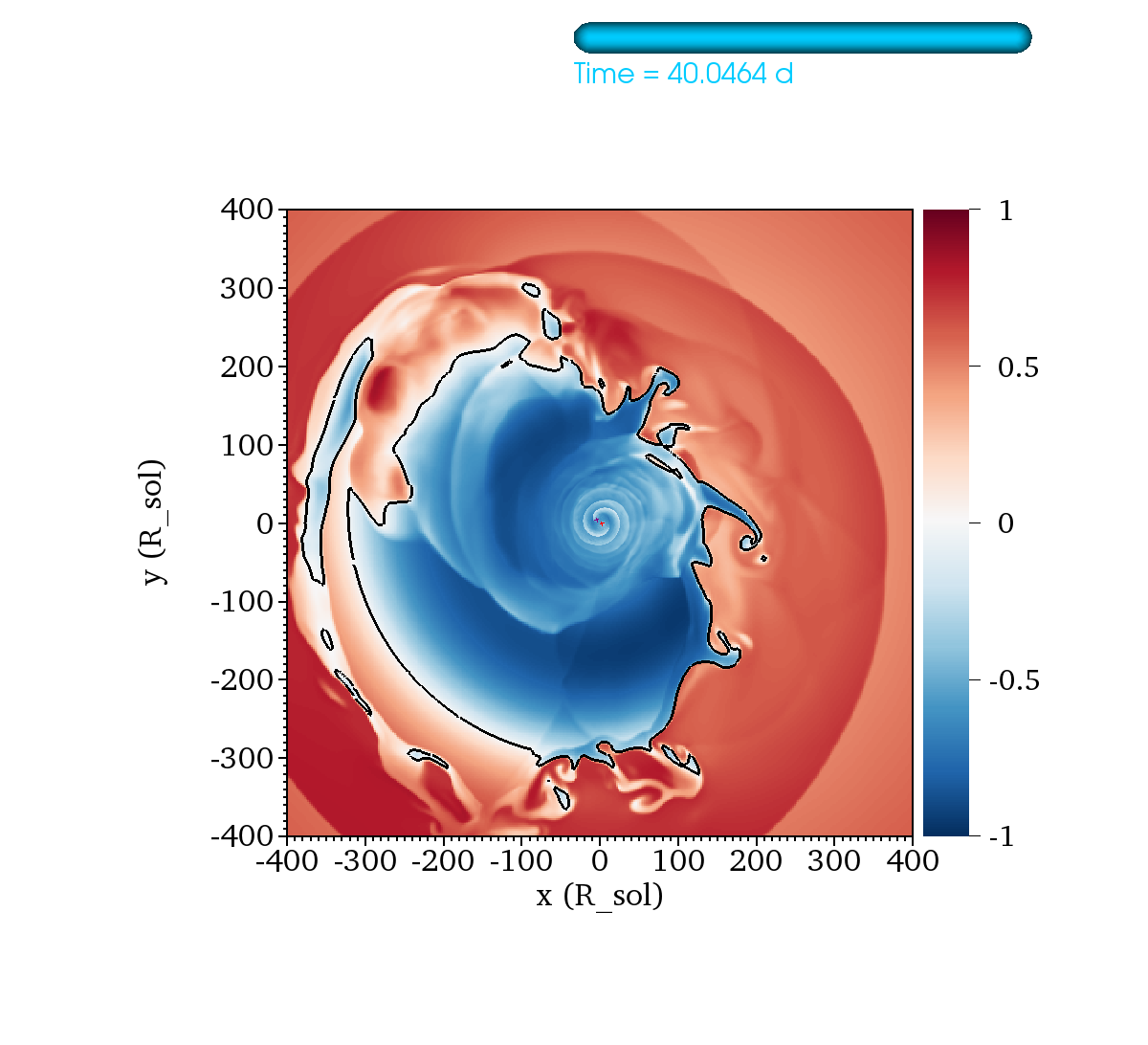}
  \caption{caption
           The normalized gas energy $\mathcal{E}_\mathrm{gas,norm}$
           with blue (red) showing bound (unbound) gas at $t=40\da$,
           with contours delineating the value zero.
           Recently formed islands of bound gas are seen near the envelope-ambient interface,
           showing how unbound envelope gas can become bound as it transfers energy to the ambient medium.
           The CM of the particles is at (0,0) and the primary and secondary particle softening spheres
           are represented by mauve and red contours, respectively 
           (near centre, barely visible due to the scale of the plot).
           \label{fig:ambient}
          }            
\end{figure}

To interpret Fig.~\ref{fig:unbound_mass} and the partial unbinding of the envelope,
we now explore the time-dependent spatial distribution of energy. 
We define a normalized energy density 
$\mathcal{E}_\mathrm{gas,norm}=\mathcal{E}\gas/\max(\mathcal{E}_\mathrm{bulk,gas}+\mathcal{E}_\mathrm{int,gas},
\;-\mathcal{E}_\mathrm{pot,gas})$
(where $\mathcal{E}_\mathrm{pot,gas}= \mathcal{E}_\mathrm{pot,gas-1} +\mathcal{E}_\mathrm{pot,gas-2} +\mathcal{E}_\mathrm{pot,gas-gas}$)
and plot snapshots  of $\mathcal{E}_\mathrm{gas,norm}$ in the orbital plane 
in the top row of Fig.~\ref{fig:spatial} for $t=5$, $10$, $20$ and $30\da$.
The quantity $\mathcal{E}_\mathrm{gas,norm}$ is  the local gas energy density 
normalized by either the magnitude of the local gas kinetic energy density (bulk and internal) or 
the magnitude of the local gas potential energy density, whichever is greater.
For our fiducial definition of unbound, $\mathcal{E}\gas\ge0$, 
blue corresponds to bound material while red corresponds to unbound material,
and $-1$ ($1$) means maximally bound (unbound).
Much of the ambient material is initially unbound due to its large internal energy density 
and large distance from the central mass concentration.
Contours show the gas density while the component of the velocity in the orbital plane is shown with arrows.

The second row of Fig.~\ref{fig:spatial} shows 
$(\mathcal{E}_\mathrm{int,gas}-\mathcal{E}_\mathrm{bulk,gas})/\max(\mathcal{E}_\mathrm{int,gas},\mathcal{E}_\mathrm{bulk,gas})$.
This is the difference between the local internal energy density and the local bulk kinetic energy density,
normalized by whichever of the two is largest.
Magenta (green) represents gas for which $\mathcal{E}_\mathrm{int,gas}$ is larger (smaller)
than $\mathcal{E}_\mathrm{bulk,gas}$ and the limits are $1$ (all internal)
and $-1$ (all bulk kinetic).
These plots correspond closely with similar plots for Mach number (not shown);
magenta (green) regions correspond to subsonic (supersonic) gas.
The third and fourth rows of Fig.~\ref{fig:spatial} show the same quantities as the top two rows,
but in a slice through the plane orthogonal to the orbital plane that intersects the particles
(the $x'$-$z$ plane).
Fig.~\ref{fig:spatial_zoom}  shows a sequence of eight snapshots for each quantity,
spaced by  $\sim 0.9\da$, between $t=12.0\da$ and $18.5\da$, 
now zoomed in by a factor of two compared with those of Fig.~\ref{fig:spatial}.

During the plunge-in, 
material is torn away from the envelope by the secondary,
forming a tidal bulge that wraps around in a spiral morphology, trailing the secondary in its orbit (density contours).
Gas closest to particle~2 in this spiral wake moves supersonically in a direction 
in between radially outward and tangential to the path of particle~2.
The wake contains highly supersonic unbound gas extending from particle~2 
(red and green in the topmost and second-from-top rows, respectively),
surrounding a bound region (blue and magenta/yellow) trailing particle~2.  Where the spiral wake
encounters the low density ambient medium, a spiral shock forms. 
This does not greatly effect the motion of the outward moving unbound gas 
(see Sec.~\ref{sec:efficiency}) which slows as it climbs out of the potential well.

At $t\approx11\da$, after roughly the first half-orbital revolution, 
newly unbound material near particle~2, 
followed  by particle~2 itself and the dense bulge  trailing it,
violently collide with dense gas in the bulk of the relatively undisturbed RG envelope.
This occurs as the inter-particle separation shrinks rapidly 
from dynamical friction while the secondary, with its near-side tidal bulge in tow, 
catches up to the tidal bulge of the RG on the far side of the RG from the secondary
that lags the particles in their orbit.
\footnote{\citet{Macleod+18a,Macleod+18b} see a similar morphology at a comparable stage in their simulations,
which have more realistic initial conditions than our own.}

From the collision, an almost radial spiral shock forms near the secondary 
that connects to the primarily azimuthal shock farther out.  
This can be seen in the top two rows of Fig.~\ref{fig:spatial_zoom},
showing the evolution of $\mathcal{E}_\mathrm{gas,norm}$ 
and $(\mathcal{E}_\mathrm{int,gas}-\mathcal{E}_\mathrm{bulk,gas})/\max(\mathcal{E}_\mathrm{int,gas},\mathcal{E}_\mathrm{bulk,gas})$
between $12.0\da$ and $14.8\da$.
Note that the velocity of the gas immediately left of the shock 
(shown by vector arrows) decreases as the shock forms.
The shock structure widens in time as more gas gets shocked.
Correspondingly, bulk kinetic energy is converted to internal energy at $t\approx13\da$,
consistent with the quantitative evolution of the  contributions to the global energy budget shown in Fig.~\ref{fig:energy_time}.

However, the total energy density $\mathcal{E}\gas$ in the central part of the spiral wake 
near the secondary decreases between $t\approx13\da$ and $t\approx15\da$,
causing unbound material to become bound once again.
This is visible in the top part of Fig.~\ref{fig:spatial_zoom}, 
where red material near the secondary becomes blue.
We can see that at $t\approx13\da$ (termination of plunge-in), 
the unbound part of the wake ``detaches'' from the secondary because the secondary no longer 
supplies enough energy to unbind the material in its immediate surroundings.
This material lies deep in the potential well and is surrounded by dense overlying layers that impede its outward motion.
The part of the spiral wake that transitions from unbound to bound between $t\approx13\da$ and $t\approx15\da$ 
explains the peak and subsequent dip in  total unbound mass in Fig.~\ref{fig:unbound_mass}
at $t\approx13\da$ (blue solid curve).

The subsequent rise in the unbound mass starting at $t\approx15\da$ and lasting for a few days 
can be explained with reference to the bottom row of Fig.~\ref{fig:spatial_zoom}, 
which shows the time frame $t=15.7\da$ to $t=18.5\da$.
At this time (about 1.5 orbital revolutions after the start of the simulation)
the shocked spiral structure trailing the secondary moves at an angle $<90^\circ$
with respect to the far-side tidal bulge gas, and their relative velocity
is much smaller than when they first collided.
The inertia of the dense spiral wake 
and the negative pressure gradient (nearly aligned  to the density gradient) 
allow the wake to accelerate  up to a nearly constant speed  toward larger $y$.
This happens in spite of the work  done by gravity so the overall energy density of the wake increases.

This process repeats  during the next orbital revolution, 
resulting in a third layer of unbound gas that can be seen as the innermost strip of red on the upper-left of the rightmost panel 
in the top row of Fig.~\ref{fig:spatial} at $t=30\da$.
By this time, a separate spiral wake  trails behind particle~1, 
but this wake does not gain  enough energy as it moves outward to become unbound.
In subsequent orbital revolutions, 
a smaller amount of material transitions from bound to unbound,
now toward positive $x$; an example is visible in the same panel at $t=30\da$ (right of centre in the plot).
In this snapshot, Rayleigh-Taylor (RT) instability-produced ``fingers'' are  visible
at large distances from the centre.
Such features are formed as the outward-moving interface between inner dense gas and outer diffuse gas decelerates.

After $t\approx19\da$, 
pockets of gas can be seen to transition from unbound to bound near the edge
of the expanding envelope.
This is most obvious late in the simulation. 
In Fig.~\ref{fig:ambient} we show a snapshot at $t=40\da$. 
In addition to the RT fingers of bound material mentioned above,
recently formed isolated blue ``islands'' of bound material are visible.

\subsection{Efficiency of partial envelope removal}
\label{sec:efficiency}
Since unbound material will never have exactly zero energy density, 
there is always an efficiency associated with the energy transfer process.
To get an idea about how much energy is ``wasted'' by increasing the energy density of already unbound material, 
we plot various energy components with time for gas that is \textit{unbound} in the bottom panel of Fig.~\ref{fig:energy_time}.
From the orange line, we see that during the simulation
a net amount of about $0.2\times10^{47}\erg$ of energy is gained by the unbound gas.
As the change in gas energy during the simulation is $1.3\times10^{47}\erg$,
the fraction that ends up in already unbound gas is about $15\%$.
This gives us an estimate of how much of the particle energy is wasted.

How does the energy transfer to unbound material take place and what happens subsequently?
We see from the bottom panel of Fig.~\ref{fig:energy_time} that most of the increase in energy of the unbound material
occurs in the first $13\da$.
This is consistent with the change in the unbound mass $\rmD M\unb$ also peaking at $t\approx13\da$.
The energy transferred is mainly in the form of kinetic energy,
as material is launched outward during plunge-in.
Subsequently, the unbound gas, whose mass remains almost constant after $t=13\da$,
sees much of its bulk kinetic energy get converted to internal energy and potential energy.

There is another way in which energy transfer to the envelope is inefficient.
To expand, the envelope must displace ambient material, which has significant pressure and mass in our simulation.
Work must be done by the envelope against thermal pressure of the ambient material,
ram pressure as the envelope expands into ambient gas,
and also to displace ambient material against gravity.
These terms can respectively be estimated as $\sim (4\pi/3)P\amb r\final^3$,
$\sim (4\pi/3)\rho\amb v_r^2 r\final^3$ and $\sim (4\pi/3)\Gn r\final^2(M\1+M\2)\rho\amb$,
where $P\amb=1\times10^5\dynecmcm$ is the ambient pressure, 
$\rho\amb=7\times10^{-9}\gcmcmcm$ is the ambient density, 
$M\1=2\Msun$ is the primary mass, $M\2=1\Msun$ is the secondary mass, 
$r\final\sim3\times10^{13}\cm$ is the radius of the envelope at $t=40\da$,
and $v_r\sim40\kms$ is a typical speed at which the envelope expands into the surroundings.
With these expressions we obtain $\sim0.1\times10^{47}\erg$ for each work term.
Thus, $\sim0.3\times10^{47}\erg$ may have been transferred from the envelope to the ambient medium
during the course of the simulation.
This is a small amount compared to the total envelope energy, 
but indicates that the expansion of the envelope would have been slightly faster
within a less dense or lower pressure ambient medium.
It also explains the decrease in unbound mass after $t\approx19\da$.
A circumbinary torus is likely to remain from the RLOF stage
preceding CEE, and this material would shape the envelope and redirect its expansion 
\citep{Metzger+Pejcha17,Macleod+18b,Reichardt+18}.

\subsection{Timescale for ejecting the envelope and final separation}
\label{sec:ejection}
The average rate of energy transfer 
from the particles to the gas is approximately constant at the end of the simulation,
and equal to about $0.03\times10^{47}\erg\da^{-1}$ 
(final average slope of orange curve of top panel of Fig.~\ref{fig:energy_time}).
Of this transfer rate, about $0.001\times10^{47}\erg\da^{-1}$, a negligible fraction, 
is being transferred from the particles to gas that is already unbound 
(final slope of orange curve of bottom panel of Fig.~\ref{fig:energy_time}).
Thus, although the envelope continues to gain energy at a relatively high rate, 
this energy is being gained by material that is still bound by the end of the simulation.
Assuming that this energy transfer rate of $0.03\times10^{47}\erg\da^{-1}$ remains constant, 
one can estimate how long it would take for the gas to attain zero total energy, 
and we find it would take an additional $38\da$.
However, this calculation uses the total gas energy, which includes the energy of the ambient medium,
equal to $E\amb\sim0.5\times10^{47}\erg$.
Thus, to obtain a more accurate estimate, 
we subtract this ambient energy from the value of $E\gas$ at $t=40\da$ given in Tab.~\ref{tab:energy_terms},
which gives an envelope gas energy of $E\env\sim-1.65\times10^{47}\erg$. 
Then the additional time needed for the envelope to attain $E\gas=0$ would be about $55\da$ after the end of the simulation at $t=40\da$.

Now, from Sec.~\ref{sec:efficiency} we know that not all of the liberated particle orbital energy 
will be transferred to bound material, and that this leads to an efficiency factor $\epsilon$,
found to be about $85\%$ in the first $40\da$ (that is, $15\%$ of the energy gets wasted).
Assuming an efficiency of $\epsilon=0.85$ for the remainder of the evolution,
the time calculated above must be divided by $\epsilon$, giving $\sim65\da$.
As the system continues to evolve, less and less gas would remain bound,
so we would expect that more of the released orbital energy would go into unbound gas,
resulting in reduced efficiency.
With an efficiency of only $10\%$, the released orbital energy would have to be about $16.5\times10^{47}\erg$,
and the timescale for ejecting the envelope would be $\sim550\da$, 
or about $1.5\yr$, which is small enough to be consistent with observations of post-CE binary systems,
for which the envelope has already been ejected.
In \citet{Ohlmann+16a}, the  orbital energy decay rate of the particles 
decreases to become much smaller by the end of their simulation at $t\sim130\da$ than at $t=40\da$.
The assumption that this decay rate remains constant is therefore probably too optimistic.

The orbital energy of the particles at $t=40\da$ is about $E_{1-2}(40\da)=-0.95\times10^{47}\erg$ (Tab.~\ref{tab:energy_terms}).
Then, equating the gas energy with the difference in particle energy between $t=40\da$ and envelope ejection, 
multiplied by the efficiency $\epsilon$, we derive the following expression for the final separation:
\begin{equation}
  a\final \sim \frac{\Gn M\core M\2}{2}\left(\frac{E\amb-E\gas(40\da)}{\epsilon} -E_{1-2}(40\da)\right)^{-1}.
\end{equation}
This estimate is independent of the rate of energy transfer 
(and foreshadows our discussion of the CE EF in Sec.~\ref{sec:energy_formalism}).
Putting $\epsilon=1$ gives the upper limit $a\final\sim2.9\Rsun$, 
while for $\epsilon=0.85$ we obtain $a\final\sim2.6\Rsun$ and for $\epsilon=0.1$ we obtain $a\final\sim0.4\Rsun$.

\section{Particle centre of mass motion and planetary nebula--central star offsets}
\label{sec:particle_CM_motion}

\begin{figure}
  \includegraphics[width=0.985\columnwidth,clip=true,trim= 0 0 0 0]{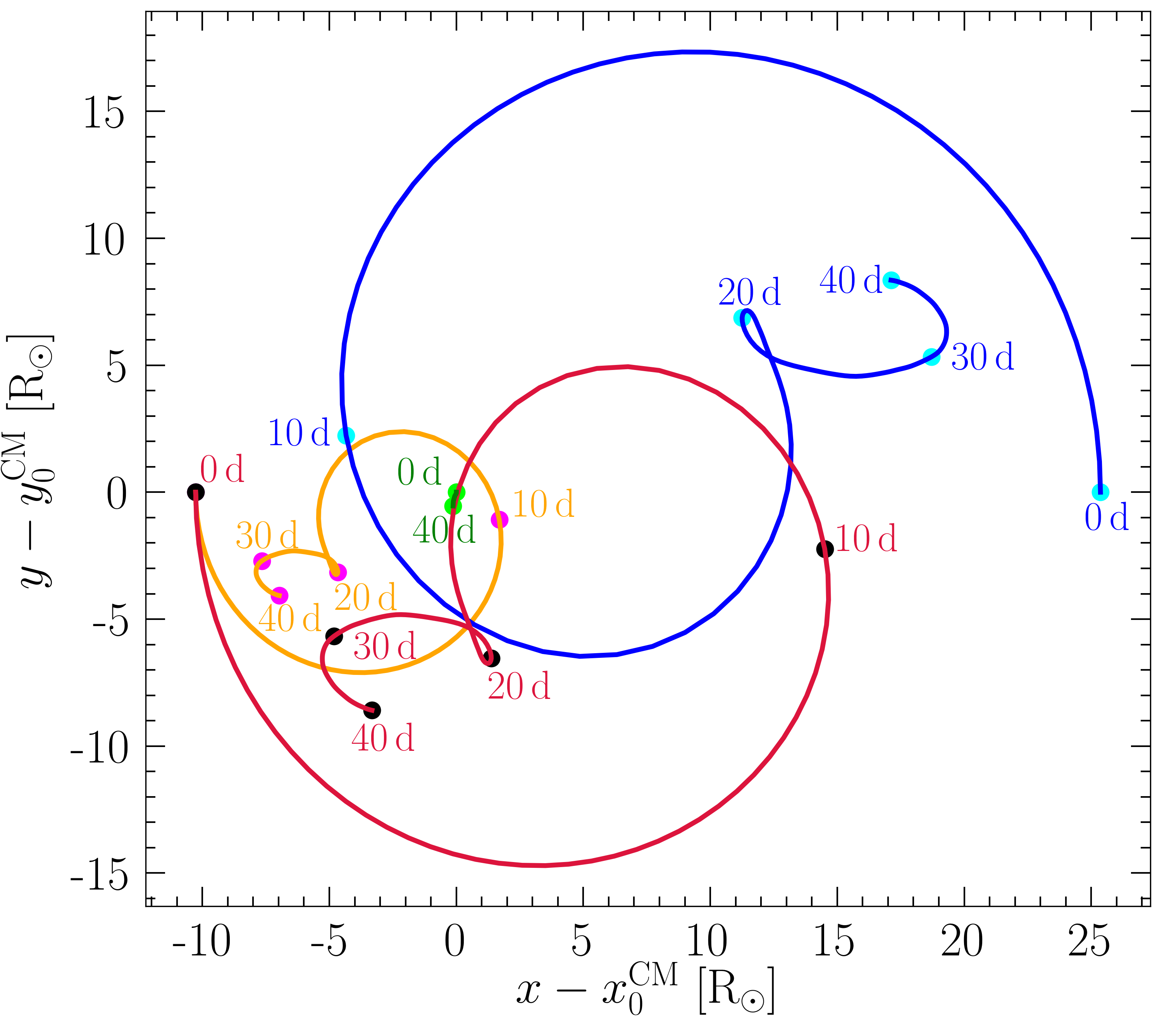}
  \includegraphics[width=\columnwidth,clip=true,trim= 0 0 0 0]{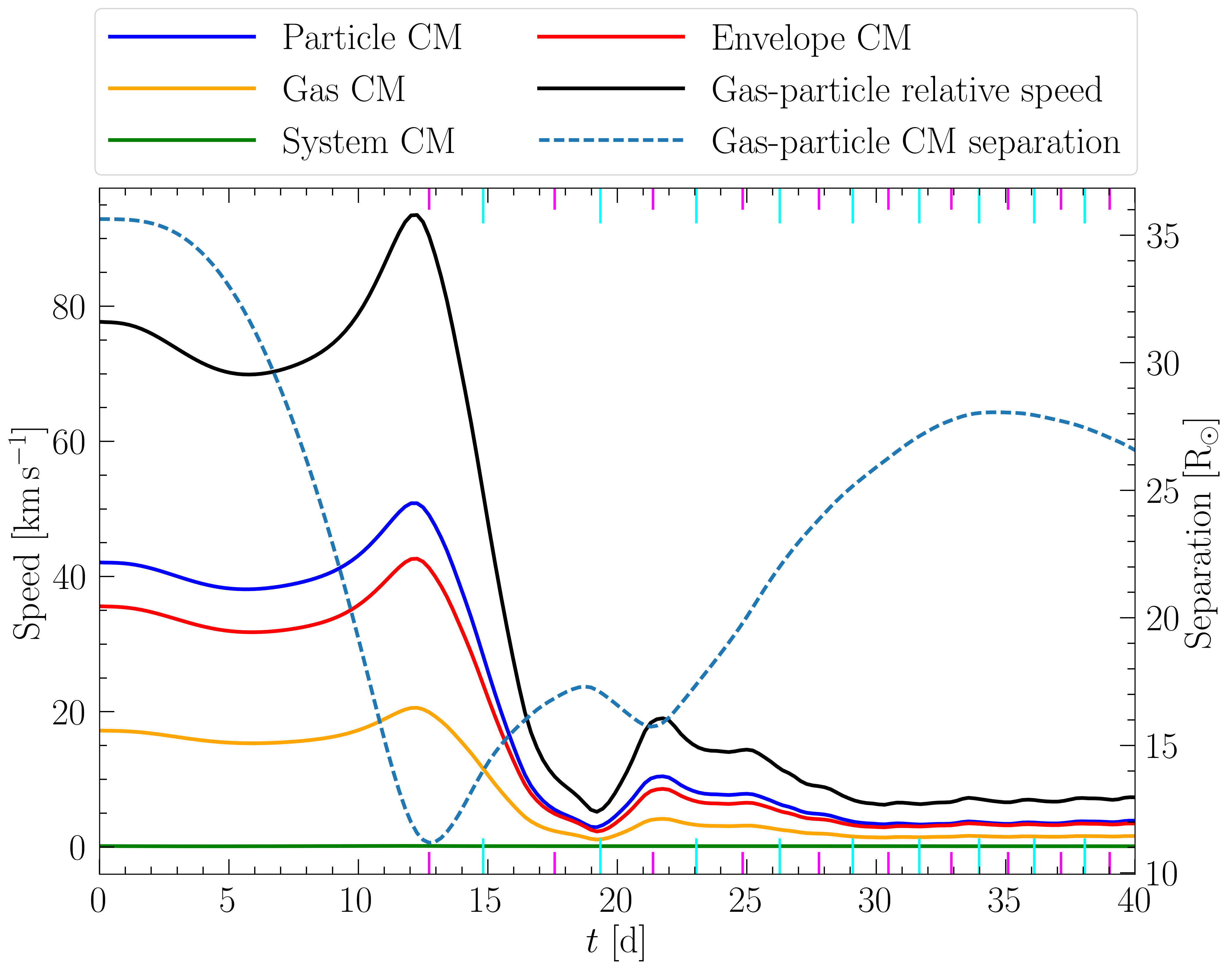}
  \caption{\textit{Top:} Motion in orbital plane of the particle CM (blue), 
           gas CM (orange), envelope CM (red) and net system CM (green).
           The system CM moves slowly and gradually downward in the plot during the course of the simulation.
           \textit{Bottom:} Evolution of the speed relative to the reference frame in which the simulation is carried out,
           for the particle CM (blue), gas CM (orange), envelope CM (red) and net system CM (green).
           The relative speed between the envelope CM and particle CM is shown in black.
           Also plotted, using the right vertical axis, is the separation between the envelope CM and particle CM (dashed).
           \label{fig:CM_motion}
          }            
\end{figure}
The asymmetry of the gas distribution in the orbital plane rapidly evolves, moving the particle and gas centres of mass (CM)
oppositely in the simulation frame,  
while nearly conserving  linear momentum 
\citep[see also][]{Sandquist+98,Ohlmann+16a}.
The top panel of Fig.~\ref{fig:CM_motion},
shows the paths in the orbital plane traced  by the particle CM (blue), 
gas CM (accounting for both envelope and ambient gas, orange), 
envelope CM (red), and system CM (green). 
The position and motion of the envelope CM are obtained by assuming that the CM of ambient gas
remains fixed during the simulation.
The system CM position remains relatively fixed with small speed ($<0.2\kms$);
deviations in its position are caused by small errors in linear momentum conservation.

The bottom panel of Fig.~\ref{fig:CM_motion} shows the speed of the particle CM versus time in blue, 
that of the gas CM in orange, that of the envelope CM in red, 
and that of the system CM in green, as well as the relative speed between envelope CM and particle CM in black.
Until the end of the plunge-in phase at $t\approx13\da$, 
the relative speed between the envelope CM and particle CM is $70$-$94\kms$,
and the final relative speed is $7$ to $8\kms$.
\citet{Sandquist+98} found the speed of the particle CM relative to the 
system CM frame at the late stages of CEE to be $3\kms$,
for their simulation of a $3\Msun$ AGB star with $0.4\Msun$ companion.

Our standard definition of `unbound,' $\mathcal{E}\gas\ge0$, 
does not account for relative motion of the particle binary system CM and that of the disrupted envelope gas. 
The particles' kinetic energy  and bulk kinetic energy of the gas are given in the inertial frame of the simulation, 
which  is the system CM frame if small deviations from linear momentum conservation are neglected.
To account for the relative CM motion 
while neglecting non-inertial effects in the frame of the particle CM, 
the bulk kinetic energy of gas would be 
$E^\mathrm{CM_{1-2}}_\mathrm{bulk,gas}= \tfrac{1}{2}\int\rho(\bfx)|\bfv(\bfx)-\bfv^{\mathrm{CM}_{1-2}}|^2\rmd V$.
This refined definition implies that gas moving faster (slower) 
with respect to the particle CM than the system CM 
is more (less) unbound than for the standard definition.
As seen in Fig.~\ref{fig:unbound_mass},
the change in unbound mass using this modified definition (dashed-double-dotted blue) has a higher maximum 
but lower final value than the standard definition (solid blue).
The opposite motion of the envelope CM and particle CM leads to an increase in the total mass of unbound gas at early times.
However, the particles eventually carry 
their own individual gas ``envelopes''  \citepalias{Chamandy+18},
which likely explains the reduction in the unbound mass seen at late times. 

\subsection{PN central star offsets}
\label{sec:offsets}
Several bipolar PNe exhibit an offset between the binary central star and PN centre 
(e.g. MyCn~18: \citealt{Sahai+99,Clyne+14,Miszalski+18}; Hen~2-161: \citealt{Jones+15}; Abell~41: \citealt{Jones+10}), 
and the Etched Hourglass Nebula MyCn~18 is the best studied among them. 
The direction of the offset seen in MyCn~18 matches the direction of the proper motion,
which suggests that the offset is caused by the proper motion \citep{Miszalski+18}.
Such proper motion could be produced by asymmetric mass loss, as seen in our simulation.
\citet{Miszalski+18} argue that previous explanations to explain the offsets in PNe are inadequate.
We note, however, that \citet{Mclean+01} speculated that asymmetric mass loss might induce such offsets.
\citet{Soker99} speculated that asymmetric mass loss during the onset of a CE phase 
could explain the offset of the outer rings of the supernova SN 1987A relative to the central star.

The observed distance to MyCn~18 is $618\pm101\au$ \citep{Miszalski+18} 
and the estimated time since the end of the CE phase is $\sim2700\yr$ \citep{Clyne+14,Miszalski+18}.
This requires  a mean  relative velocity of $\sim1\kms$ 
between the PN central star and  nebula in the plane of the sky to explain the offset
if the motion started at the end of the CE phase. 
The speeds  of $\sim4$-$8\kms$ that we obtain for the particle CM 
relative to the inertial frame and envelope CM
are of the required order of magnitude at the end of our simulation. 
The direction of the observed offset is within $5^\circ$ of the PN minor axis, 
and likely parallel to the orbital plane of the binary \citep{Hillwig+16}.
This agrees with the motion of the particle CM in our simulation,
whose velocity in the $z$-direction perpendicular to the orbital plane has magnitude $\leq 0.3\kms$ during the simulation,
with average $z$-velocity only $-6\times10^{-3}\kms$ between $t=30\da$ and $t=40\da$.
Our simulation time is short compared to the age of PNe. Nevertheless,  
 we  plot the separation between the particle CM and envelope CM as a function of time
(dashed line in the bottom panel of Fig.~\ref{fig:CM_motion}).
This separation first decreases with time as the secondary plunges toward the primary core and bulk of the envelope, 
and then increases as a result of the asymmetric mass loss.

For the MyCn~18 system,
\citealt{Miszalski+18} obtain primary and secondary masses of $0.6\pm0.1\Msun$ and $0.19\pm0.05\Msun$ respectively, 
whereas our particle masses are $M\core=0.4\Msun$ and $M\2=1\Msun$.
Observations provide only a plane-of-the-sky projection, 
and thus a minimum of the full 3D offset, which would require a larger offset velocity. 

More realistic simulation initial conditions--such as those which start from the RLOF--may 
result in somewhat more symmetric mass ejection \citep{Reichardt+18} 
and hence somewhat smaller relative CM speeds than we find. 
Nevertheless, because the relative motion between the particle CM and envelope seen in our simulation
is consistent with observations, our proposed mechanism for such offsets warrants further study.
    
\section{Energy formalism}
\label{sec:energy_formalism}
A common approach for quantifying envelope
unbinding in CEE is the so called ``energy formalism'' (EF). 
As expressed  in Eq. 3 of \citet{Ivanova+13a}, this is
\begin{equation}
  \label{alpha}
  \frac{\Gn M\1 M\oneenv}{\lambda R\1} = 
                                         \alpha\CE \frac{\Gn M\2}{2}\left(\frac{M\core}{a\final} 
                                        -\frac{M\1}{a\init}\right)
\end{equation}
where $M\oneenv= M\1 -M\core$, the quantities $\alpha\CE$ and $\lambda$ are parameters,
$a_i$ and $a_f$ are the initial and final orbital radial. 
The formula applies only when 
`final' refers to the time at which the envelope becomes completely unbound
such that drag is eliminated and the inspiral halts.
The left-hand-side (LHS) is the envelope `binding energy,' 
which includes the \textit{negative} of the potential energy due to the gas--particle~1 gravitational interaction 
as well as that due to gas self-gravity.
The parameter $\lambda$ can be calculated from first principles for a known envelope density profile.%
\footnote{Alternatively, $\lambda$ can be combined with $\alpha\CE$, resulting in a single parameter $\lambda\alpha\CE$.}
Following convention,  the`binding energy' 
also includes the negative of the envelope internal energy,
so the equation of state must also be known to compute $\lambda$.
The right-hand-side (RHS) of equation (\ref{alpha}) is the energy used to unbind envelope gas,
and  equals the negative of the change in the orbital energy of the system between $t=t\init$, 
when $a=a\init$, and $t=t\final$, when $a=a\final$, multiplied by an efficiency factor $\alpha\CE$.
The value of $\alpha\CE$ estimated from population synthesis studies  is  $0.1\le \alpha\CE\le 0.3$ 
\citep{Davis+10,Zorotovic+10,Cojocaru+17,Briggs+18}, 
though it is still largely unknown and could also vary between different types of objects.
Equation~\eqref{alpha} can be  inverted to give an expression for $a\final$.
In the limit $\alpha\CE\ll1$,
equation~\eqref{alpha} leads to  
$a\final\sim(\alpha\CE\lambda/2)(M\core/M\1)(M\2/M\1)(1-M\core/M\1)^{-1}R\1$.
This asymptotic expression produces larger values than equation~\eqref{alpha}
by only $\sim10\%$ for $\alpha=0.25$ and the other choices of parameter ranges used in this work,
and is convenient for  estimating how $a\final$ depends on various parameters, 
although we use the full expression to obtain numerical values.

\subsection{Applying the energy formalism}
\label{sec:applying}

\begin{table*}
  \begin{center}
  \caption{Similar to Tab.~\ref{tab:energy_terms}, but now showing the initial energy, in units of $10^{47}\erg$,
           for the initial condition of the simulation (RGB primary with $a\init=49\Rsun$, fourth column from left) as well
           as for three other initial conditions involving the same secondary but a different initial separation
           and/or an AGB, rather than RGB, primary (columns~5-7).
           Unlike for Tab.~\ref{tab:energy_terms}, values do not include the contribution of the ambient medium.
           Further, we assume a Newtonian potential for $|\bm{r}-\bm{r}\1|<r\soft$
           rather than the spline potential as in Table~1.
           This increases the magnitude of the envelope-particle~1 potential energy term by $\sim0.02\times10^{47}\erg$.
          \label{tab:initial_energy_terms}
          }
  \begin{tabular}{lllrrrr}
    \hline
				&				&
&\multicolumn{2}{c}{Red giant}			&\multicolumn{2}{c}{Asymptotic giant}		\\	
Energy component at $t=0$	&Symbol				&Expression					
&$a\init=49\Rsol$	&$a\init=109\Rsol$	&$a\init=124\Rsol$	&$a\init=284\Rsol$	\\
\hline                  	 
Particle~1 kinetic		&$E_\mathrm{bulk,1,i}$		&$\tfrac{1}{2}M\core v\core^2$			
&$0.05$			&$0.02$		&$0.03$		&$0.01$	\\
Particle~2 kinetic		&$E_\mathrm{bulk,2,i}$		&$\tfrac{1}{2}M\2v\2^2$				
&$0.49$			&$0.22$		&$0.17$		&$0.07$	\\
Particle-particle potential	&$E_\mathrm{pot,1-2,i}$		&$-\Gn M\core M\2/a$				
&$-0.28$		&$-0.12$	&$-0.16$	&$-0.07$\\
\hline
Envelope bulk kinetic		&$E_\mathrm{bulk,e,i}$		&$\tfrac{1}{2}m\env v_\mathrm{1,i}^2$		
&$0.20$			&$0.09$		&$0.07$		&$0.03$	\\
Envelope internal		&$E_\mathrm{int,e,i}$		&$4\pi\int_0^{R\1}\tfrac{P}{\gamma-1} r^2dr$	
&$1.81$			&$1.81$		&$0.71$		&$0.71$	\\
Envelope-envelope potential	&$E_\mathrm{pot,e-e,i}$		&$-(4\pi)^2\Gn \int_0^{R\1}\rho(r)r\int_0^r\rho(r') r'^2 dr'dr$	
&$-2.13$	&$-2.13$		&$-0.57$	&$-0.57$\\
Envelope-particle~1 potential	&$E_\mathrm{pot,e-1,i}$		&$-4\pi\Gn m\core\int_0^{R\1}\rho r dr$		
&$-1.56$		&$-1.56$	&$-0.88$	&$-0.88$\\
Envelope-particle~2 potential	&$E_\mathrm{pot,e-2,i}$		&$-\tfrac{\Gn m\2 m\env}{a\init}$		
&$-1.20$		&$-0.54$	&$-0.37$	&$-0.16$\\
\hline
Particle total			&$E_\mathrm{1-2,i}$		&$E_\mathrm{bulk,1,i}+E_\mathrm{bulk,2,i}+E_\mathrm{pot,1-2,i}$		
&$0.26$			&$0.12$		&$0.04$		&$0.02$	\\
Envelope total			&$E_\mathrm{e,i}$		&$E_\mathrm{bulk,e,i}+E_\mathrm{int,e,i}+\sum_j E_\mathrm{e-j,i}$	
&$-2.87$		&$-2.32$	&$-1.05$	&$-0.88$\\
\hline
Total particle and envelope	&$E_\mathrm{1-2-e,i}$		&$E_\mathrm{1-2,i}+E_\mathrm{e,i}$		
&$-2.61$		&$-2.21$	&$-1.01$	&$-0.86$\\
\hline
  \end{tabular}
  \end{center}
\end{table*}

\begin{table}
  \begin{center}
  \caption{The left and right sides of the EF, 
           given by equation~\eqref{alpha} from \citet{Ivanova+13a} 
           for the case where the final inter-particle separation $a\final$ 
           is equal to the mean orbital separation $\approx7\Rsun$ at the end of the simulation at $t=40\da$, 
           and for different assumptions about the initial separation $a\init$.
           The envelope is predicted to be fully unbound when the left-hand and right-hand sides become equal.
           Examination of the entries leads directly to the conclusion that the envelope is not expected to be unbound at $a=7\Rsun$,
           since this would require $\alpha\CE>1$, which is not physical.
           \label{tab:required_alpha}
          }
  \begin{tabular}{lccc}
    \hline
			&$a\init$	&LHS		&RHS$(a\final=7\Rsun)$ 	\\
			&$[\!\Rsun]$	&$[10^{47}\erg]$&$[10^{47}\erg]$	 	\\
\hline                 	 
Eq.~\eqref{alpha}	&$49$		&$1.9$		&$0.2\alpha\CE$			\\
Eq.~\eqref{alpha}	&$109$		&$1.9$		&$0.6\alpha\CE$			\\
\hline
  \end{tabular}
  \end{center}
\end{table}

Equation~\eqref{alpha} only applies if $a\final$ 
corresponds to the inter-particle separation after the envelope is completely unbound.
Since this is \textit{not} the case at $t=t\final$ in the simulation, 
we cannot  use the simulation data
to obtain $\alpha\CE$,
but we can check whether we should \textit{expect} 
the envelope to be unbound at $a=7\Rsun$,
given a reasonable estimate for $\alpha\CE$.

To assess the consistency between simulation results and theoretical expectations, we  evaluated the various energy terms
of Tab.~\ref{tab:energy_terms} at $t=0$ for the envelope alone, excluding the ambient medium.
The values are listed in the fourth column of Tab.~\ref{tab:initial_energy_terms}.
We have verified that small differences between a value from Tab.~\ref{tab:initial_energy_terms} 
and the corresponding value in the fourth column of Tab.~\ref{tab:energy_terms}, 
is accounted for by  energy in the ambient medium.

We next evaluate the left and right sides of equations~\eqref{alpha} 
for $a\init=49\Rsun$ and $a\final=7\Rsun$, which is the approximate mean inter-particle separation at $t=40\da$.
For the RG in our model, $\lambda$ evaluates to $1.31$.
The first and third rows of Tab.~\ref{tab:required_alpha} respectively
show the LHS and RHS of equation~\eqref{alpha} for the simulation.
For the LHS and RHS to be equal, 
$2\le \alpha_\mathrm{CE}\le 5$ would be required.
Since $\alpha\CE>1$ is unphysical, 
we should \textit{not expect} the envelope to be unbound at $a=7\Rsun$,
in agreement with the simulation results.

Realistically, the initial state at $t=t\init$ might be the RLOF stage, 
just prior to CEE \citep[e.g.][]{Nandez+Ivanova16,Macleod+18a,Reichardt+18}.
We can estimate the orbital separation in the RLOF phase
as the Roche-lobe radius \citep{Eggleton83},
\begin{equation}
  \label{Roche}
  r_\mathrm{L}= \frac{0.49q^{2/3}a}{0.6q^{2/3} +\ln(1+q^{1/3})}.
\end{equation}
For the system studied in this work $q=M\1/M\2=2$, 
and this gives $a\init\approx109R_\odot$,
which would reduce $a\init$-dependent terms by more than a factor of two.

Values of the initial energy terms for $a\init=109\Rsun$ are given in the fifth column of Tab.~\ref{tab:initial_energy_terms}.  
The second and fourth rows of Tab.~\ref{tab:required_alpha},
show the LHS and RHS of equation~\eqref{alpha} for this larger initial separation.
Increasing the initial separation somewhat increases the orbital energy 
that can be tapped  thereby reducing $\alpha\CE$,
but the difference from the case where $a\init=49\Rsun$ is small,
and $\alpha\CE>1$ would still be required.
Failure to unbind the envelope at $a=a\final\approx7\Rsun$ is not simply
overcome by starting with $a=a\init=109\Rsun$ instead of $49\Rsun$.
Instead, envelope unbinding requires the binary to tighten to separation
$a\ll 7\Rsun$ in the absence of other energy sources.

\subsection{Predicting the final inter-particle separation}
\label{sec:a_final}

\begin{table}
  \begin{center}
  \caption{Final inter-particle separations $a\final$ predicted 
           by the EF \eqref{alpha} \citep{Ivanova+13a} 
           for various assumed values of $\alpha\CE$.
           Initial conditions involving an RGB primary, 
           with initial separation $a\init$ either slightly greater than the primary radius
           or equal to the Roche limit separation, are considered.
           \label{tab:a_final}
          }
  \begin{tabular}{llcrrrr}
\hline
     	&			&$\alpha\CE$:			    	&$0.1$	&0.25	&0.5	&1	\\
\hline
		&			&$a\init$ ($\!\Rsun$)		&\multicolumn{4}{c}{$a\final$ ($\!\Rsun$)}	\\
\hline
RGB		&Eq.~\eqref{alpha}	&49				&0.3			&0.8	&1.5	&2.6	\\
$\lambda=1.31$	&         		&109				&0.4			&0.9	&1.7	&3.1	\\
\hline
  \end{tabular}
  \end{center}
\end{table}

To predict $a\final$ 
for a given value of $\alpha\CE$,
we can use equation~\eqref{alpha} 
with either $a\init=49\Rsun$ (simulation) or $a\init=109\Rsun$ (Roche limit). 
The values are given in the top half of Tab.~\ref{tab:a_final} 
for values of $\alpha=0.1$, $0.25$, $0.5$ and $1$.
Tab.~\ref{tab:a_final} tells us that we cannot expect envelope ejection 
until $a$ has reduced to less than $3\Rsun$, and likely less than $1\Rsun$.
This is much smaller than the final separation in our simulation and that of \citet{Ohlmann+16a},
who used very similar initial conditions but evolved the system to $t\sim130\da$, at which time $a\approx4\Rsun$.
It is therefore consistent with the theory, 
that the envelope did not eject in the simulation of \citet{Ohlmann+16a} either.

This analysis shows that envelope ejection requires the binary separation to reduce further.
This possibility was considered in Sec.~\ref{sec:ejection}, 
where it was pointed out that even at the end of the simulation at $t=40\da$,
energy was being transferred from particles to gas at an almost steady average rate (see Fig.~\ref{fig:energy_time}),
and that if this were to continue to late times, the envelope might be ejected by $\sim10^2$--$10^3\da$,
still small enough to account for observations of PPNe, which have ages $>10^2\yr$.
The orbital separation does \textit{appear}, from Fig.~1 of \citet{Ohlmann+16a}, 
to be approaching an asymptotic value,
while the energy transfer rate reduces with time (their Fig.~2).
Therefore, running the simulation longer might not  lead to envelope unbinding.
They estimate that it would take $\sim100\yr$ to eject the envelope if unbinding were to continue at the final rate.

\citet{Clayton+17}, using idealized 1D MESA CEE simulations which include radiative transport and shock capture, 
argue that envelopes can be ejected on timescales of $\sim10^3\yr$. 
In their models, pulsations develop, some of which lead to the
dynamical ejection of shells containing up to $\sim10\%$ of the envelope mass.
Whether  these long ejection timescales are in tension with observations is presently uncertain.
In general, it is important to assess
what other additional physics could be included that would better facilitate envelope ejection.
    
\section{Summary and conclusions}
\label{sec:conclusions}
This work can be  divided into three main parts.
In the first part (Sections~\ref{sec:energy_budget} and \ref{sec:unbinding}), 
we analyze the energy budget in our simulation of CEE.
The key findings are as follows:
\begin{itemize}
  \item As with previous work, the CEE  can  be divided into 
        a plunge-in phase whose termination approximately coincides with the first periastron passage,
        and a slow spiral-in phase.
        The transition between phases occurs when the secondary and its trailing tidal tail
    collide with the posterior tidal bulge of the primary.
        Further analysis of the orbital dynamics, including a measurement of the drag force 
        and comparison with theory, is warranted.
  \item There is  little net energy transfer between the orbital energy of particles (giant core and companion)
        and the binding energy of gas through the end of the dynamical plunge-in phase ($t=0$ to $13\da$), 
        but the transfer is sufficient to unbind $14\%$ of the envelope by $t=13\da$.
        This is because the secondary gains a stronger hold on material in the inner envelope 
        but energizes and ejects material in the outer envelope. 
  \item Conversely, after the plunge-in until the end of the simulation ($t=13$ to $40\da$), 
        energy is steadily transferred from particles to gas 
        but little gas from the initially more tightly bound inner layers becomes unbound. 
        It remains inconclusive as to whether a much longer run would lead to further unbinding.
  \item We find that the choice of ambient medium is important 
        in determining the slope of the change in unbound mass with time after plunge-in.
        This is not merely a numerical issue, but highlights the importance
        of the interaction between the envelope and its environment 
        in determining how much mass gets unbound.
\end{itemize}

In the second part of the study (Sec.~\ref{sec:particle_CM_motion})
we explored the relative motion of the centres of mass of the particles and gas:
\begin{itemize}
  \item We calculate the relative motion of the particles CM and that of the envelope 
        resulting from gas ejection near the secondary as it spirals inward.
        At the end of the simulation the particle CM
        moves steadily at $4\kms$ with respect to the simulation frame
        and $8\kms$ in the envelope CM frame, almost parallel to the orbital plane. 
        This motion does not drastically change the level of unbinding compared
        to ignoring it but offers an explanation for the previously unexplained 
        observed offsets between PN binary central stars and the geometric centres of their nebulae.
\end{itemize}

In the third part of the study (Sec.~\ref{sec:energy_formalism}) 
we compare the theory to simulations and discuss the limits of simulations.
The key points are:
\begin{itemize}
  \item We show that to eject the envelope with $\alpha\CE\le0.25$, 
        we would, for our ZAMS $2\Msun$ RGB star + $1\Msun$ secondary system, 
        need to evolve the simulation to a final inter-particle separation $a\final\le0.9\Rsun$, 
        which is currently inaccessible to simulations due to numerical limitations. 
  \item That the envelope remains mostly bound at the end of our simulation 
        agrees with our theoretical expectation, 
        given the physics and numerical parameters of the simulation. 
\end{itemize}
While we cannot say for sure what much longer simulations will bring, 
exacerbating envelope unbinding may benefit from the following: 
(1) additional energy sources (e.g. recombination energy, accretion energy),
or (2) improved numerical reliability at low inter-particle separations; 
(3) improved realism of energy transport in the stellar models 
that may diffusively redistribute the envelope energy 
such that a larger fraction of mass has just enough energy to remain unbound.
Convection is the most likely possible contributor to the latter and is known to occur in RGB and AGB stars, 
though it remains to be seen whether the associated change in the unbinding efficiency 
(as embodied in the parameter $\alpha\CE$)
would be toward enhanced or reduced efficiency.
In any case, developing CE simulations that include convection is critical in our view,
and the most natural and direct path toward this goal would involve implementing more realistic equations of state
such that the temperature gradients of the giant stars are faithfully reproduced \citep{Ohlmann+17}.

Many challenges remain for simulating CEE.
Progress will require improvements in the numerics (initial conditions, resolution, refinement strategy)
as well as the inclusion of more physics (convection, radiative transport, recombination, jet feedback).

\section*{Acknowledgements}
We thank Orsola De~Marco, Paul Ricker, Sebastian Ohlmann and Thomas Reichardt for stimulating and helpful discussions.  
We thank Brian Metzger, Brent Miszalski and Noam Soker for comments.
We thank the referee for comments that led to  improvements in the manuscript.
We acknowledge support form National Science Foundation (NSF) grants AST-1515648 and AST-1813298.
This work used the Extreme Science and Engineering Discovery Environment (XSEDE), 
which is supported by   NSF grant number ACI-1548562. 
The authors acknowledge the Texas Advanced Computing Center (TACC) at The University of Texas at Austin for providing HPC resources 
(through XSEDE allocation TG-AST120060) that have contributed to the research results reported within this paper.

\footnotesize{
\noindent
\bibliographystyle{mnras}
\bibliography{refs}
}

\appendix

\section{Comparison with previous work}
\label{sec:ohlmann_comparison}

\begin{figure}
  \includegraphics[width=\columnwidth,clip=true,trim= 0 0 0 0]{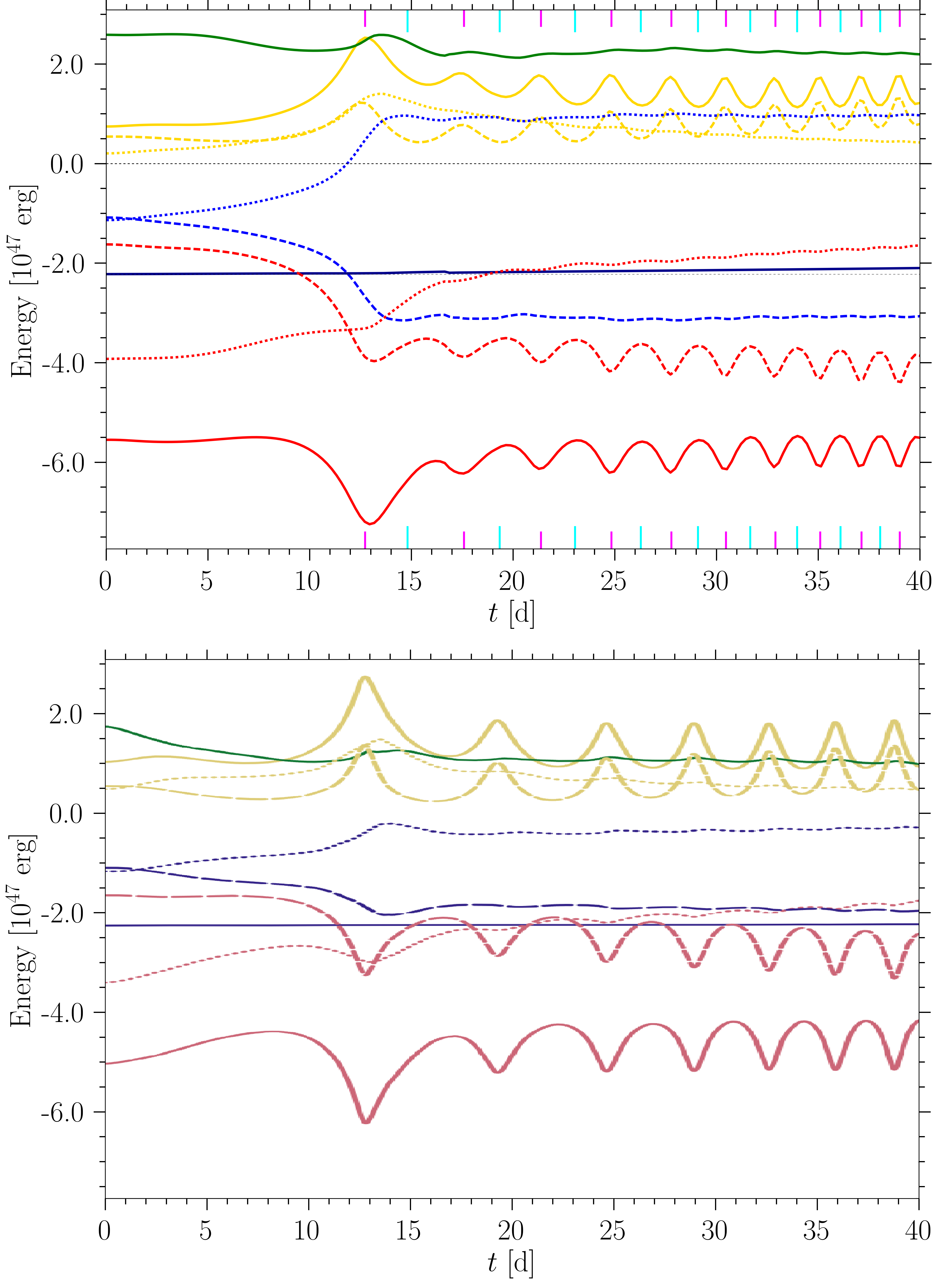}
  \caption{Comparison between energy terms (integrated over the simulation domain) in our simulation (top)
           and in the simulation of \citet{Ohlmann+16a} (bottom, adapted from the latter work).
           Legend labels are the same as those of \citet{Ohlmann+16a}:
           `total' (solid blue), `kinetic' (solid yellow), `potential' (solid red), `internal' (solid green),
           `total envelope' (dotted blue), `kinetic envelope' (dotted yellow), `potential envelope' (dotted red),
           `total cores' (dashed blue), `kinetic cores' (dashed yellow) and `potential cores' (dashed red).
           In the upper panel the contribution from $E_\mathrm{pot,gas-1}$ is included 
           in the envelope potential energy and total envelope energy
           but the contribution from $E_\mathrm{pot,gas-2}$ is not so included.
           On the other hand, the contribution from $E_\mathrm{pot,gas-2}$ 
           is included in the cores potential energy and total cores energy, 
           but these terms do not include the contribution from $E_\mathrm{pot,gas-1}$.
           \label{fig:energy_time_Ohlmann}
          }            
\end{figure}

We used almost the same parameter values and initial conditions as \citet{Ohlmann+16a} and thus it is useful
to compare directly their results and ours.
In the top panel of Fig.~\ref{fig:energy_time_Ohlmann} we plot the energy terms as in Fig.~2 of \citet{Ohlmann+16a},
and in the bottom panel we show a version of their figure with the time axis truncated at $t=40\da$.
The curves are as described in the legend but \citet{Ohlmann+16a} used a different kind of code
and it was not entirely clear to us precisely how the different energy components were divided among the various curves.
We found that close agreement was obtained if
the curves labeled as envelope potential energy and total envelope energy (dotted red and dotted blue, respectively) 
include the contribution from $E_\mathrm{pot,gas-1}$ but not from $E_\mathrm{pot,gas-2}$,
and the curves labeled as cores potential energy  and total cores energy (dashed red and dashed blue, respectively) 
include the contribution from $E_\mathrm{pot,gas-2}$ but not from $E_\mathrm{pot,gas-1}$, 

The \citet{Ohlmann+16a} setup allowed for a much lower pressure and lower density ambient medium.
Thus, to make a direct comparison with our simulation, it was necessary to subtract from each energy term the fraction
contributed by the ambient medium (or by the gravitational interaction between the ambient medium and the other components); 
these quantities involving the ambient gas were assumed to remain constant for the duration of the simulation.
The ambient energy inflow from the boundaries is measured to be negligible.
As expected from the analysis of \citetalias{Chamandy+18}, close agreement between results from the two simulations is apparent,
in spite of the very different methodologies used.
There is, however, a larger separation between the total particle energy and total gas energy curves 
(shown in dashed blue and dotted blue, respectively)
after plunge-in in the top panel of Fig.~\ref{fig:energy_time_Ohlmann} as compared with the bottom panel. 
The particle and gas potential energies also experience larger changes during plunge-in (dashed and dotted red, respectively).

Assuming that our partitioning of the energy components mimics reasonably well that of \citet{Ohlmann+16a}, 
these differences could be caused by differences in initial conditions. 
Firstly, in our simulation the RG is not rotating with respect to the inertial frame of reference of the simulation,
while in that of \citet{Ohlmann+16a} the RG is initialized with a solid body rotation of $95\%$ of the orbital angular speed.
(The reality would lie somewhere in between and can be estimated 
as $\sim30\%$ of the orbital angular speed from the results of \citealt{Macleod+18a}).
In spite of this difference, however, the inter-particle separation $a$ reaches a smaller value ($<10\Rsun$) at the first periastron passage
in the simulation of \citet{Ohlmann+16a} than in that of \citetalias{Chamandy+18} ($14\Rsun$), 
even though the time of this first periastron passage (i.e. the end of plunge-in, as we have defined it) 
occurs at about $t=13\da$ in both simulations.

Secondly, \citet{Ohlmann+16a} performed a relaxation run to set up their initial condition, while we did not,
which would have led to differences in the initial stellar profiles 
(apart from the slight differences that would have already existed due to the slightly different \textsc{mesa} models employed).
We note that some quantities, like internal energy (solid green) and total potential energy (solid red) 
remain approximately constant for the first $\sim5\da$ in our simulation, while showing more variation in that of \citet{Ohlmann+16a}.
This suggests that the RG is more stable in our simulation.
Possible reasons are that we iterated over the RG core mass to obtain a smoother initial RG profile,
we resolved the entire RG at the highest refinement level,
and we used a denser and higher pressure ambient medium to stabilize the outer layers of the RG.
The latter is a compromise since a larger ambient density and pressure complicates the analysis.
Clearly, obtaining an initial condition that is both highly stable and physically realistic in CE simulations is computationally challenging.
In any case, we are encouraged by the close agreement between the two simulations,
and take this as confirmation that our results are physical, as opposed to being dominated by numerical artefacts.

\section{Comparison of 1D and 3D stellar profiles}
\label{sec:profile_comparison}

\begin{figure}
  \includegraphics[width=\columnwidth,clip=true,trim= 0 0 0 0]{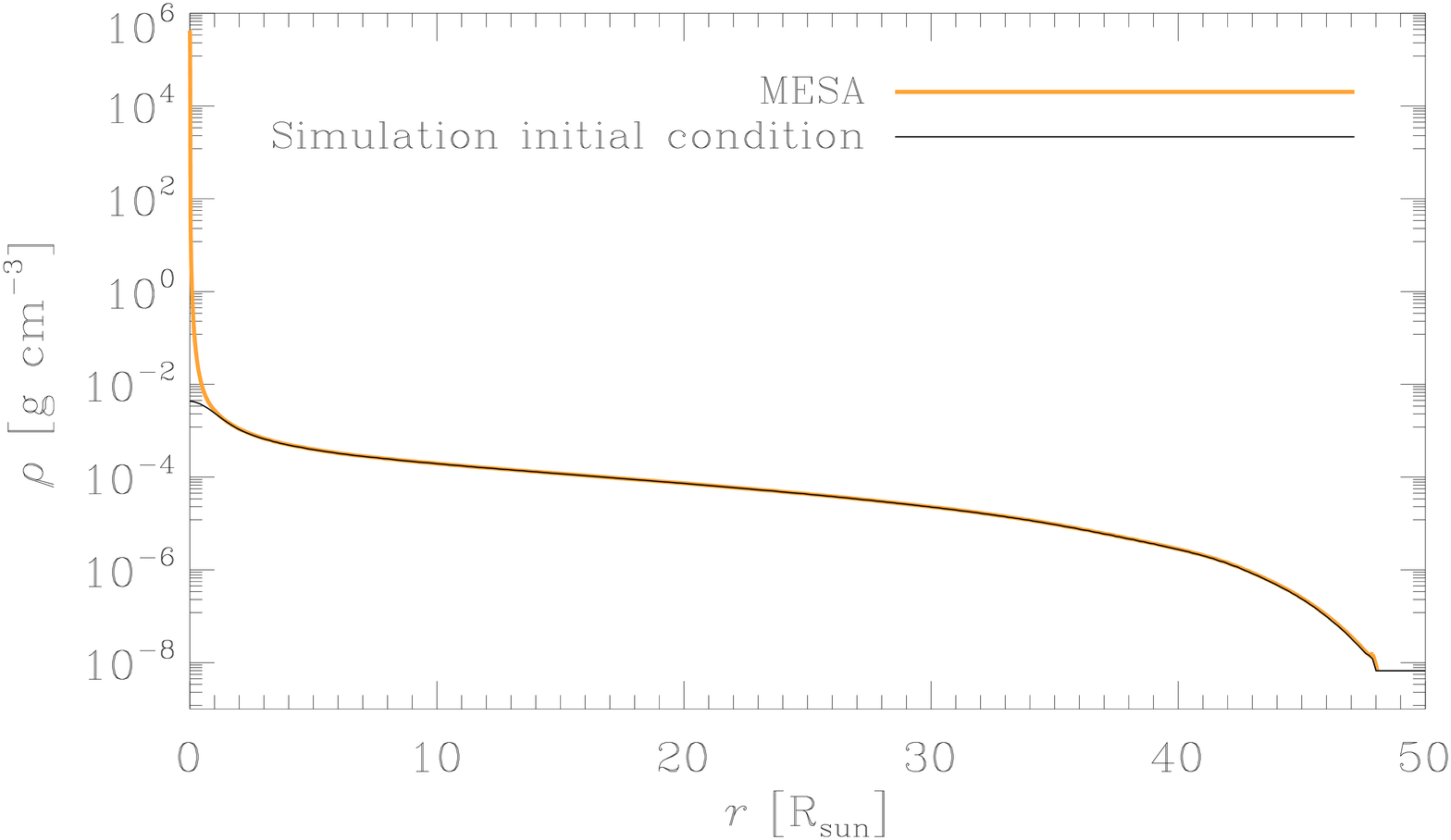}\\
  \includegraphics[width=\columnwidth,clip=true,trim= 0 0 0 0]{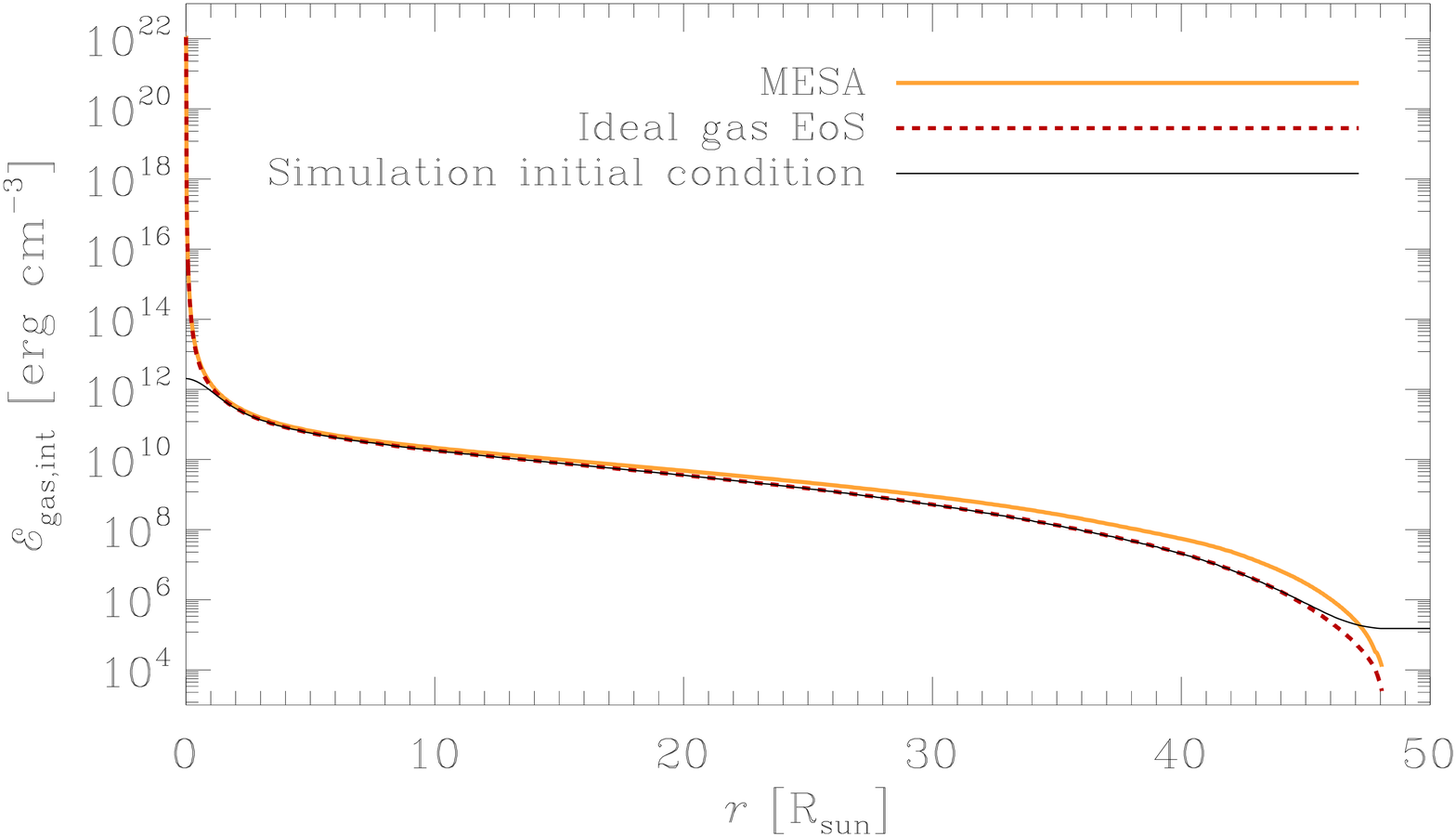}\\
  \includegraphics[width=\columnwidth,clip=true,trim= 0 0 0 0]{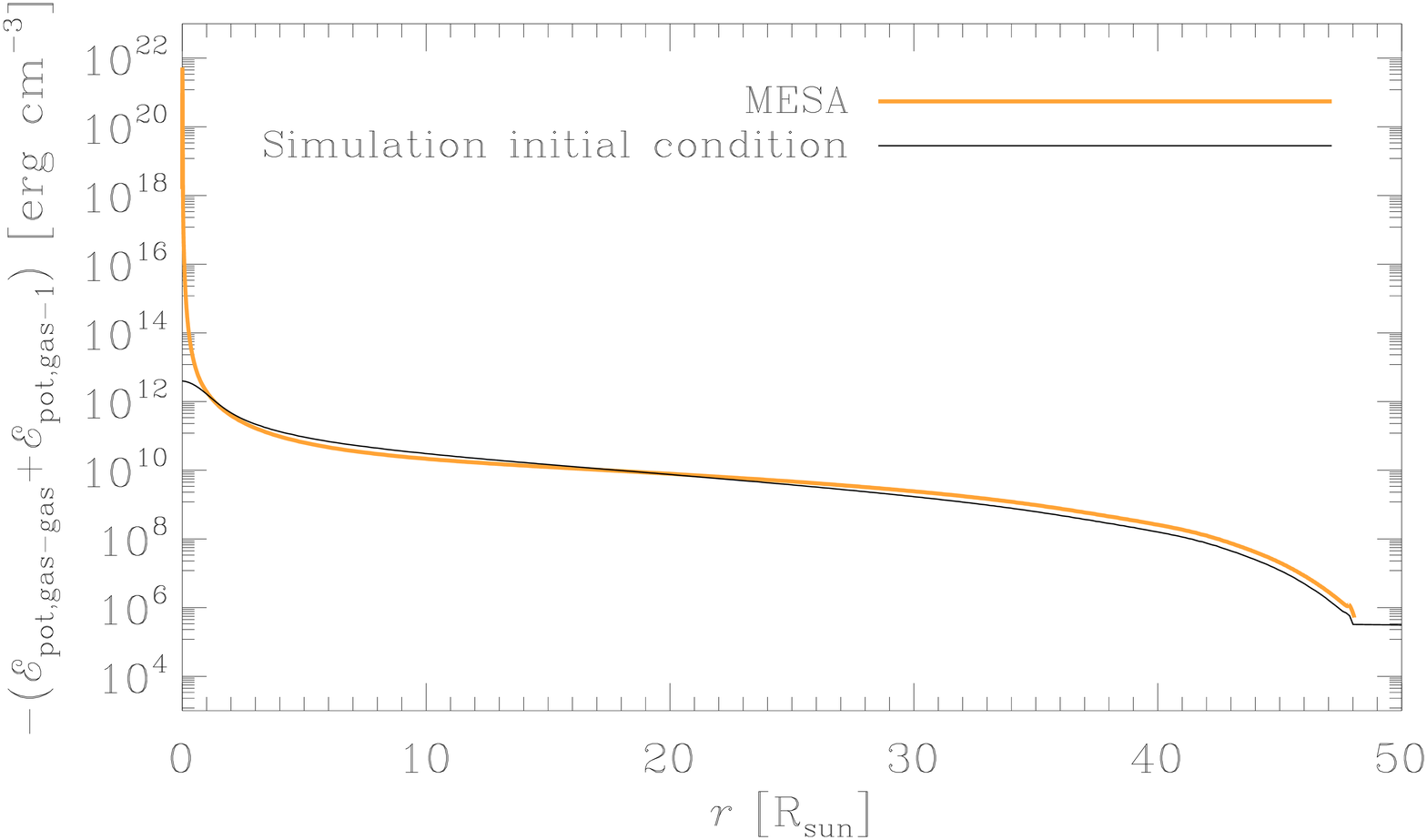}\\
  \caption{\textit{Top panel:} Radial profiles of gas density for the RGB star modeled using MESA (thick orange)
           and the profile of our 3D RGB star in the simulation at $t=0$ (thin black).
           \textit{Middle panel:} Comparison of the internal energy density profiles in the MESA model,
           simulation initial condition, and MESA model with the equation of state replaced with an ideal gas equation of state,
           as in the simulation (dashed red).
           \textit{Bottom panel:} Comparison of (negative of) potential energy density profiles 
           in the MESA model and simulation initial condition.
           \label{fig:profile_comparison}
          }            
\end{figure}

In Fig.~\ref{fig:profile_comparison} we compare the MESA solution with the profile
obtained from a radial cut of our initial RGB star at $t=0$, taken directly from the AstroBear mesh.
From top to bottom, panels show the gas density, internal energy density and potential energy density
(the latter is equal to  $\mathcal{E}_\mathrm{pot,gas-gas}+\mathcal{E}_\mathrm{pot,gas-1}$,
and thus excludes the contribution from the primary-secondary gravitational interaction 
as well as that from ambient-ambient self-gravity).
For the internal energy, we have also plotted the profile that would obtain 
if the MESA equation of state (EoS) had been that of an ideal gas with $\gamma=5/3$
(the case $\mathcal{E}_\mathrm{int,gas}=\tfrac{3}{2}P\gas$).
The agreement between our initial condition and the MESA model is very good for $r>1\Rsun$.
Our density and pressure were designed to match the MESA profile for $r>r\soft=2.4\Rsun$ \citep[see also][]{Ohlmann+17}.
Thus, internal energy density closely matches the MESA profile if the EoS is replaced by an ideal gas EoS. 
Near the stellar surface, the AstroBear internal energy density profile transitions smoothly to the ambient value.
The profile of $\mathcal{E}_\mathrm{int,gas}$, made by assuming an ideal gas EoS, is slightly lower than 
the actual MESA profile, especially at larger radius.
The small differences between the internal energy density profiles for the actual MESA model
and that assuming an ideal gas EoS can be attributed to convective energy.

The potential energy profile of our initial condition also much agrees  with the MESA profile,
being slightly larger in magnitude  for $1\lesssim r/\!\Rsun\lesssim19$, 
and slightly smaller at  larger radii.

To assess how unbound our envelope is compared with that in the MESA model,
we compute the total internal energy $E_\mathrm{int,gas}$ 
as well as the total energy $E_\mathrm{int,gas}+E_\mathrm{pot,gas-gas}+E_\mathrm{pot,gas-1}$
by integrating the profiles from $r\soft=1\Rsun$ to the stellar surface $R\1=48.1\Rsun$.
For the MESA model we obtain $E_\mathrm{int,gas}= 2.32\times10^{47}\erg$ 
and $E_\mathrm{int,gas}+E_\mathrm{pot,gas-gas}+E_\mathrm{pot,gas-1}= -1.29\times10^{47}\erg$.
For the MESA model with ideal EoS we obtain $E_\mathrm{int,gas}= 1.78\times10^{47}\erg$
and $E_\mathrm{int,gas}+E_\mathrm{pot,gas-gas}+E_\mathrm{pot,gas-1}= -1.83\times10^{47}\erg$.
Finally, for our initial condition we obtain $E_\mathrm{int,gas}= 1.76\times10^{47}\erg$
and $E_\mathrm{int,gas}+E_\mathrm{pot,gas-gas}+E_\mathrm{pot,gas-1}= -1.92\times10^{47}\erg$.
Thus, the net energy terms in our initial condition closely match those of the MESA model
if the EoS is replaced by the ideal gas EoS, as in our setup.

If, for example, we adopt $r=1\Rsun$ as the location of the core-envelope boundary, 
the envelope in our initial condition is more strongly bound than that of the actual MESA model
by about $0.6\times10^{47}\erg$.
This difference is insensitive to the location of the core-envelope boundary for $r>0.1\Rsun$.
This suggests that the unbound mass obtained in our simulation may be somewhat smaller than
that which would be obtained with a more realistic initial condition.

\section{Convergence tests}
\label{sec:convergence}

To test the dependence of the results on the parameters used in our numerical setup we performed five additional runs.
The parameter values of the runs, including the fiducial run (Model~A), are listed in Tab.~\ref{tab:runs}.

Models~B and C were designed to test the dependence of the results 
on the radius $r\refine$ of the spherical region around the particles that was refined at the highest resolution $\delta$.
Both runs employ smaller values of $r\refine$  than that of the fiducial run at any given time.
The orbit of Model~B is remarkably similar to that of Model~A until $t=16.7\da$, when $r\refine$ becomes particularly small compared to Model~A
(and perhaps slightly before that time),
but the period increases only slightly as a result.
In Model~C, the simulation is restarted from a frame of Model~A at $t=24.1\da$ and runs for $3.0\da$
with a smaller value of $r\refine$.
The orbit is almost identical to that of Model~A.
These results suggest that the refinement region in the fiducial run is somewhat larger than what is actually required,
and we could afford to be less conservative in the future.

Models~D-F were designed to test the effects of changing the softening length $r\soft$ 
(always equal for both particles) and/or the smallest resolution element $\delta$.
The results for the orbital separation $a$ as a function of time are shown in Fig.~\ref{fig:r_soft}.
In Model~A, both the softening length $r\soft$ and the smallest resolution element $\delta$ 
are halved at $t=16.7\da$.
In Model~D (dashed-dotted), the softening length is halved at $t=16.7\da$ as in Model~A,
but the size of the smallest resolution element is kept constant.
Before $t\approx22\da$, the orbit matches closely to that of Model~A.
Thereafter, the amplitude of the separation curve reduces compared to that of Model~A,
and the mean separation increases at late times, while the period also slightly increases.
By contrast, in Model~E (dashed), where $\delta$ is halved as in Model~A but $r\soft$ is not, 
the amplitude is slightly larger than in Model~A while the period and mean separation
deviate only modestly from those of Model~A (though Model~E was terminated earlier than the other runs).
Finally, in Model~F (dotted) neither $r\soft$ nor $\delta$ is halved, unlike in Model~A, where both are halved.
This run leads to an orbit that deviates from Model~A more than Model~E but less than Model~D.
Model~F has a smaller period than Model~A but a comparable amplitude,
while the mean separation at late times is slightly smaller than for Model~A.

Combining the results from Models~A, D, E and F, we  infer that:
(i) at later times (and smaller separations), 
the mean separation and separation amplitude are sensitive to the number of resolution elements per softening length.
A smaller value of $r\soft/\delta$ (e.g. $\sim9$ in Model~D compared to $\sim17$ in Model~A) 
can lead to a larger mean separation, smaller separation amplitude, and larger period;
(ii) For a given value of $r\soft/\delta$, larger values of $r\soft$ and $\delta$ 
tend to reduce the period and reduce the mean separation slightly,
while hardly affecting the amplitude. 
We caution, however, that effects of changing the softening length cannot be attributed to numerics only
because decreasing the softening length increases the gravitational force near the particles,
which can result in a higher concentration of mass around the particles \citepalias{Chamandy+18}.

\begin{table}
  \begin{center}
  \caption{Parameters of the various models. 
           Model~A is the main model studied in this work while Models~B through F are for testing.
           Parameters are the start or restart time of the simulation $t\init$, 
           the end time of the simulation $t\final$,
           the spline softening length $r\soft$,
           the size of the resolution element at the highest refinement level $\delta$,
           and the radius of the spherical volume within which cells are refined at the highest level $r\refine$. 
           For $t\le16.7\da$, this sphere is centred on 
           particle~1 while for $t>16.7\da$ it is centred on particle~2 (in all models).
           For Models~A and B, values of $r\soft$ and $\delta$ are given for $t>16.7\da$.
           For $t\le16.7\da$ $r\soft=2.4\Rsun$ and $\delta=0.14\Rsun$ for those models.
           \label{tab:runs}
          }
  \begin{tabular}{@{}ccccccc@{}}
    \hline
    Model  &$t\init$    &$t\final$  &$r\soft$     &$\delta$       &$r\soft/\delta$  &$r\refine$    \\
           &$[\!\da]$   &$[\!\da]$  &$[\!\Rsun]$  &$[\!\Rsun]$    &                 &              \\
    \hline
    A      &$0$         &$40.0$     &$1.2$        &$0.07$         &$17$           &Fiducial      \\
    B      &$0$         &$22.0$     &$1.2$        &$0.07$         &$17$           &Small         \\
    C      &$24.1$      &$27.1$     &$1.2$        &$0.07$         &$17$           &Small         \\
    D      &$16.7$      &$35.6$     &$1.2$        &$0.14$         &$9$            &Fiducial      \\
    E      &$16.7$      &$32.4$     &$2.4$        &$0.07$         &$34$           &Fiducial      \\
    F      &$16.7$      &$37.3$     &$2.4$        &$0.14$         &$17$           &Fiducial      \\
    \hline
  \end{tabular}
  \end{center}
\end{table}

\begin{figure}
  \includegraphics[width=\columnwidth,clip=true,trim= 0 0 0 0]{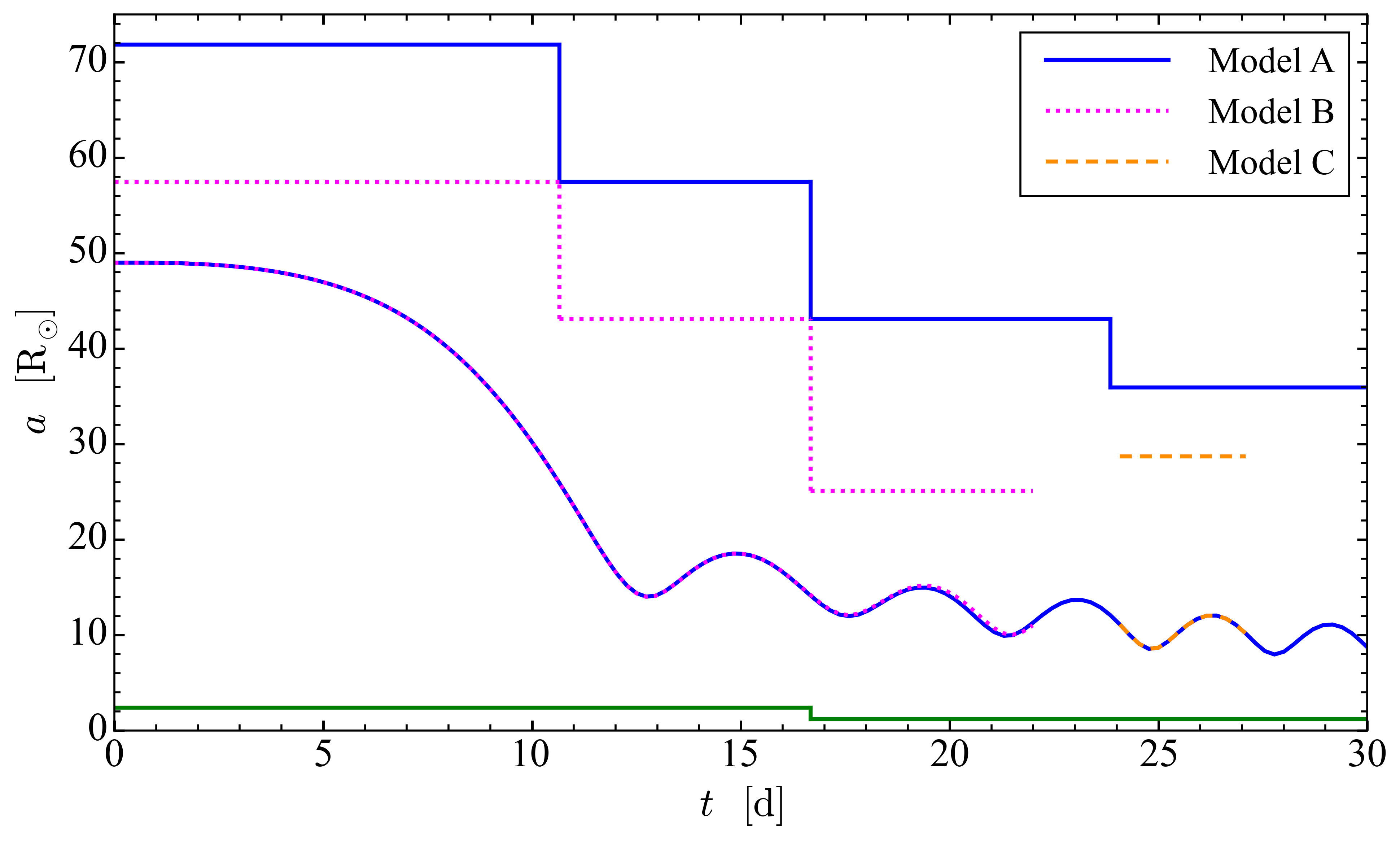}
  \caption{Orbital separation $a$ between particles~1 and 2,
           now plotted along with the radius $r\refine$ of the spherical volume refined at the highest level
           (jagged line with colour the same colour as the separation curve for that run).
           Plotted are the fiducial run, Model~A (solid blue),
           as well as Models~B (dotted magenta) and C (dashed orange), 
           which both have smaller refinement radius $r\refine$ than Model~A. 
           The softening radius $r\soft$ (same for all runs) is also shown, for reference (solid green).
           \label{fig:r_refine}
          }            
\end{figure}

\begin{figure}
  \includegraphics[width=\columnwidth,clip=true,trim= 0 0 0 0]{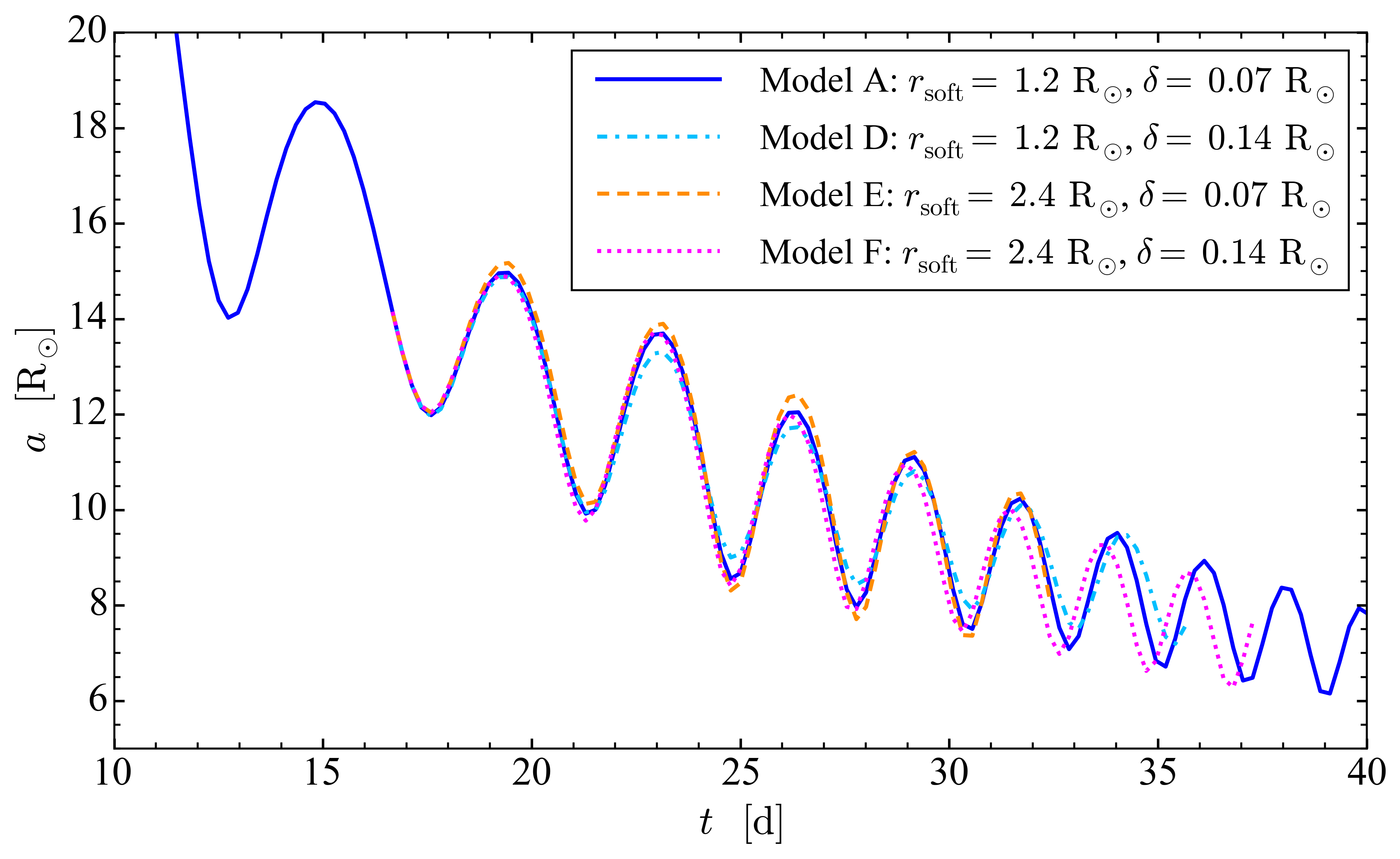}
  \caption{Orbital separation $a$ between particles~1 and 2 for the fiducial run, Model~A (solid blue), 
           Model~D, restarted from Model~A from $t=16.7\da$ but without halving $\delta$ (dashed-dotted light blue),
           Model~E, restarted from Model~A from $t=16.7\da$ but without halving $r\soft$ (dashed orange),
           and Model~F, restarted from Model~A from $t=16.7\da$ but without halving either $\delta$ or $r\soft$ (dotted magenta).
           \label{fig:r_soft}
          }            
\end{figure}

\end{document}